\documentclass{aa}

\newcommand{\mc}[3]{\multicolumn{#1}{#2}{#3}}

\usepackage[varg]{txfonts}
\usepackage{color}
\usepackage{multirow}
\usepackage{hyperref}
\usepackage{enumitem}
\usepackage{nicefrac}
\usepackage{caption}
\usepackage{subcaption}
\hypersetup{
    colorlinks = true,
    allcolors = {blue},
}

\begin{document}

\title{Full-Stokes polarimetry with circularly polarized feeds} 

\subtitle{Sources with stable linear and circular polarization in the GHz regime}

\author{I. Myserlis\inst{1}
 \and E. Angelakis\inst{1}
 \and A. Kraus\inst{1}
 \and C. A. Liontas\inst{2}
 \and N. Marchili\inst{3}
 \and M. F. Aller\inst{4}
 \and H. D. Aller\inst{4}
 \and\\ V. Karamanavis\inst{1}
 \and L. Fuhrmann\inst{1}
 \and T. P. Krichbaum\inst{1}
 \and J. A. Zensus\inst{1}
}


\institute{Max-Planck-Institut f\"ur Radioastronomie, Auf dem Huegel 69, 53121, Bonn, Germany
               \and Fraunhofer Institute for High Frequency Physics and Radar Techniques FHR, Fraunhoferstraße 20, 53343, Wachtberg, Germany
               \and IAPS-INAF, Via Fosso del Cavaliere 100, 00133, Roma, Italy
               \and Department of Astronomy, University of Michigan, 311 West Hall, 1085 S. University Avenue, Ann Arbor, MI:48109, USA               }

\date{Received / Accepted }

\abstract{We present an analysis pipeline that allows recovering reliable information for all four Stokes 
  parameters with high accuracy. Its novelty relies on the effective treatment of the instrumental 
  effects already prior to the computation of the Stokes parameters contrary to conventionally used methods, 
  such as the M\"uller matrix one. For instance, the instrumental linear polarization is corrected across the 
  whole telescope beam and significant Stokes $Q$ and $U$ can be recovered even when the recorded signals are 
  severely corrupted by instrumental effects. The accuracy we reach in terms of polarization degree is of the order 
  of 0.1--0.2~\%. The polarization angles are determined with an accuracy of almost 1\degr. The presented methodology 
  was applied to recover the radio linear and circular polarization of around 150 Active Galactic Nuclei. The sources 
  were monitored between July 2010 and April 2016 with the Effelsberg 100-m telescope at 4.85~GHz and 8.35~GHz with a 
  cadence of around 1.2 months. The polarized emission of the Moon was used to calibrate the polarization angle 
  of the monitored sources. Our analysis showed a small system-induced rotation of about 1\degr at both observing frequencies.
  Finally, we find five sources with significant and stable linear polarization; three sources
  remain constantly linearly unpolarized over the period we examined; a total of 11 sources have stable circular
  polarization degree $m_\mathrm{c}$ and four of them with non-zero $m_\mathrm{c}$. We also identify eight sources that
  maintain a stable polarization angle over the examined period. All this is provided to the community for
  future polarization observations reference. We finally show that our analysis method is conceptually
  different from the traditionally used ones and performs better than the M\"uller matrix method. Although it has 
  been developed for a system equipped with circularly polarized feeds it can easily be generalized to 
  systems with linearly polarized feeds as well.}


\maketitle

\section{Introduction}
\label{sec:intro}
  As an intrinsic property of non-thermal emission mechanisms, the linear and circular polarization from
  astrophysical sources carry information about the physical conditions and processes in the radiating regions
  \citep[e.g.][]{Laing1980a,Wardle1998, Homan2009}. Propagation through birefringent material -- such as the
  intergalactic and interstellar medium -- can further generate, modify, or even eliminate the polarized part
  of the transmitted radiation \citep[e.g.][]{Pacholczyk1970a,Jones1977a,Huang2011}. Consequently, processes
  that introduce variability in the emitting or transmitting regions induce dynamics in the observed
  polarization parameters \citep[e.g.][]{Marscher2008,Myserlis2014}. Although these processes increase the
  complexity, they carry information about the mechanisms operating at the emitting regions.

  The measured degree of linear and especially circular polarization of extragalactic sources in the
  radio window, is usually remarkably low. \citet{Klein2003} studied the B3-VLA sample in the range from 2~GHz
  to 10~GHz to find that the average linear polarization degree ranges from $\sim3.5$~\% to
  5~\%. \citet{Myserlis2015} studied the circular polarization of almost 45 blazars in the GHz regime and
  found a population median of around 0.4~\%. 
  Consequently, despite its importance, the reliable detection of polarized emission is particularly
  challenging especially when propagation and instrumental effects as well as variability processes are
  considered.

  In the following, we present a pipeline for the reconstruction of the linear and circular polarization
  parameters of radio sources. The pipeline includes several correction steps to minimize the effect of 
  instrumental polarization, allowing the detection of linear and circular polarization degrees as low as 0.3~\%.
  The instrumental linear polarization is calculated across the whole telescope beam and hence it can be corrected 
  for the observations of both point-like and extended sources. The methodology was developed for the 4.85~GHz 
  and 8.35~GHz receivers of the 100 m Effelsberg telescope. Although these systems are equipped with circularly 
  polarized feeds, our approach can be easily generalized for telescopes with linearly polarized feeds, as well.

  The consistency of our method is tested with the study of the most stable sources in our sample (in terms of
  both linear and circular polarization). Their stability indicates that physical conditions such as the 
  ordering, magnitude or orientation of their magnetic field remain unchanged over long timescales.
  The corresponding polarization parameters are reported for the calibration of polarization observations. We
  report both polarized and randomly polarized (unpolarized) sources. Conventionally, the latter are used for
  estimating the instrumental effects and the former to calibrate the data sets and quantify their
  variability.

  The paper is structured as follows. In Section~\ref{sec: observations} an introduction to the technical
  aspects of the observations is presented. In Section~\ref{sec: FullS_polarimetry} we present the methodology
  we developed to extract the polarization parameters with high accuracy. Our approach relies mainly on the 
  careful treatment of the instrumental linear and circular polarization, discussed in 
  Sect.~\ref{subsec:instrument model} and \ref{subsec:LR_gain_corr}, respectively, as well as the correction 
  of instrumental rotation presented in Sect.~\ref{subsec:instrumental_evpa}. In 
  Section~\ref{sec:M-matrix_comparison} we perform a qualitative comparison between our method and the 
  M\"uller matrix one. In Section~\ref{sec:standards_polarization} we describe the statistical analysis 
  of the results obtained with our methodology and report on sources with stable linear and circular 
  polarization. Finally, a discussion and the conclusions of our work are presented 
  in Section~\ref{sec:discussion}.

  Throughout the manuscript we use the conventions adopted by Commissions 25 and 40 at the 15th
  General Assembly of the IAU in 1973:
\begin{enumerate}[label=\alph*)]
\item the reference frame of Stokes parameters $Q$ and $U$ is that of right ascension and declination with the
  polarization angle starting from north and increasing through east, and
\item positive circular polarization measurements correspond to right handed circular polarization,
\end{enumerate}
This circular polarization convention is also in agreement with the Institute of Electrical and Electronics
Engineers (IEEE) standard, according to which the electric field of a positive or right handed circularly
polarized electromagnetic wave rotates clockwise for an observer looking in the direction of propagation
\citep{IEEEpol}.

\section{Observations}
\label{sec: observations}

  The data set we discuss here was obtained with the Effelsberg 100 m telescope at 4.85~GHz and 8.35~GHz. 
  The corresponding receivers are equipped with circularly polarized feeds (Table~\ref{tab:receivers}). 
  A short description of the Stokes parameter measurement process using such systems is provided in 
  Sect.~\ref{subsec:system description}.
  
  The data set covers the period between July 2010 and April 2016. Until January 2015 the
  observations were conducted within the framework of the F-GAMMA monitoring
  program\footnote{\url{http://www.mpifr-bonn.mpg.de/div/vlbi/fgamma}} \citep{Fuhrmann2016}; beyond January
  2015, data were obtained as part of multi-frequency monitoring campaigns on selected sources. The median
  cadence is around 1.2 months.  The average duration of the observing sessions is 1.3 days.

\begin{table}[!ht]
  \caption{Receiver characteristics.}     
  \label{tab:receivers}  
  \centering                    
  \begin{tabular}{lccc} 
    \hline\hline                 
    Receiver & &4.85 GHz &8.35 GHz \\ 
    \hline\\
    Bandwidth           &(GHz)     &0.5         &1.1      \\       
    System Temperature  &(K)       &27          &22       \\
    FWHM                &(arcsec)  &146         &81       \\
    Number of feeds     &          &2           &1        \\
    Polarization        &          &LCP, RCP    &LCP, RCP \\
    Sensitivity         &(K/Jy)    &1.55        &1.35     \\
    \\\hline                                  
  \end{tabular}
\end{table}

  The observations were conducted with ``cross-scans'', that is by slewing the telescope beam over the
  source position in two perpendicular directions. For the data set considered here those passes (hereafter
  termed ``sub-scans'') were performed along the azimuth and elevation directions. The advantage of the
  cross-scan method is that it allows correcting for the power loss caused by possible telescope pointing
  offsets. A detailed description of the observing technique is given by \citet{2015A&A...575A..55A}.
 
  In Sect.~\ref{sec: FullS_polarimetry} we give a detailed description of the methodology followed for
  the reconstruction of the total flux density $I$, the degree of linear and circular polarization
  $m_{\mathrm{l}}$ and $m_{\mathrm{c}}$, and the polarization angle, $\chi$. The number of sources with at
  least one significant measurement (signal-to-noise ratio, SNR$\ge3$) of any of $I$, $m_{\mathrm{l}}$, $\chi$
  or $m_{\mathrm{c}}$, are shown in Table~\ref{tab:observations}, where we also list the mean uncertainty and
  cadence of the corresponding data sets at both frequencies.

\begin{table}[!ht]
  \normalsize
  \caption{Number of sources with at least one significant data point (SNR$\ge3$), mean uncertainty and cadence for $I$, $m_{\mathrm{l}}$, $m_{\mathrm{c}}$, and  $\chi$.}
  \label{tab:observations}  
  \centering                    
  \begin{tabular}{llccc}
    \hline\hline                 
    Data set & &Units & 4.85 GHz & 8.35 GHz \\ 
    \hline\\
    $I$              & sources       &                                  & 155  & 150 \\
                     & uncertainty   &  (mJy)                           &  19  & 15  \\
                     & cadence       &  $\left(\mathrm{months} \right)$ & 1.3  & 1.3 \\
    $m_{\mathrm{l}}$ & sources       &                                  & 90   & 96  \\
                     & uncertainty   &  ($\%$)                          & 0.1  & 0.1 \\
                     & {cadence}     &  $\left(\mathrm{months} \right)$ & 1.7  & 1.7 \\
    $\chi$           & sources       &                                  & 90   & 96  \\
                     & uncertainty   &  ($\degr$)                       &  1   & 0.6 \\
                     & {cadence}     &  $\left(\mathrm{months} \right)$ & 1.7  & 1.7 \\
    $m_{\mathrm{c}}$ & sources       &                                  & 63   & 54  \\
                     & uncertainty   &  ($\%$)                          & 0.1  & 0.1 \\
                     & {cadence}     &  $\left(\mathrm{months} \right)$ & 6    & 12  \\
    \\\hline
  \end{tabular}
\end{table}

\section{Full-Stokes polarimetry}
\label{sec: FullS_polarimetry}

In the current section we present the steps taken for reconstructing the circular and linear
polarization parameters of the incident radiation from the observables delivered by the telescope. 
Our approach aims at recovering the polarization state outside the terrestrial atmosphere.

Our methodology is readily applicable to systems with circularly polarized feeds and it can be easily modified
for systems with linearly polarized feeds. The latter are sensitive to the horizontal, $E_{\mathrm{h}}(t)$, and vertical, 
$E_{\mathrm{v}}(t)$, linearly polarized electric field components of the incident radiation. The Stokes parameters in terms 
of these components can be written as:
\begin{align}
I &= \left<E^{*}_{\mathrm{h}}(t) E_{\mathrm{h}}(t)\right> + \left<E^{*}_{\mathrm{v}}(t) E_{\mathrm{v}}(t)\right> = \left< E_{\mathrm{H}}^{2} \right> + \left< E_{\mathrm{V}}^{2} \right>,  \label{eq:I_lin}\\
Q &= \left<E^{*}_{\mathrm{h}}(t) E_{\mathrm{h}}(t)\right> - \left<E^{*}_{\mathrm{v}}(t) E_{\mathrm{v}}(t)\right> = \left< E_{\mathrm{H}}^{2} \right> - \left< E_{\mathrm{V}}^{2} \right>,  \label{eq:Q_lin}\\
U &= 2 \mathrm{Re}\left(\left<E^{*}_{\mathrm{h}}(t) E_{\mathrm{v}}(t)\right>\right) = 2 \left< E_{\mathrm{H}}  E_{\mathrm{V}} \cos\delta \right>,\label{eq:U_lin}\\
V &= 2 \mathrm{Im}\left(\left<E^{*}_{\mathrm{h}}(t) E_{\mathrm{v}}(t)\right>\right) = 2 \left< E_{\mathrm{H}}  E_{\mathrm{V}} \sin\delta \right>,\label{eq:V_lin}
\end{align}
where,
\[
\begin{array}{lp{0.8\linewidth}}
  E_{\mathrm{H,V}}  & the amplitudes of the two orthogonal linearly polarized electric field components \\ 
  \delta            & the phase difference between $E_{\mathrm{h}}(t)$ and $E_{\mathrm{v}}(t)$.\\
\end{array}
\]
A comparison between the Stokes parameterizations for linear (Eqs.~\ref{eq:I_lin}--\ref{eq:V_lin}) and circular 
bases (Eqs.~\ref{eq:I}--\ref{eq:V}) shows that for systems with linearly polarized feeds the treatment of Stokes 
$I$ needs not to be changed while Stokes $Q$, $U$ and $V$ should be treated as Stokes $V$, $Q$ and $U$ for systems 
with circularly polarized feeds, respectively. As an example, for systems with linearly polarized feeds, 
the instrumental polarization correction scheme presented in Sect.~\ref{subsec:instrument model} should be applied 
in the $U$-$V$ instead of the $Q$-$U$ space, while the analysis of Sect.~\ref{subsec:LR_gain_corr} should be applied 
to Stokes $Q$ instead of $V$.

\begin{figure}[!ht]
 \begin{center}
   \includegraphics[trim =300 100 250 30,clip,width=.45\textwidth]{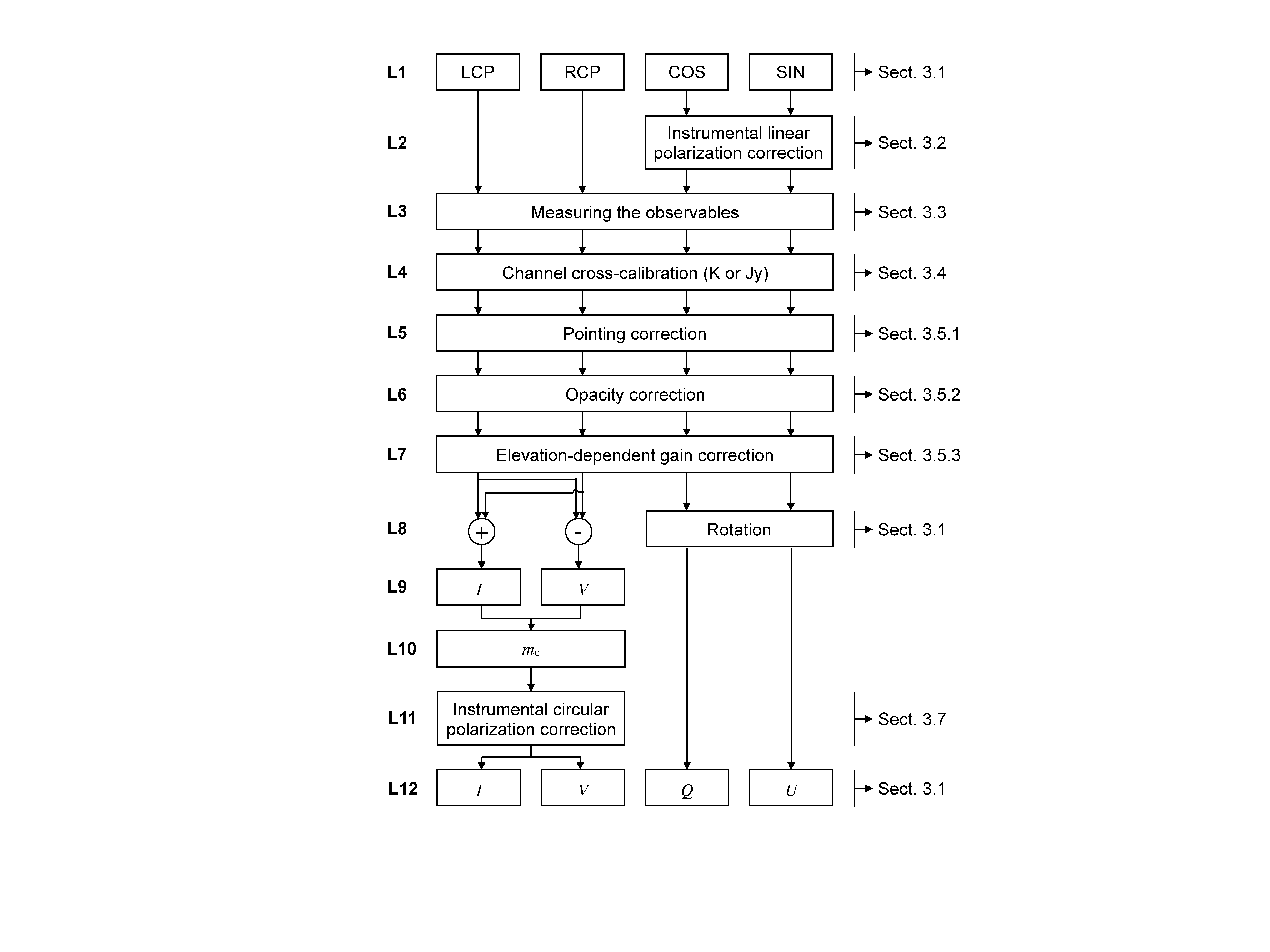}
   \caption{{A schematic summary of the analysis sequence. Each analysis level is labeled with an index
       on the left and is discussed in the section noted on the right. The mean effect of each correction step
       is listed in Table~\ref{tab:corr_lvls}.}}
  \label{fig:block_diagram}
 \end{center}
\end{figure}

Figure~\ref{fig:block_diagram} serves as a schematic summary of the analysis sequence. Each analysis
level is labeled with an index (e.g. L1, L2 etc.) and is discussed in the section noted in that flow
chart. The mean effect of each correction step is listed in Table~\ref{tab:corr_lvls}.

\begin{table}[!ht]
\normalsize
\caption{{Average percentage effect of each correction step on all Stokes parameters. Left columns refer to
  4.85~GHz while right ones to 8.35~GHz.}}
  \label{tab:corr_lvls}  
  \centering                    
  \begin{tabular}{p{2.5cm}@{\hskip 0.2cm}r@{\hskip 0.2cm}r@{\hskip 0.2cm}r@{\hskip 0.1cm}r@{\hskip 0.2cm}r@{\hskip 0.1cm}r@{\hskip 0.2cm}r@{\hskip .1cm}r}
  \hline\hline   
                      &\mc{2}{c}{$I$} &\mc{2}{c}{$Q$}      &\mc{2}{c}{$U$}      &\mc{2}{c}{$V$}\\           
\hline \\
Instrumental LP       &\ldots &\ldots &$\le$0.5  &$\le$0.5 &$\le$0.5  &$\le$0.5 &\ldots &\ldots \\
Pointing              &0.4    &0.9    &0.4       &0.9      &0.4       &0.8      &3.6    &1.4    \\  
Opacity               &4.0    &4.0    &4.0       &4.0      &4.0       &4.0      &4.0    &4.0    \\                 
Gain curve            &1.0    &0.9    &1.0       &0.9      &1.0       &0.9      &1.0    &0.9    \\
Instrumental CP       &0.0    &0.0    &\ldots    &\ldots   &\ldots    &\ldots   &89.0   &118.5  \\
\\\hline
  \end{tabular}
\end{table}

\subsection{Measuring the Stokes parameters}
\label{subsec:system description}

Because our receivers are sensitive to the left- and right-hand circularly polarized components of the
electric field, it is convenient to express the incident radiation in a circular basis:
\begin{align}
E_{\mathrm{l}}(t) &= E_{\mathrm{L}} e^{i \omega t}, \label{eq:E_l}\\
E_{\mathrm{r}}(t) &= E_{\mathrm{R}} e^{i (\omega t + \delta)} \label{eq:E_r}
\end{align}
where,
\[
\begin{array}{lp{0.8\linewidth}}
  E_{\mathrm{L,R}}  & the amplitudes of the two (orthogonal) circularly polarized electric field components \\ 
  \omega                 & the angular frequency of the electromagnetic wave \\
  \delta                    & the phase difference between $E_{\mathrm{l}}(t)$ and $E_{\mathrm{r}}(t)$.\\
\end{array}
\]

The four Stokes parameters can then be written in terms of $E_{\mathrm{l}}(t)$ and $E_{\mathrm{r}}(t)$ (omitting the
impedance factors), as:
\begin{align}
I &= \left< E_{\mathrm{L}}^{2} \right> + \left< E_{\mathrm{R}}^{2} \right>,  \label{eq:I}\\
Q &= 2 \left< E_{\mathrm{L}}  E_{\mathrm{R}} \cos\delta \right>, \label{eq:Q}\\
U &= 2 \left< E_{\mathrm{L}}  E_{\mathrm{R}} \sin\delta \right> = 2 \left< E_{\mathrm{L}}  E_{\mathrm{R}}  \cos(\delta-90\degr) \right>,\label{eq:U}\\
V &= \left< E_{\mathrm{R}}^{2} \right> - \left< E_{\mathrm{L}}^{2} \right>, \label{eq:V} 
\end{align}
where $\left< \right>$ denotes averaging over time to eliminate random temporal fluctuations of
$E_{\mathrm{L}}$, $E_{\mathrm{R}}$ and $\delta$ \citep[e.g.][]{Cohen1958, Kraus1966}. A detailed
discussion of the Stokes parametrization is given by \citet{Chandrasekhar1950,Kraus1966,Jackson1998}.

The system measures the four Stokes parameters by correlation operations, i.e. multiplication and time averaging 
of the signals $E_{\mathrm{l}}(t)$ and $E_{\mathrm{r}}(t)$, based on the parametrization of Eqs.~\ref{eq:I}--\ref{eq:V}:
\begin{align}
I &= \left<E^{*}_{\mathrm{l}}(t) E_{\mathrm{l}}(t)\right> + \left<E^{*}_{\mathrm{r}}(t) E_{\mathrm{r}}(t)\right>,  \label{eq:I_corr}\\
Q &= 2 \left<E^{*}_{\mathrm{l}}(t) E_{\mathrm{r}}(t)\right>, \label{eq:Q_corr}\\
U &= 2 \left<E^{*}_{\mathrm{l}}(t) E_{\mathrm{r}}(t)\right>_{90\degr},\label{eq:U_corr}\\
V &= \left<E^{*}_{\mathrm{r}}(t) E_{\mathrm{r}}(t)\right> - \left<E^{*}_{\mathrm{l}}(t) E_{\mathrm{l}}(t)\right>, \label{eq:V_corr} 
\end{align}
where the ``*'' denotes the complex conjugate and the subscript ``$90\degr$'' of Eq.~\ref{eq:U_corr} denotes 
that the correlation is performed after an additional phase difference of $90\degr$ is introduced between 
$E_{\mathrm{l}}(t)$ and $E_{\mathrm{r}}(t)$. The auto-correlations of $E_{\mathrm{l}}(t)$ and $E_{\mathrm{r}}(t)$ 
-- needed for $I$ (Eq.~\ref{eq:I_corr}) and $V$ (Eq.~\ref{eq:V_corr}) -- are processed separately in two receiver 
channels labeled LCP and RCP, respectively. On the other hand, the two cross-correlations of $E_{\mathrm{l}}(t)$ 
and $E_{\mathrm{r}}(t)$ -- needed for $Q$ (Eq.~\ref{eq:Q_corr}) and $U$ (Eq.~\ref{eq:U_corr}) --
are delivered in yet another pair of channels labeled COS and SIN. The LCP, RCP, COS and SIN channel 
data sets constitute the input for our pipeline (Fig.~\ref{fig:block_diagram}, level L1).

The alt-azimuthal mounting of the telescope introduces a rotation of the polarization vector in the $Q$-$U$ 
plane by the parallactic angle, $q$ (Fig.~\ref{fig:block_diagram}, level L8). 
In the general case, a potential gain difference between the COS and SIN channels introduces an additional 
rotation, $\phi$. The angle $\phi$ vanishes once we balance the COS and SIN channel gains
(Sect.~\ref{subsec:X-channel calibration}; Fig.~\ref{fig:block_diagram}, level L4) but we need to take it
into account when we calculate Stokes $Q$ and $U$ using the COS and SIN signals before the channel 
cross-calibration:
\begin{equation}
 \label{eq:lp_domains}
 \left[
 \begin{array}{c}
  Q\\U
 \end{array}
 \right] = 
 \left[
 \begin{array}{cr}
 \cos (2q + \phi) & -\sin (2q + \phi) \\
 \sin (2q + \phi) &  \cos (2q + \phi) 
 \end{array}
 \right] \cdot
 \left[
 \begin{array}{c}
  \mathrm{SIN}\\-\mathrm{COS}
 \end{array}
 \right]
\end{equation}
Throughout the following analysis we occasionally express $Q$ and $U$ in either of the default north-east or
the azimuth-elevation (azi-elv) reference frames. We differentiate the latter case by explicitly using the notation
$Q_{\mathrm{azi,elv}}$ or $U_{\mathrm{azi,elv}}$, which can be calculated by setting $q=0$ in
Eq.~\ref{eq:lp_domains}.

\begin{figure}[!ht]
 \begin{center}
  \includegraphics[trim =0 0 0 0,clip,width=.47\textwidth ]{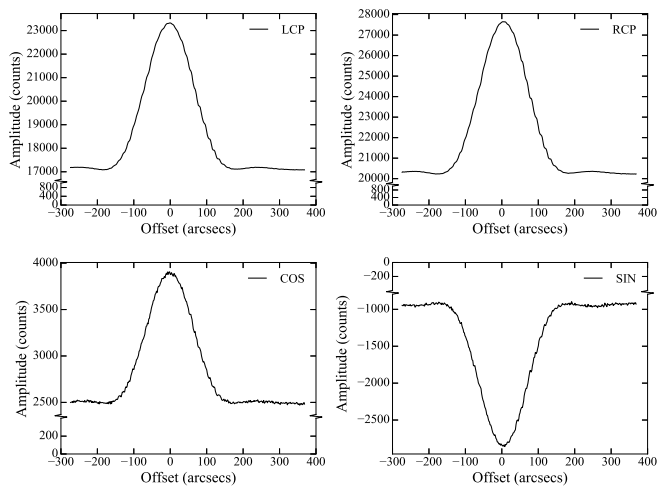}
  \caption[LCP, RCP, COS and SIN channel data sets]
  {An example of the same sub-scan in all channels on the source 3C\,286 at 4.85 GHz. The abscissa is the 
  offset from the commanded position of the source. The LCP and RCP channel data sets are shown in the upper row 
  while the COS and SIN data sets in the lower row. The target source is $\sim$11\% linearly polarized. }
  \label{fig:pol_channels_data}
 \end{center}
\end{figure}

In Fig.~\ref{fig:pol_channels_data} we show an example of an $\sim$11\% linearly polarized point-like source
in LCP, RCP, COS and SIN channels. The Stokes $I$, $Q$, $U$ and $V$ are computed
from Eqs.~\ref{eq:I}--\ref{eq:V} and \ref{eq:lp_domains} once the source amplitude in those channels is
known (Sect.~\ref{subsec:fitting}; Fig.~\ref{fig:block_diagram}, level L12).

\subsubsection{Feed ellipticity and the measurement of Stokes parameters}
\label{subsubsec:feed_ellipticity}
Equations~\ref{eq:E_l} to \ref{eq:lp_domains} describe the measurement of Stokes parameters for systems with ideal 
circularly polarized feeds. For such systems, the recorded left- and right-hand circularly polarized electric field 
components are perfectly orthogonal. In reality, instrumental imperfections lead to a (slight) ellipticity of the circular 
feed response. In this case, a small fraction of the incident left-hand circularly polarized electric field component is recorded 
by the right-hand circularly polarized channel of the system and vice versa.

The feed ellipticity can lead to deviations of the measured Stokes parameters from the incident ones.
In Appendix~\ref{app:feed_ellipticity} we provide an elementary approach to derive a rough estimate of the effect for the systems we used.
A thorough study instead can be found in e.g. \citet{McKinnon1992} or \citet{Cenacchi2009}.
For Stokes $I$, we estimate that those deviations are at the level of 1~mJy 
for our dataset, which is much less than the average uncertainty of our measurements (15--20 mJy, Table~\ref{tab:observations}).
Stokes $Q$ and $U$ on the other hand can be significantly modified by the feed ellipticity effect. A novel methodology to 
correct for the system-induced linear polarization across the whole telescope beam is described in Sect.~\ref{subsec:instrument model}.
Finally, Stokes $V$ is practically not affected in that simplified approach because the additional LCP and RCP terms introduced by the feed ellipticity 
(Eqs.~\ref{eq:LCP_rec_R_k_final} and \ref{eq:RCP_rec_R_k_final}) cancel out. Nevertheless, as described in Sect.~\ref{subsec:LR_gain_corr} 
our measurements suffer from instrumental circular polarization, which is most likely caused by a gain imbalance between 
the LCP and RCP channels. Two independent methodologies for the instrumental circular polarization correction 
are presented in Sect.~\ref{subsec:instrumental_org_pol_remove}.

\subsection{Correcting for instrumental linear polarization}
\label{subsec:instrument model}

\begin{figure}[!ht]
\centering
\begin{tabular}{ccc}
\includegraphics[trim={20 20 20 20}, clip, width=0.22\textwidth,angle=0]{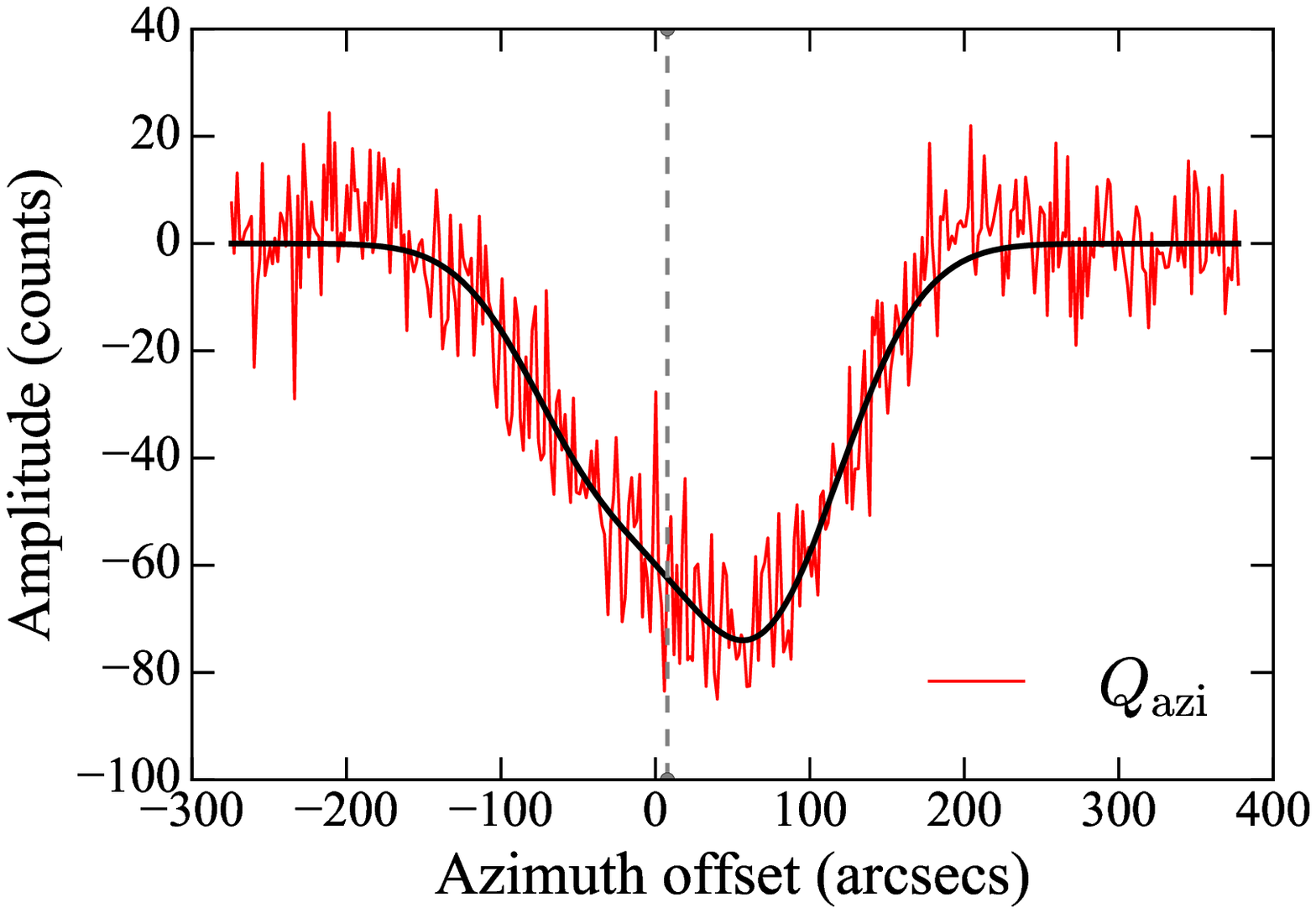} & \includegraphics[trim={20 20 20 20}, clip, width=0.22\textwidth,angle=0]{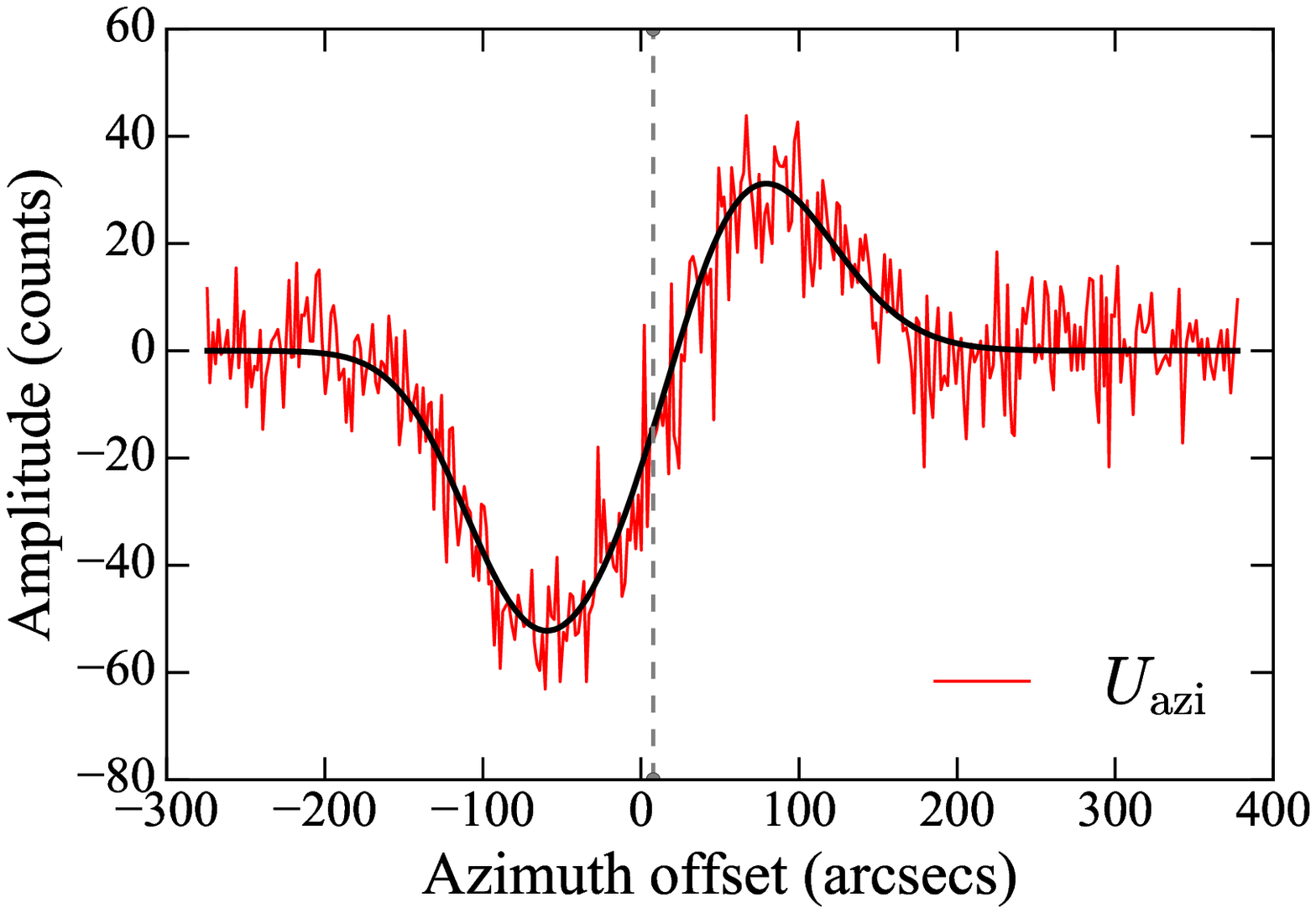} \\ 
\includegraphics[trim={20 20 20 20}, clip, width=0.22\textwidth,angle=0]{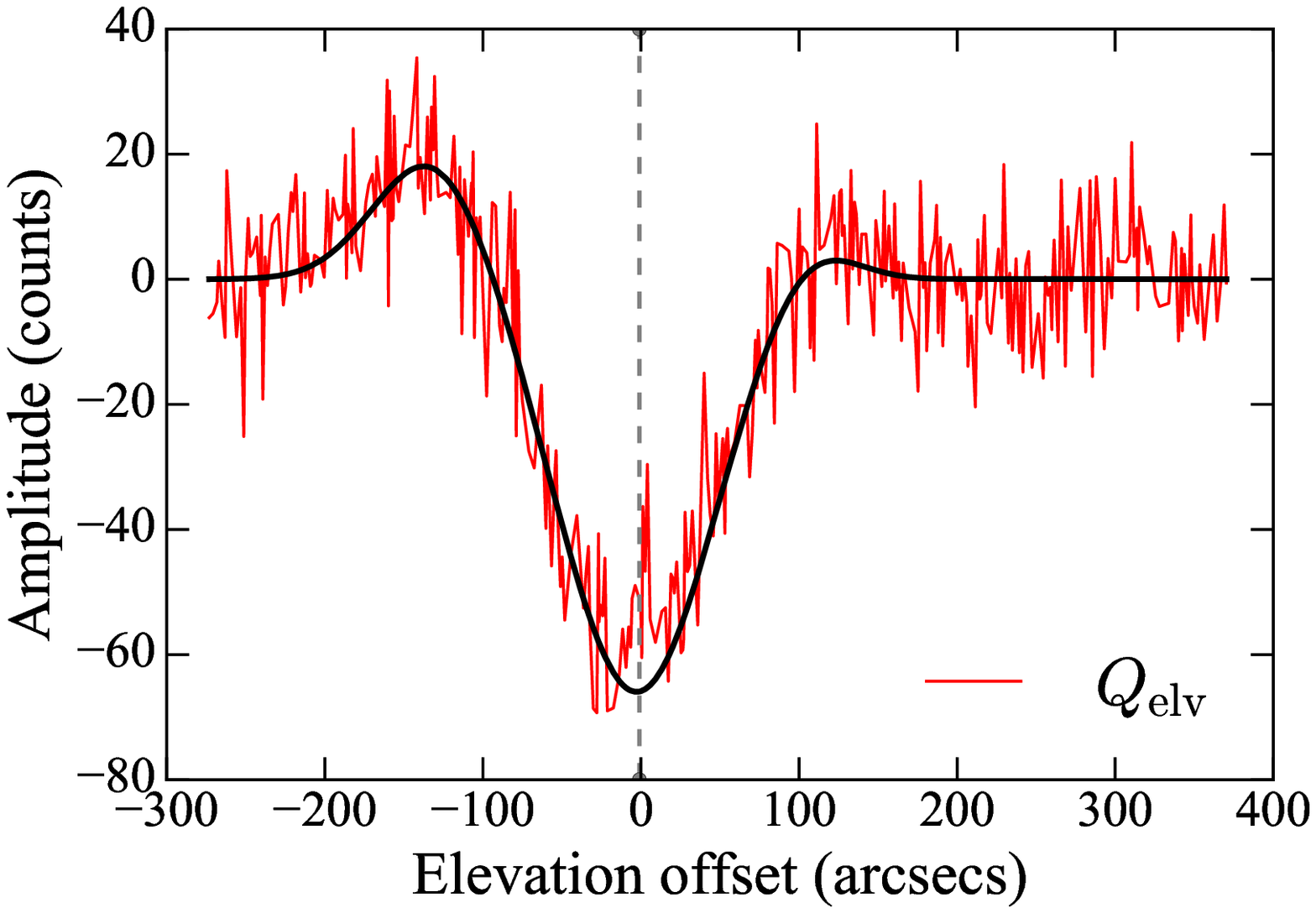} & \includegraphics[trim={20 20 20 20}, clip, width=0.22\textwidth,angle=0]{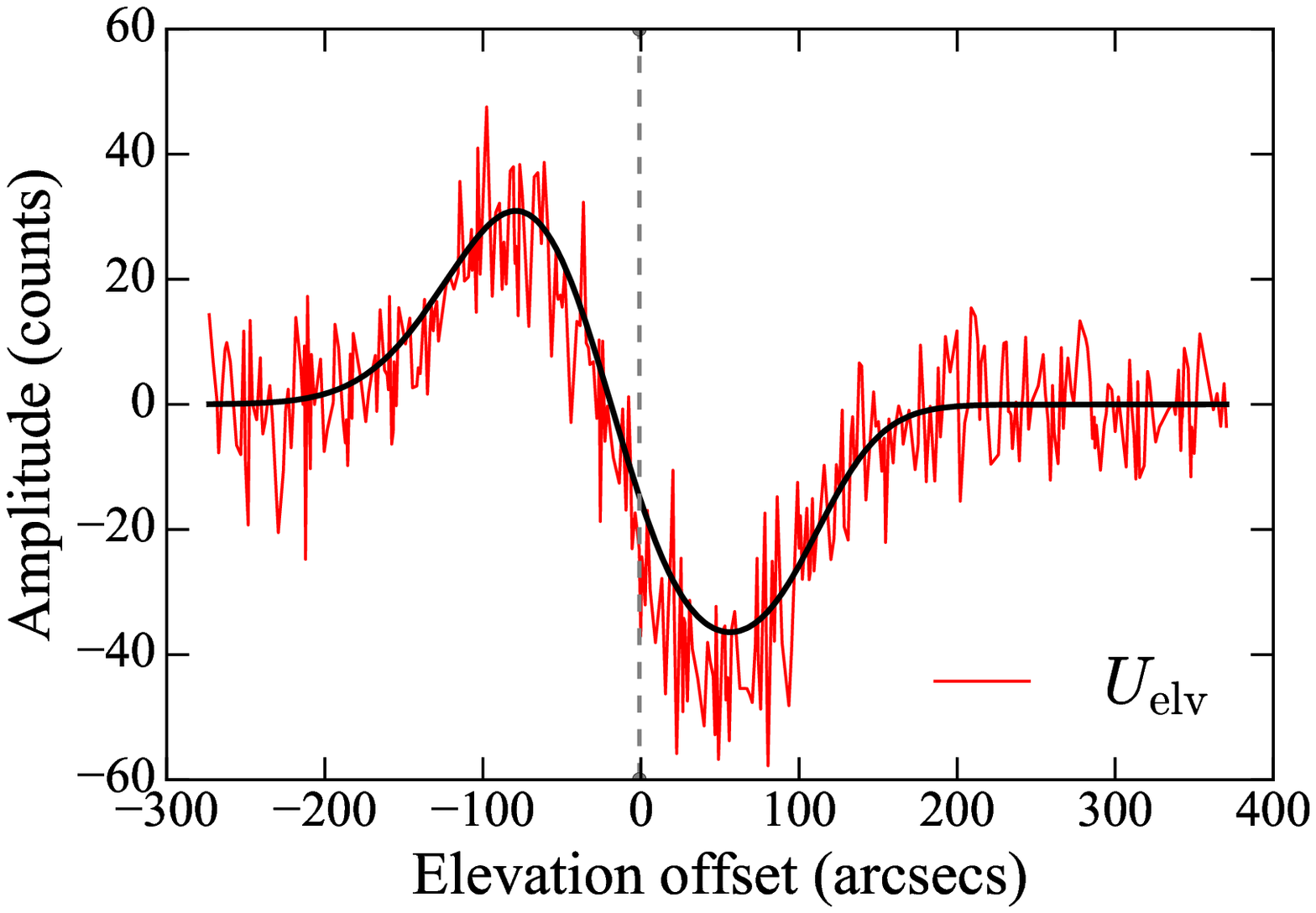} \\ 
\end{tabular}
\caption{Stokes $Q_{\mathrm{azi,elv}}$ (left column) and $U_{\mathrm{azi,elv}}$ (right column) data sets 
recorded at 4.85~GHz for the unpolarized, point-like source NGC\,7027 in the two scanning directions, the azimuth in 
the top and the elevation in the bottom row. Instead of the expected flat, noise-like pattern, the presence 
of spurious signals are clearly visible in both scanning directions. The instrumental polarization, calculated 
by the instrument model $M$, is shown in each panel with a smooth black line. The dashed grey lines mark 
the source position along the scanning direction.}
\label{fig:artifacts}
\end{figure}

\begin{figure}[!ht]
  \centering 
        \includegraphics[width=\hsize]{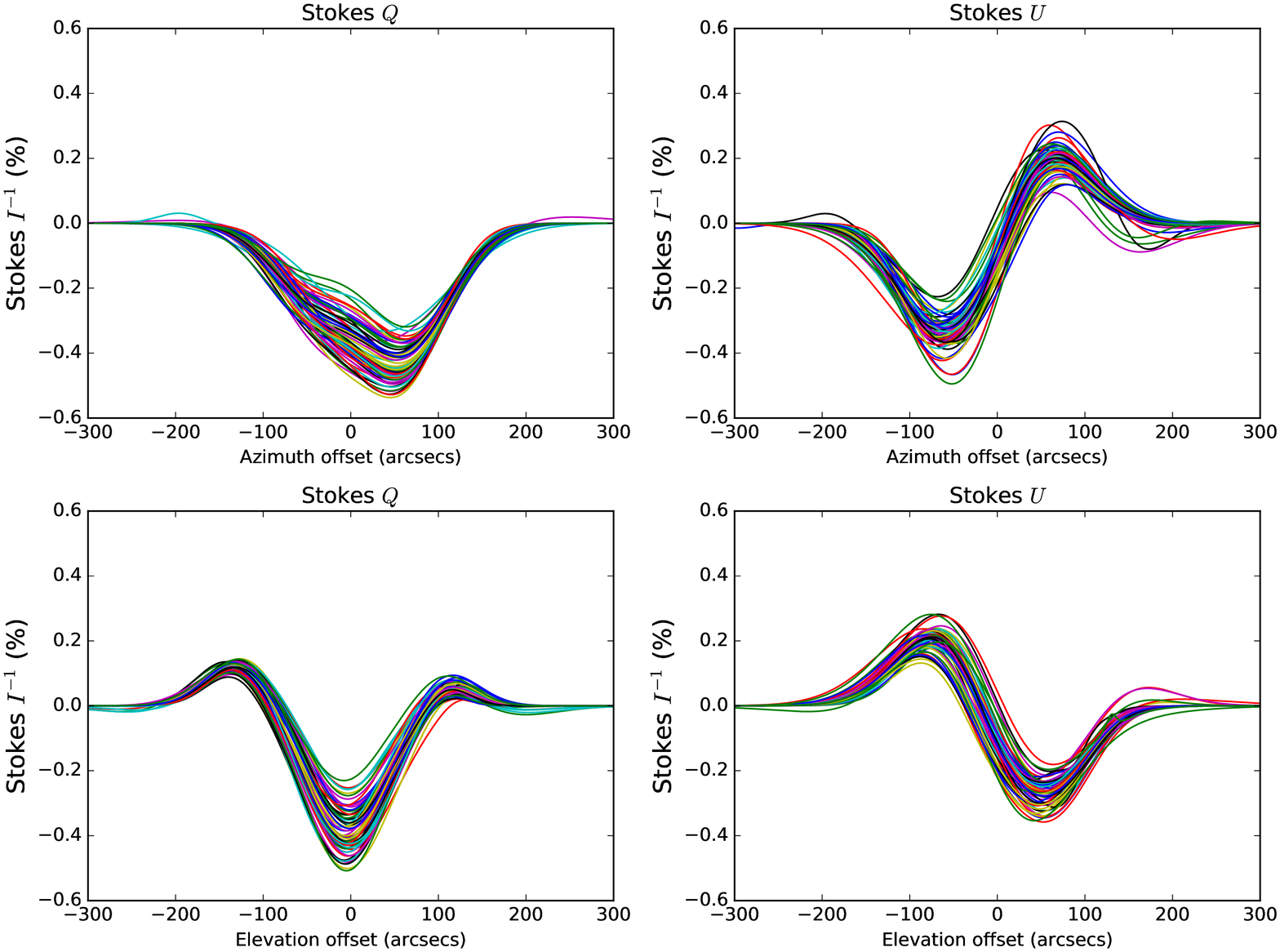}\\
        \includegraphics[width=\hsize]{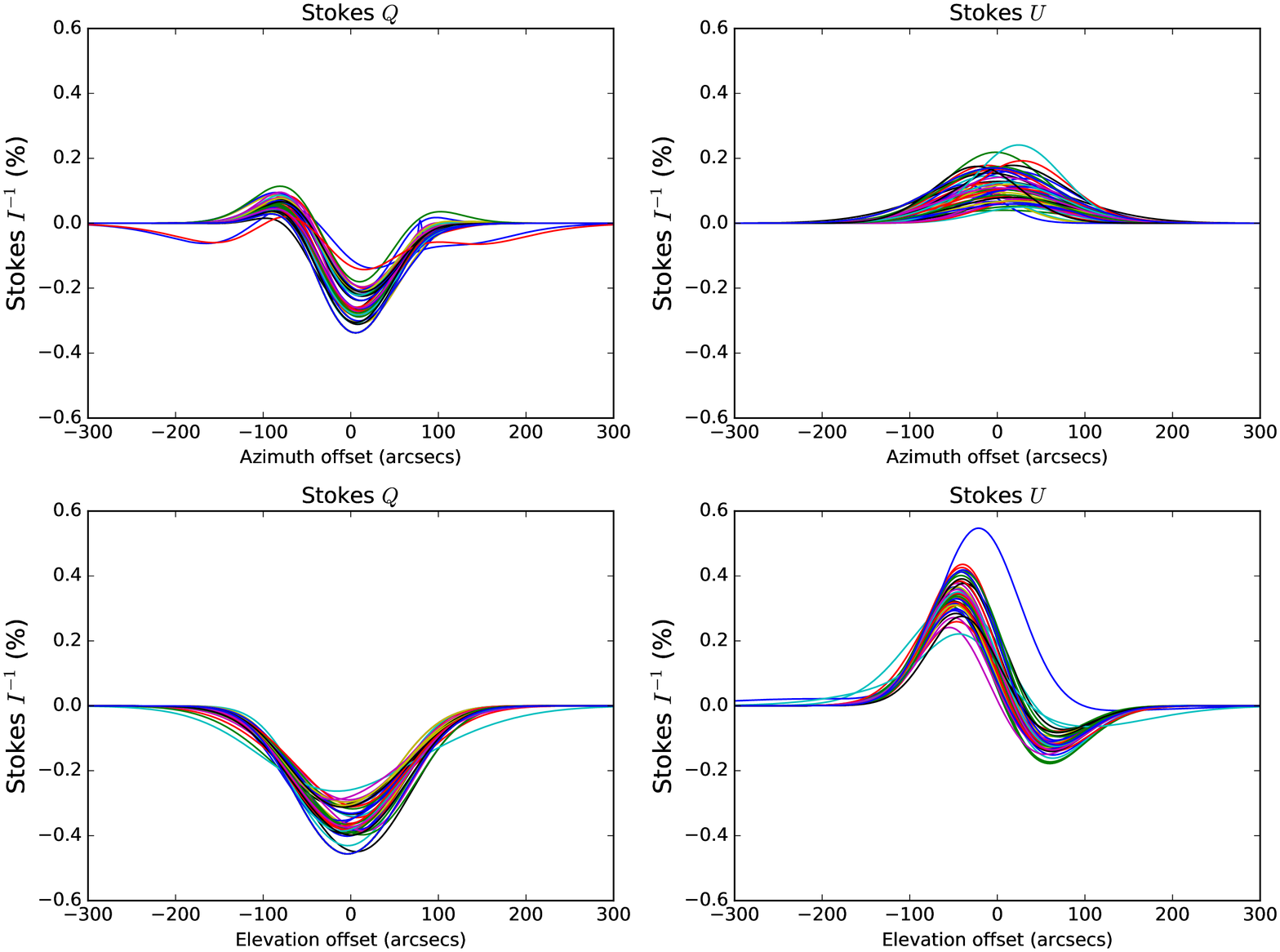}
        \caption{The generated Stokes $Q$ and $U$ instrument models in the two scanning directions, the
          azimuth in the top and the elevation in the bottom panels for (a) the 4.85~GHz (upper two rows) and
          (b) 8.35~GHz (lower two rows) receivers. The models were generated for 65 observing sessions. 
          The variability of the plotted models is comparable to the respective errors of the fit.}
  \label{fig:all_models}
\end{figure}

Instrumental imperfections manifest themselves as a cross-talk between the signals $E_{\mathrm{l}}(t)$ and
$E_{\mathrm{r}}(t)$. This can be best seen in unpolarized sources for which their cross-correlation is not
null contrary to what is theoretically expected (Eq.~\ref{eq:Q} and \ref{eq:U}). Figure~\ref{fig:artifacts}
shows an example of $Q_{\mathrm{azi,elv}}$ and $U_{\mathrm{azi,elv}}$ data sets of a linearly unpolarized
source. Instead of the expected constant, noise-like signal, spurious patterns are clearly visible.  Their
amplitudes can be up to $\sim0.5\%$ of the total flux density (Fig.~\ref{fig:all_models}). Those 
signals can also be interpreted as ``slices'' of the polarized beam patterns over the azimuth and elevation 
directions.

To correct for the instrumental polarization (Fig.~\ref{fig:block_diagram}, level L2), we describe the
telescope response to unpolarized sources with one instrument model $M$ for each of the $Q_{\mathrm{azi,elv}}$ 
and $U_{\mathrm{azi,elv}}$ along azimuth and elevation. Each $M$ is written as a sum of $j$ Gaussians or first 
derivatives of Gaussians selected empirically:
\begin{equation}
\label{eq:instr_model}
M = \sum^{\le3}_{j=1} F_j \left(\alpha_j I,(\mu-\beta_j),\gamma_j \sigma \right)
\end{equation}
where, $F_{j}$ is a Gaussian or first derivative of Gaussian with amplitude $\alpha_j I$, peak offset
$(\mu-\beta_j)$ and full width at half maximum (FWHM) $\gamma_j \sigma$. 
The explicit functional form of the models we used for the 4.85~GHz and 8.35~GHz receivers are given in 
Appendix~\ref{app:instr_models}. The parameters $I$, $\mu$ and $\sigma$ in Eq.~\ref{eq:instr_model} are 
the measured mean amplitude, peak offset and FWHM in the LCP and RCP channels, respectively. The identification 
of the optimal instrument model for a given observing session from this family of models requires the evaluation 
of the parameters $\alpha_j$, $\beta_j$ and $\gamma_j$.

For the evaluation of $\alpha_j$, $\beta_j$ and $\gamma_j$ we fit all observations on linearly upolarized
sources simultaneously. We first concatenate all sub-scans (index $i$ in Eq.~\ref{eq:instr_model_fit}) on all
sources (index $k$). Subsequently, each of the four data sets is fitted with a function of the
form:
\begin{equation}
A=\sum_k \sum_i M_{ik}
\label{eq:instr_model_fit}
\end{equation}
In these terms, $A$ is simply a concatenation of a total of $i\cdot k$ instrument models of the form $M$.
In Fig.~\ref{fig:all_models} we show the fitted instrument models for 65 observing sessions.  The variability of the 
plotted models is comparable to the respective errors of the fit, which indicates that the instrumental polarization remained 
fairly stable throughout the period of 5.5 years we examined.

\begin{figure*}[!ht]
 \begin{center}
  \includegraphics[width=\hsize]{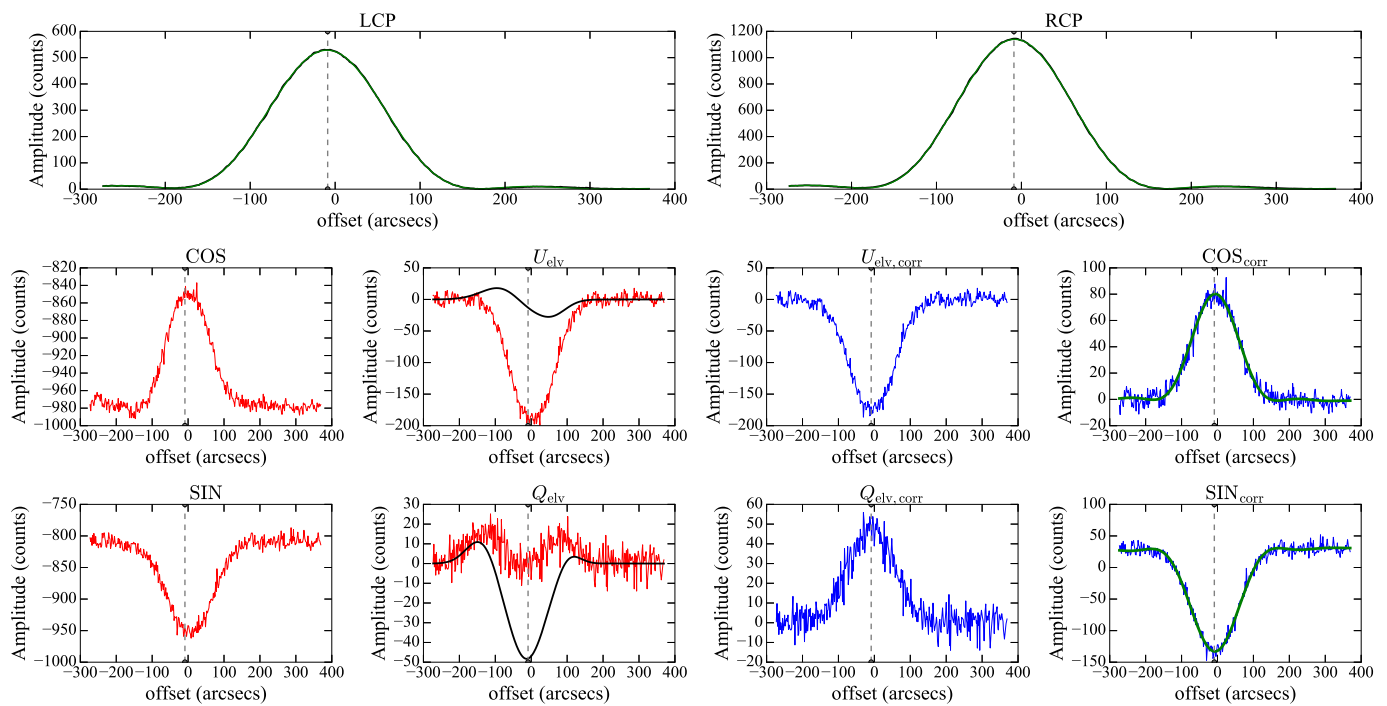}
  \caption{Example of the instrumental linear polarization correction for an elevation sub-scan at 4.85~GHz. 
  The observed COS/SIN (or equivalently $Q_{\mathrm{elv}}$/$U_{\mathrm{elv}}$) signals are shown in the middle and 
  bottom rows, right below the LCP singal (with red), while the corrected signals are shown right below the RCP singal 
  (with blue). The correction is performed by subtracting the expected instrumental polarization signals (smooth black 
  lines) from the observed $Q_{\mathrm{elv}}$ and $U_{\mathrm{elv}}$ data sets (2nd and 3rd column of the middle and 
  bottom rows). The instrumental polarization signals are calculated by substituting the measured mean amplitude, 
  $I$, peak offset, $x$ and FWHM, $\sigma$, of the LCP and RCP signals (top row) in the instrument model, $M$, 
  created for the given observing session.}
  \label{fig:artifact_corr}  
 \end{center}
\end{figure*}

\begin{figure}[!ht]
 \includegraphics[width=\hsize]{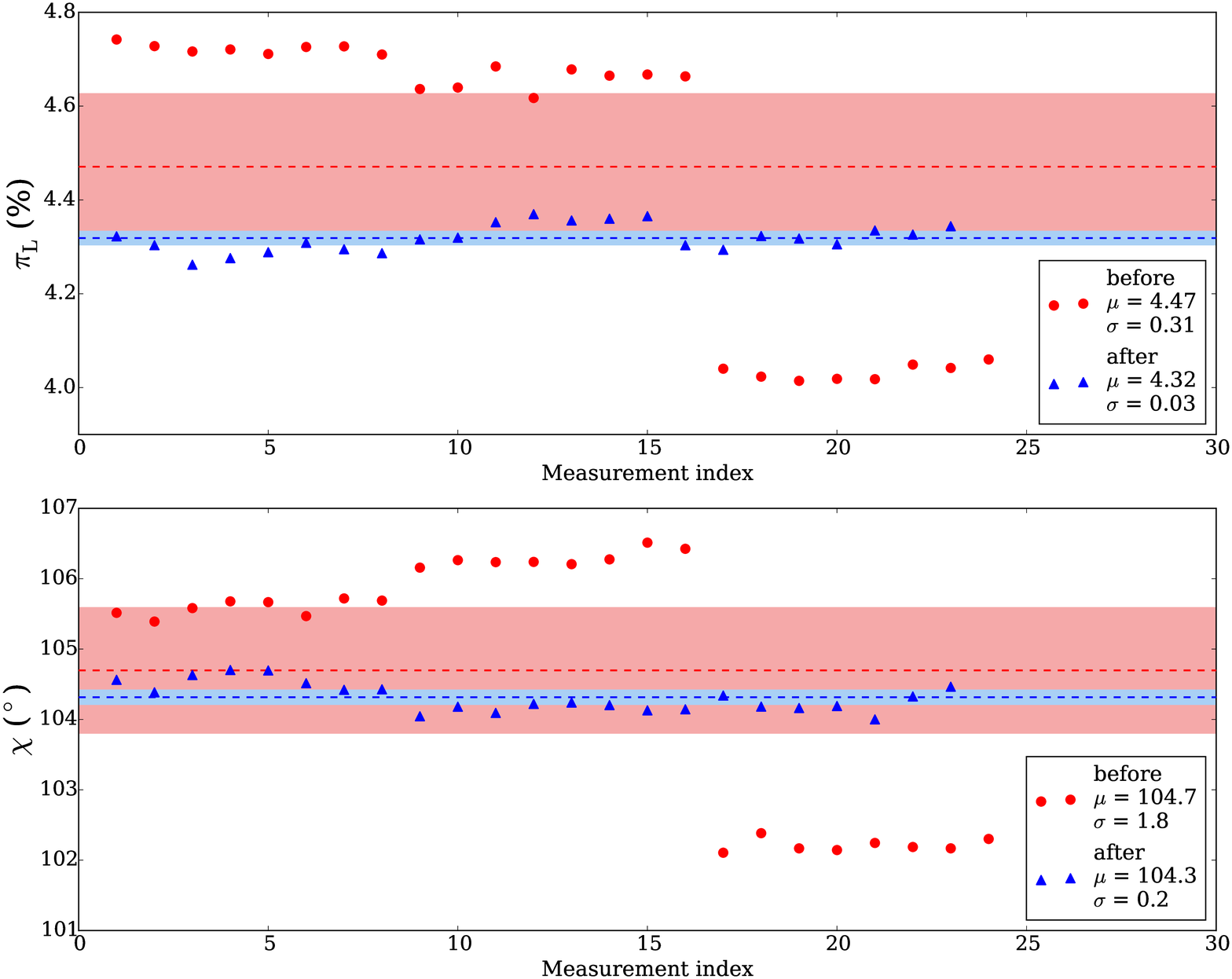} \\
 \caption{Degree of linear polarization (top panel) and polarization angle (bottom panel) of the source 3C\,48 at 4.85~GHz 
 before (red circles) and after (blue triangles) applying the instrumental linear polarization correction. The data correspond 
 to 24 sub-scans of the source within a single observing session. The average values are marked with dotted lines and the 
 highlighted areas indicate the 1-$\sigma$ region around them. The respective values are shown in the legend.}
 \label{fig:3C48_LP_improvement}
\end{figure}

After having optimized $\alpha_j$, $\beta_j$ and $\gamma_j$ for a given session, we remove the instrumental
polarization from each sub-scan, in two steps: 
\begin{enumerate}
\item we first substitute the measured $I$, $\mu$ and $\sigma$ in Eq.~\ref{eq:instr_model}
  to determine the explicit form of the instrumental polarization in that sub-scan; 
\item we then subtract this instrumental effect from the observed $Q_{\mathrm{azi,elv}}$ and
  $U_{\mathrm{azi,elv}}$. 
\end{enumerate}
An example is shown in Fig.~\ref{fig:artifact_corr}. With this approach, the scatter of the measured polarization 
parameters can be dramatically decreased owing to the fact that each sub-scan is treated separately 
(Fig.~\ref{fig:3C48_LP_improvement}).

\subsection{Measuring the observables}
\label{subsec:fitting}

Once $Q_{\mathrm{azi,elv}}$ and $U_{\mathrm{azi,elv}}$ have been corrected for instrumental polarization the
process of measuring the Stokes parameters requires first the precise determination of the source amplitudes
in the LCP, RCP, COS and SIN channels (Fig.~\ref{fig:block_diagram}, level L3). As an example, given the low 
degree of circular polarization $m_\mathrm{c}$ expected for our sources and because $V$ is the difference 
between their amplitudes in LCP and RCP (Eq.~\ref{eq:V}), an accuracy of at least 0.1~\% to 0.3~\% is required 
for an uncertainty of no more than about 0.1~\% to 0.2~\% in the $m_\mathrm{c}$. This precision would correspond 
to a 3--5$\sigma$ significance for a 0.5~\% circularly polarized source.

\begin{figure}[!ht]
\centering
\begin{tabular}{ccc}
\includegraphics[trim={20 20 10 20}, clip, width=0.22\textwidth,angle=0]{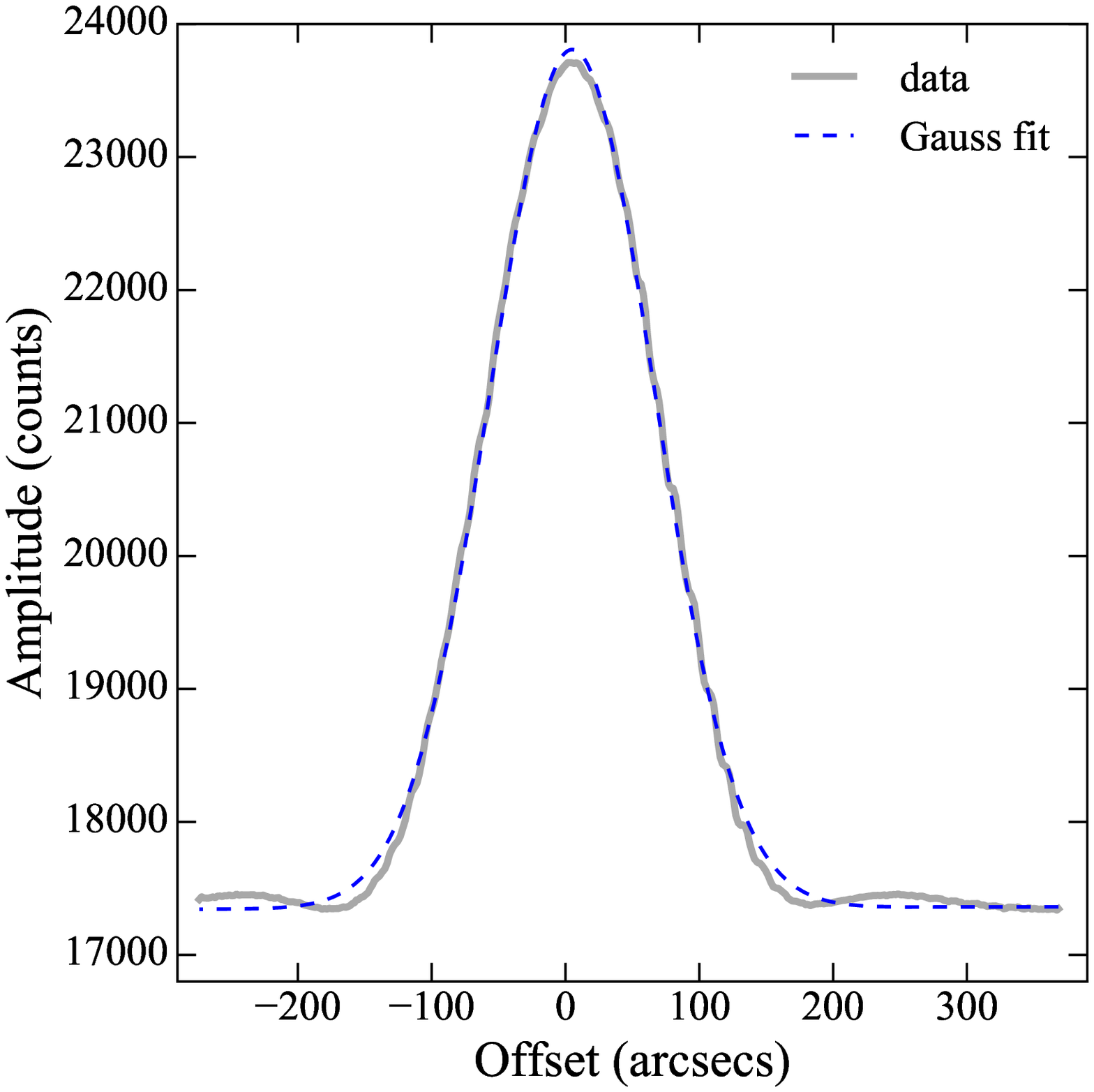} & \includegraphics[trim={20 20 10 20}, clip, width=0.22\textwidth,angle=0]{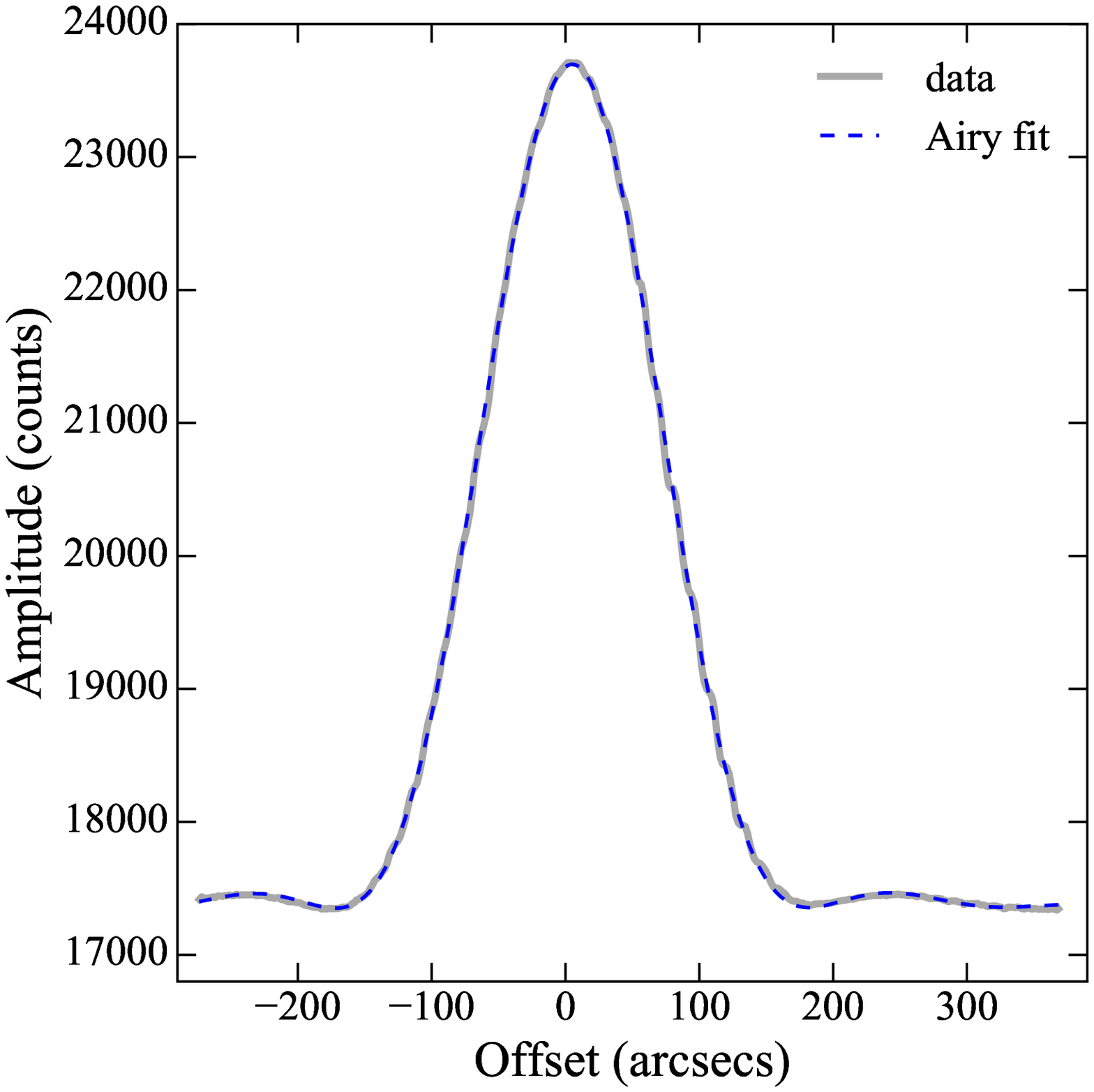} \\ 
\includegraphics[trim={20 20 10 20}, clip, width=0.22\textwidth,angle=0]{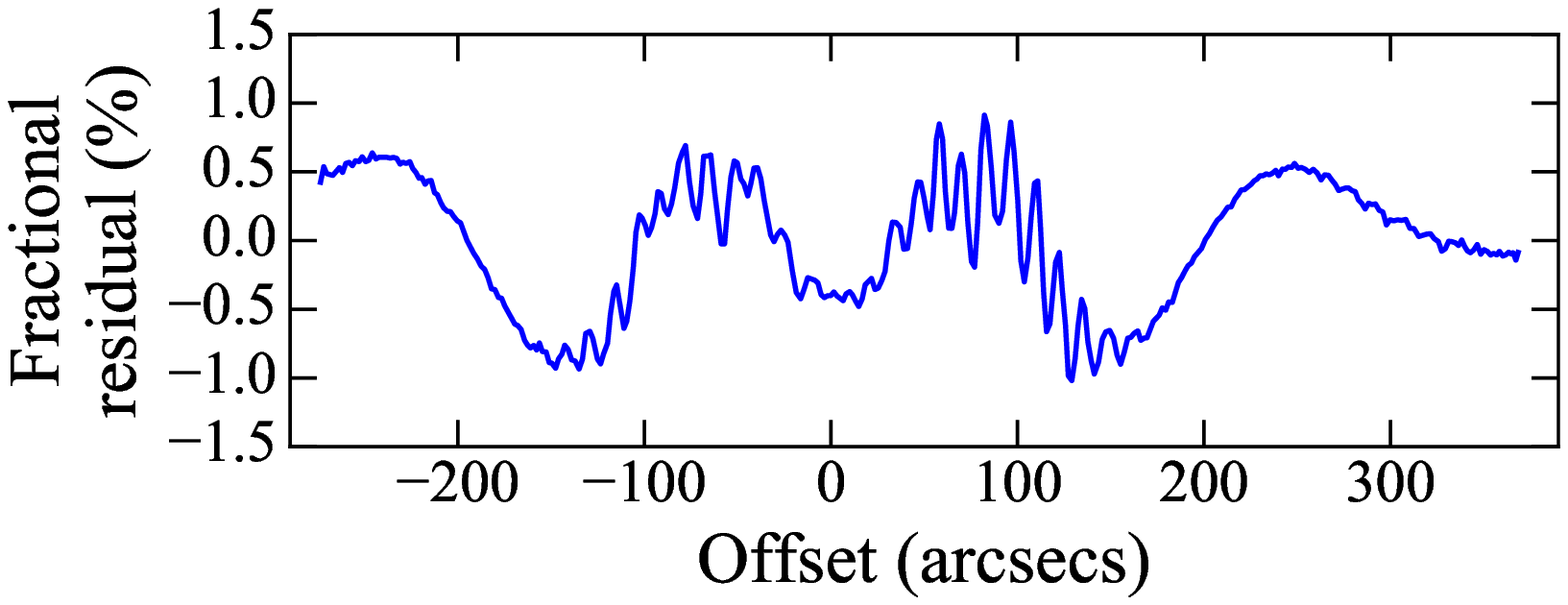} & \includegraphics[trim={20 20 10 20}, clip, width=0.22\textwidth,angle=0]{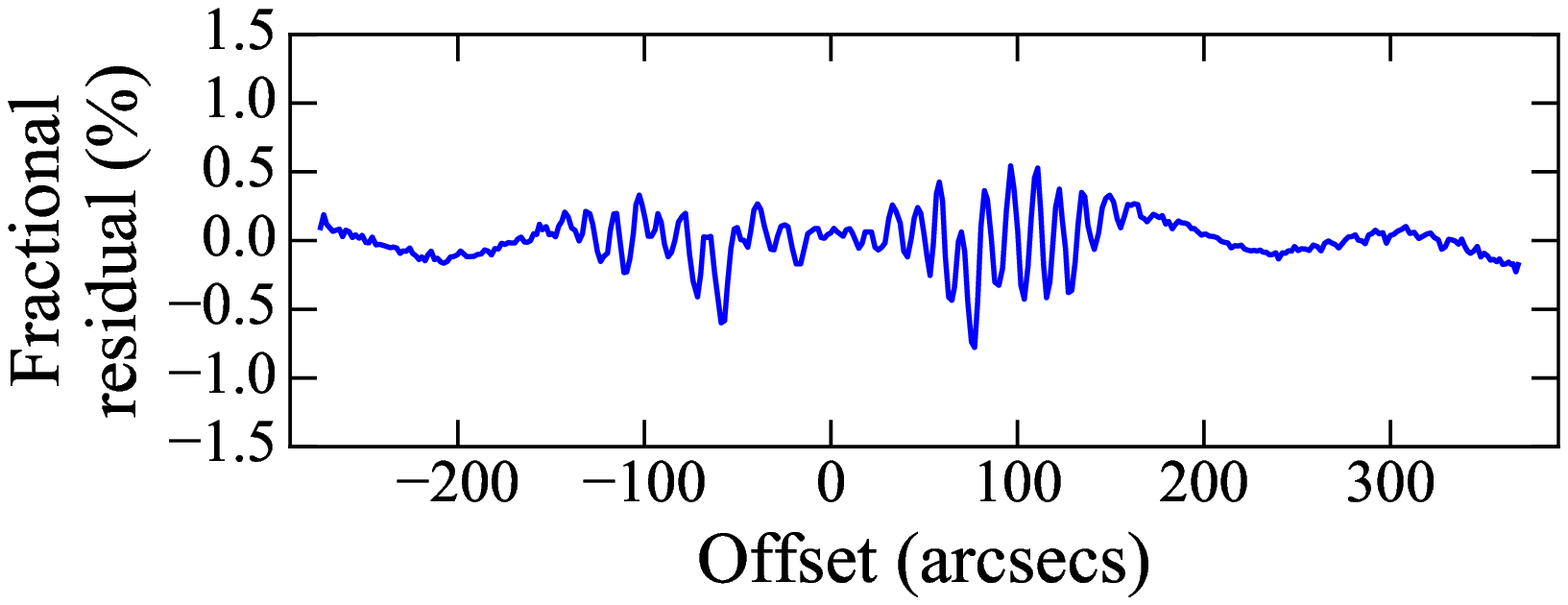} \\ 
\end{tabular}
\caption{A comparison between the Gaussian (left column) and Airy disk (right column) beam pattern 
models. The recorded data sets are shown in the top row with thick, grey lines and the model fits with dashed, 
blue lines. The corresponding fractional residuals, \emph{(data-model)/data}, are given in the bottom row 
for a direct comparison between the models.}
\label{fig:fit_methods}
\end{figure}

Our tests showed that the most essential element for the amplitude measurement is the accurate knowledge of
the telescope response pattern and particularly the accurate determination of the baseline level. 
We found that the antenna pattern for a uniformly illuminated circular aperture which is described by 
the Airy disk function:
\begin{equation}
 \label{eq:airy_pattern}
 I = I_{0} \left[ \frac{2J_{1}(x)}{x} \right]^{2} ,
\end{equation}
delivers significantly more accurate results that the commonly used Gaussian function, mainly because the 
latter fails to provide a precise description of the response beyond the FWHM. In Eq.~\ref{eq:airy_pattern}, 
$I_{0}$ is the maximum response level of the pattern at the center of the main lobe and $J_{1}$ is the Bessel 
function of the first kind. In reality, the Effelsberg 100m telescope beam is described by a more complex 
expression since its aperture is not uniformly illuminated, mainly due to the supporting structure of the 
secondary reflector. Nevertheless, the amplitude uncertainties using the Airy disk antenna pattern approximation 
(0.1~\%--0.2~\%) are small enough to accommodate reliable low circular polarization degree measurements.
Figure~\ref{fig:fit_methods} demonstrates the effectiveness of the Gaussian and the Airy disk beam pattern models 
in terms of the fractional residuals when we fit the observed data. Those are clearly minimized in the case of 
the Airy disk pattern.

Ideally the Airy disk could also be used for modeling the instrumental polarization
(Eq.~\ref{eq:instr_model}) instead of Gaussians. This however would cause only an insignificant improvement (a
small fraction of a percent) in the knowledge of the instrumental polarization magnitude. It would require a
several-hundred-Jy source to cause a measurable effect.

\subsection{Channel cross-calibration}
\label{subsec:X-channel calibration}

Because of inevitable gain differences between LCP, RCP, COS and SIN receiver channels, their response to the
same photon influx is generally different (different number of ``counts''). The level balancing --
cross-calibration -- of their signals is necessary before accurate polarization measurements can be conducted 
(Fig.~\ref{fig:block_diagram}, level L4).

The cross-calibration is performed by the periodical injection of a known-polarization signal at the feed point of the 
receiver every 64\,ms with a duration of 32\,ms. For the used receivers this signal is generated by a noise
diode designed to be: (a) circularly unpolarized, and (b) completely linearly polarized at a given polarization angle.
The noise diode amplitude for each channel can be estimated as the average difference of the telescope response with 
the noise diode ``on'' and ``off'' in that channel. The cross-channel calibration is then achieved by expressing the 
LCP, RCP, COS and SIN amplitudes in noise diode units. The noise diode signal can be further calibrated to physical 
units, e.g. Jy, by comparison with reference sources \citep[e.g.][]{Ott1994,Baars1977,Zijlstra2008}.

\subsection{Post-measurement Corrections}
\label{subsec:corrections}

Before the final calculation of the Stokes parameters, the source amplitudes in channels LCP, RCP, COS and SIN
are subjected to a list of post-measurement corrections which are discussed in detail in
\citet{Angelakis2007,2009A&A...501..801A,2015A&A...575A..55A,Myserlis2015}.

\subsubsection{Pointing correction}
\label{subsubsec:pointing correction}

This step corrects for the power loss caused by offsets between the true source position and the
cross-section of the two scanning directions (Fig.~\ref{fig:block_diagram}, level L5). Imperfect pointing may
potentially also increase the scatter of the amplitudes from different sub-scans as, in the general case,
the offset depends on the scanning direction. Assuming an Airy disk beam pattern the amplitude corrected 
for a pointing offset $p_{\mathrm{off}}$, is 
\begin{equation}
 A_{\mathrm{poi}}^{\mathrm{azi},\mathrm{elv}} = A^{\mathrm{azi},\mathrm{elv}} \cdot \left( \frac{2J_{1}(w \cdot p_{\mathrm{off}}^{\mathrm{elv},\mathrm{azi}})}{w \cdot p_{\mathrm{off}}^{\mathrm{elv},\mathrm{azi}}} \right)^{-2}
 \label{eq:poi_corr}
\end{equation}
where,
\[
\begin{array}{lp{0.8\linewidth}}
  \mathrm{azi},\mathrm{elv} & denotes the scanning direction, \\
  A_{\mathrm{poi}}          & the source amplitude corrected for pointing offset, \\
  A                         & the uncorrected source amplitude, \\
  p_{\mathrm{off}}          & the average absolute offset in arcsecs on the other scanning direction, \\
  w                         & is calculated as
                            \begin{equation}
                             \label{eq:LP}
                             w = \frac{3.23266}{\mathrm{FWHM}}
                            \end{equation} where $\mathrm{FWHM}$ is the full width at half maximum of the telescope at the observing frequency in arcsecs.\\
\end{array}
\]
It is important to note that the offset in the one direction (e.g. elevation) is used for correcting the
amplitude in the other direction (e.g. azimuth).

\begin{figure}[!ht]
 \begin{center}
  \includegraphics[width=0.9\hsize]{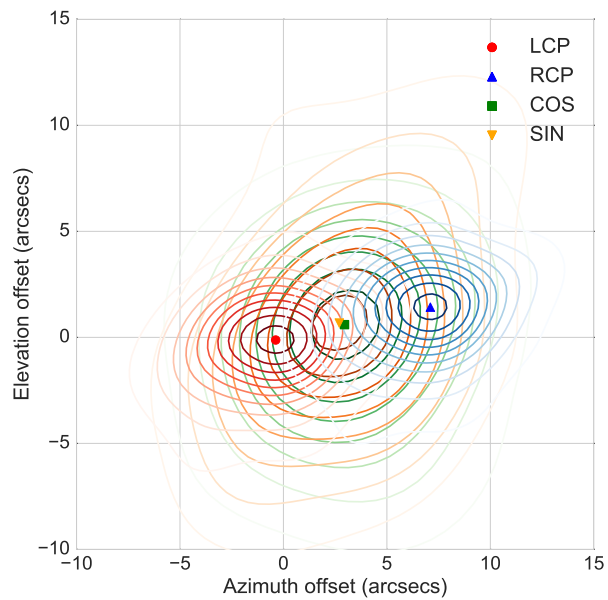}
  \caption{Density plots of the measured offsets between the source position and the LCP, RCP, COS and SIN beam patterns for the 4.85 GHz receiver. The beam-squint can be clearly seen by the miss-alignment of the contour peaks (LCP: circle, RCP: upward-looking triangle, COS: square, SIN: downward-looking triangle).}
  \label{fig:LRCSpoi_contours}
 \end{center}
\end{figure}

The pointing correction is performed independently for each channel as the beam patterns are generally
separated on the plane of the sky due to the miss-alignment between the feeds and the main axis of the telescope 
\citep[``beam-squint'' effect, e.g. ][]{Heiles2002}. In Fig.~\ref{fig:LRCSpoi_contours}, we plot the density 
contours of all measured offsets from the source position separately for each channel. The beam-squint is 
directly evident as the miss-alignment of the contour peaks. For polarimetric observations, the beam-squint 
can introduce fake circular polarization, since the LCP and RCP beam patterns measure the source with different 
sensitivities.

\subsubsection{Opacity correction}
\label{subsubsec:opacity correction}
The opacity correction corrects for the signal attenuation caused by the Earth's atmosphere and relies on
the calculation of the atmospheric opacity at the source position, $\tau_{\mathrm{atm}}$ (Fig.~\ref{fig:block_diagram}, level L6).

Given the amplitude $A$ of a measurement at an elevation $ELV$, the amplitude corrected for atmospheric
opacity will be
\begin{equation}
 A_{\mathrm{opc}} = A \cdot e^{\tau_{\mathrm{atm}}}
 \label{eq:Sopc}
\end{equation}
where, $\tau_{\mathrm{atm}}$ the atmospheric opacity at $ELV$. Under the assumption of a simple atmosphere model, 
$\tau_{\mathrm{atm}}$ can be computed as a simple function of the zenith opacity $\tau_{z}$ 
\begin{equation}
\label{eq:tau_atm}
\tau_{\mathrm{atm}} = \tau_{\mathrm{z}}\cdot AM = \tau_{z} \frac{1}{\sin(ELV)}
\end{equation}
with $AM$ the airmass at $ELV$.

\begin{figure}[!ht]
 \begin{center}
  \includegraphics[width=\hsize]{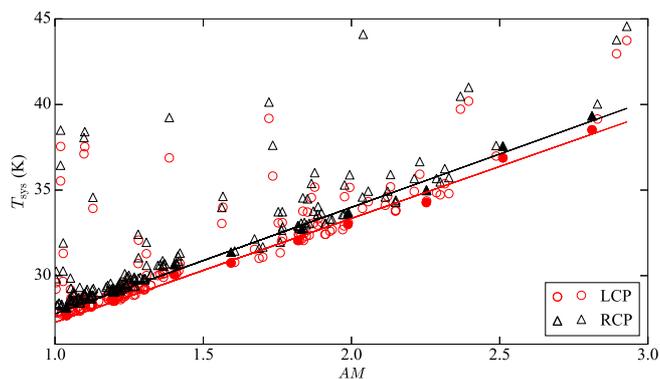}
  \caption{LCP and RCP system temperature ($T_{\mathrm{sys}}$) versus the airmass ($AM$) for 
  one observing session at 4.85~GHz. The $AM$ range is split in a given number of bins and the data points 
  with the lowest $T_{\mathrm{sys}}$ within each bin (filled markers) are used to fit the two lower envelopes
  (solid lines). Their slopes are practically identical, indicating that the atmospheric absorption does not 
  influence the polarization of the transmitted radiation.}
  \label{fig:Tsys_am}
 \end{center}
\end{figure}

For any given observing session, a linear lower envelope is fitted to the airmass ($AM$) - system temperature
($T_{\mathrm{sys}}$) scatter plot (Fig.~\ref{fig:Tsys_am}) with $T_{\mathrm{sys}}$ taken from the off-source
segment of all sub-scans. It can be shown that the inferred slope is a direct measure of the atmospheric
opacity at the zenith, $\tau_{\mathrm{z}}$ \citep{Angelakis2007,2009A&A...501..801A} .

As we show in the example session of Fig.~\ref{fig:Tsys_am}, $\tau_{\mathrm{z}}$ is independent of LCP and
RCP channels implying that the atmospheric absorption does not influence the polarization of the transmitted
radiation. Hence, we applied the same $\tau_{\mathrm{atm}}$ values to correct the amplitudes in all
channels.

\subsubsection{Elevation-dependent gain correction}
\label{subsubsec:gain-curve correction}
The last correction accounts for the dependence of the telescope gain on elevation caused by the gravitational
deformation of the telescope's surface (Fig.~\ref{fig:block_diagram}, level L7).

The amplitude corrected for this effect given a value $A$ measured at elevation $ELV$, will be  
\begin{equation}
 A_{\mathrm{gc}} = \frac{A}{G(ELV)}
 \label{eq:gc_corr}
\end{equation}
where, $G(ELV)$ the gain at elevation $ELV$. The gain is assumed to be a second order polynomial function
of $ELV$. The parameters of the parabola used here have been taken from the Effelsberg website
\footnote{\url{https://eff100mwiki.mpifr-bonn.mpg.de}}.

\begin{figure}[!ht]
\begin{center}
 \includegraphics[width=\hsize]{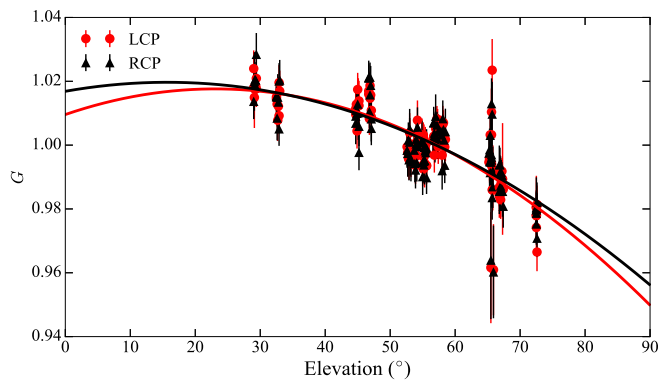}
 \caption{{Elevation - Gain curves separately for LCP and RCP data sets at 4.85~GHz. The similarity
   of the fitted parabolas indicate that the gravitational deformations do not affect the polarization
   measurements.}}
 \label{fig:LR_gcs}
\end{center}
\end{figure}

Figure~\ref{fig:LR_gcs} shows the gain curve computed for one selected session. As it is seen there the
least-square-fit parabolas for the LCP and RCP data sets are very similar. Hence, the gravitational deformations
do not affect the polarization measurements giving us the freedom to use the same correction factors $G$ for
all channels.

As it was already stated, the correction factors used for the opacity and elevation-dependent gain corrections
were identical for all channels. Consequently, they do not affect fractional expressions of the Stokes parameters 
such as the polarization degree or angle (Eqs.~\ref{eq:ml}, \ref{eq:mc} and ~\ref{eq:chi}). Yet they are included 
formally in this step of the analysis as they affect the values of the total and polarized flux densities.

\subsection{Correcting for instrumental rotation}
\label{subsec:instrumental_evpa}

Imprecise knowledge of the noise diode polarization angle potentially leads to poor knowledge of the power
to be expected in the COS and SIN channels. Consequently, this leads to imperfect channel cross-calibration
which will manifest itself as an instrumental rotation. To study this effect we conducted observations of
the Moon which has a stable and well understood configuration of the polarization orientation.

The lunar black body radiation is linearly polarized. The polarization degree maximizes close to the limb
while the polarization angle has an almost perfect radial configuration \citep[e.g.][]{Heiles1963,Poppi2002,Perley2013a}.

We first performed the usual azimuth and elevation cross-scans centered on the Moon. Before estimating the
polarization angle -- which was the objective of this exercise -- the observed $Q_{\mathrm{azi,elv}}$ and
$U_{\mathrm{azi,elv}}$ were corrected for instrumental polarization. As we discuss in
Sect.~\ref{subsec:instrument model} the instrument model for a sub-scan on a point source depends on the
parameters measured in that sub-scan: $I$, $\mu$ and $\sigma$. For extended sources the brightness
distribution is needed instead.  For this reason, for each sub-scan we first recovered the lunar brightness
distribution by de-convolving the observed $I$ with an Airy disk beam pattern. The de-convolution is then
used to evaluate the explicit form of the instrumental polarization for that sub-scan by convolving the
corresponding instrument model $M$ with the calculated brightness distribution. $Q_{\mathrm{azi,elv}}$ and
$U_{\mathrm{azi,elv}}$ were then corrected for the instrumental polarization, which was calculated 
across the whole extend of the source.

We restricted the comparison of the observed and the expected polarization angle at the four points of the
Moon's limb that we probed. That is north, south, east and west.  To quantify the instrumental rotation, we
compared the median polarization angle around those four limb points with the values expected for the radial
configuration. The east and west limb points are expected to be at 90$\degr$ (or -90$\degr$), while
the north and south limbs at 0$\degr$ (or 180$\degr$). On the basis of 62 measurements at 4.85 GHz and
40 at 8.35 GHz, our analysis yields an average offset of $1.26\degr\pm0.11\degr$ for the former and
$-0.50\degr\pm0.12\degr$ for the latter. These are the values we consider the best-guess for the
instrumental rotation. All polarization angles reported in this paper have been corrected for this
rotation.

There is evidence that the instrumental rotation depends on elevation. A Spearman's test over all points
on the Moon's limb yielded a $\rho$ of 0.69 ($p=0.01$) and 0.79 ($p=0.02$) for the 4.85 GHz and 8.35 GHz,
respectively. Aside from this being a low-significance result it is also based on a narrow elevation range
($\sim$32.5$\degr$--50$\degr$). Yet, it is an indication that the instrumental rotation may have a more
complex behavior.

\subsection{Correcting for instrumental circular polarization}
\label{subsec:LR_gain_corr}

\begin{figure}[!ht]
  \centering 
    \includegraphics[trim =0 0 0 0,clip,width=.45\textwidth]{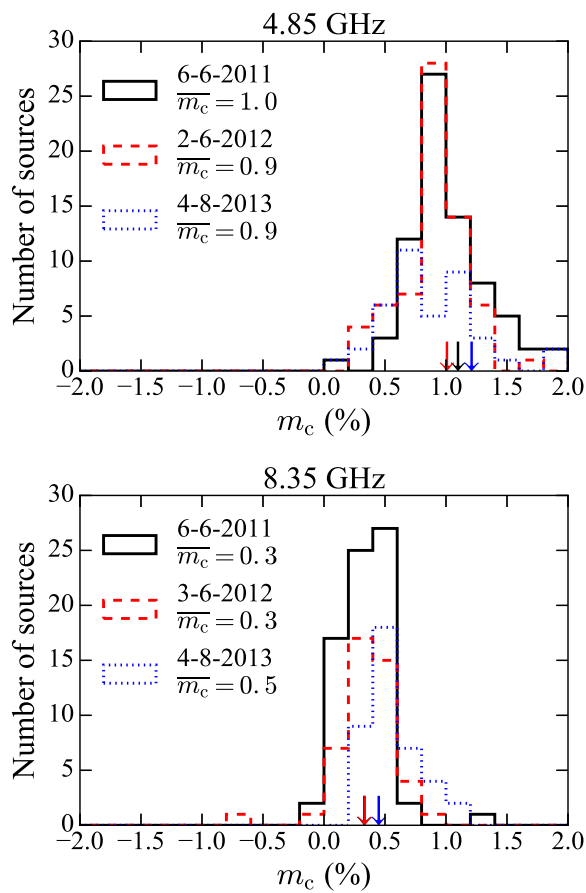} 
    \caption{Distributions of the circular polarization degree, $m_{\mathrm{c}}$, measurements at 4.85 (top)
      and 8.35~GHz (bottom) for three observing sessions. The non-zero average values, shown in the legend,
      indicate the presence of instrumental circular polarization. The $m_{\mathrm{c}}$ measurements of the
      planetary nebula NGC\,7027, in the corresponding sessions, are marked with arrows (in the bottom plot,
      two arrows overlap).}
  \label{fig:mc_distros}
\end{figure}

Imbalances between LCP and RCP channels similar to the ones discussed in Sect.~\ref{subsec:instrumental_evpa} 
for COS and SIN, can introduce instrumental circular polarization. Two effects with which the instrumental 
circular polarization is manifested are:
\begin{enumerate}
\item 
  As we show in Fig.~\ref{fig:mc_distros}, the distributions of the circular polarization degree,
  $m_{\mathrm{c}}$, measurements are centered around a non-zero value.
\item We measure systematically non-zero circular polarization from circularly unpolarized sources. An example
  is the case of the planetary nebula NGC\,7027 (point-like at the two frequencies we consider), a free-free
  emitter expected to be circularly and linearly unpolarized, for which non-zero circular polarization is
  measured (Fig.~\ref{fig:mc_distros}).
\end{enumerate}

\begin{figure}[!ht]
  \centering 
    \includegraphics[width=\hsize]{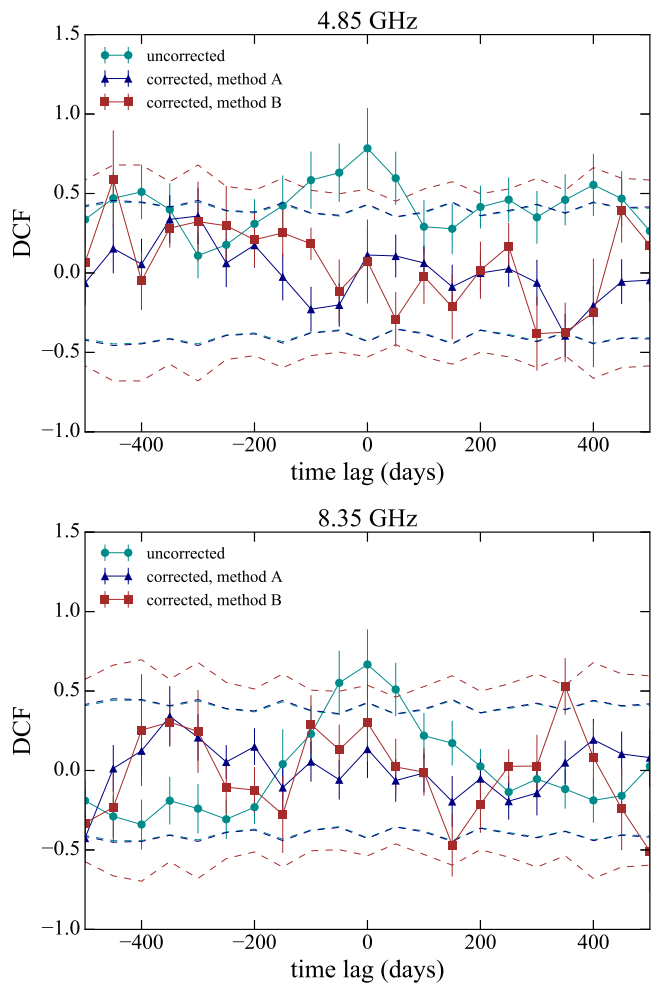} \\ 
    \caption{Discrete correlation function (DCF) between the $m_{\mathrm{c}}$ light curves of 4C\,+38.41 and
      CTA\,102 before (cyan circles) and after the instrumental circular polarization correction using either method A
      (blue triangles) or method B (red squares) as described in the text. The corresponding 3$\sigma$ significance levels are
      shown with dashed lines.}
  \label{fig:dcfs}
\end{figure}

The instrumental circular polarization may be as high as $\sim$0.5~\% to 1~\% -- comparable to the 
mean population values at those frequencies -- and shows significant variability. The latter is indicated 
by the significantly correlated, concurrent variability we observe in $m_{\mathrm{c}}$ light curves of 
different sources. For example, in Fig.~\ref{fig:dcfs}, we plot the locally normalized discrete correlation
function \citep[DCF,][]{Lehar1992,Edelson1988} between the $m_{\mathrm{c}}$ light curves of two randomly
chosen bright sources in our sample, namely 4C\,+38.41 and CTA\,102 before and after the instrumental
polarization correction. For the uncorrected data, the most prominent maxima appear at zero time lag where
the correlation factor is $0.8\pm0.2$ and $0.7\pm0.2$ for the 4.85~GHz and 8.35~GHz data, respectively. 
These are also the only DCF maxima above the 3$\sigma$ significance level.

\subsubsection{Correction methods}
\label{subsec:instrumental_org_pol_remove}

If $L_{\mathrm{D}}$ and $R_{\mathrm{D}}$ is the diode signal in the LCP and RCP channels,
the measured amplitudes of a source $i$ in a session $j$ expressed in diode units, will be:
\begin{align}
L'_{ij} &= \frac{L_{ij}}{L_{\mathrm{D},j}} \label{eq:Lobs} ~~\mathrm{and} \\
R'_{ij} &= \frac{R_{ij}}{R_{\mathrm{D},j}} \label{eq:Robs} ,
\end{align}
where $L_{ij}$ and $R_{ij}$ are the source incident signals modulated only by the channel gain imbalance. The
degree of circular polarization of the incident radiation can be recovered from the measured amplitudes
$L'_{ij}$ and $R'_{ij}$ by estimating the ratio $r=R_{\mathrm{D}}/L_{\mathrm{D}}$. Under the assumption that the
noise diode is truly circularly unpolarized $r$ becomes unity. In reality that is not the case and
instrumental circular polarization emerges.

Using Eqs.~\ref{eq:I}, \ref{eq:V}, \ref{eq:Lobs} and \ref{eq:Robs}, the corrected circular polarization degree 
can be written as:
\begin{equation}
m_{\mathrm{c},ij} =  \frac{R_{ij} - L_{ij}}{L_{ij} + R_{ij}} = \frac{m'_{\mathrm{c},ij} r_{j} + m'_{\mathrm{c},ij} + r_{j} - 1}{m'_{\mathrm{c},ij} r_{j} - m'_{\mathrm{c},ij} + r_{j} + 1} ,
\label{eq:mc_corr}
\end{equation}
where,
\[
\begin{array}{lp{0.8\linewidth}}
  m'_{\mathrm{c},ij}    &the measured circular polarization degree estimated using $L'_{ij}$ and $R'_{ij}$
                                     \begin{equation}
                                     m'_{\mathrm{c},ij} = \frac{R'_{ij} - L'_{ij}}{L'_{ij} + R'_{ij}} ~.
                                     \end{equation}
\end{array}
\]
Thus, in order to recover the corrected circular polarization degree $m_{\mathrm{c},ij}$, we need to determine 
the ratio $r_{j}$. In the following we show two independent methods to compute it (Fig.~\ref{fig:block_diagram}, level L11).

\paragraph{Method A: Zero-level of $m_{\mathrm{c}}$}\mbox{}\\\\
The first method relies on the determination of the circular polarization degree $m'_{\mathrm{c},ij}$ that we
would measure if the incident radiation was circularly unpolarized (zero level), i.e. $m_{\mathrm{c},ij} =
0$. We consider two estimates of the zero level:
\begin{enumerate}
\item the circular polarization degree of unpolarized sources (e.g. NGC\,7027), and
\item the average circular polarization degree of a sufficiently large, unbiased collection of sources.
\end{enumerate}
We then compute $r_{j}$ by using either of these measures as $m'_{\mathrm{c},ij}$ in Eq.~\ref{eq:mc_corr} and
setting $m_{\mathrm{c},ij} = 0$. To avoid biases caused by small number statistics, we used the second measure
only for sessions where at least 20 sources were observed. To make sure that the average circular 
polarization degree is not affected by sources which are significantly polarized, we applied an iterative process
to exclude them from the calculation similar to the \textit{gain transfer} technique presented in 
\citet{Homan2001} and \citet{Homan2006}.

\begin{figure}[!ht]
  \centering 
    \includegraphics[width=\hsize]{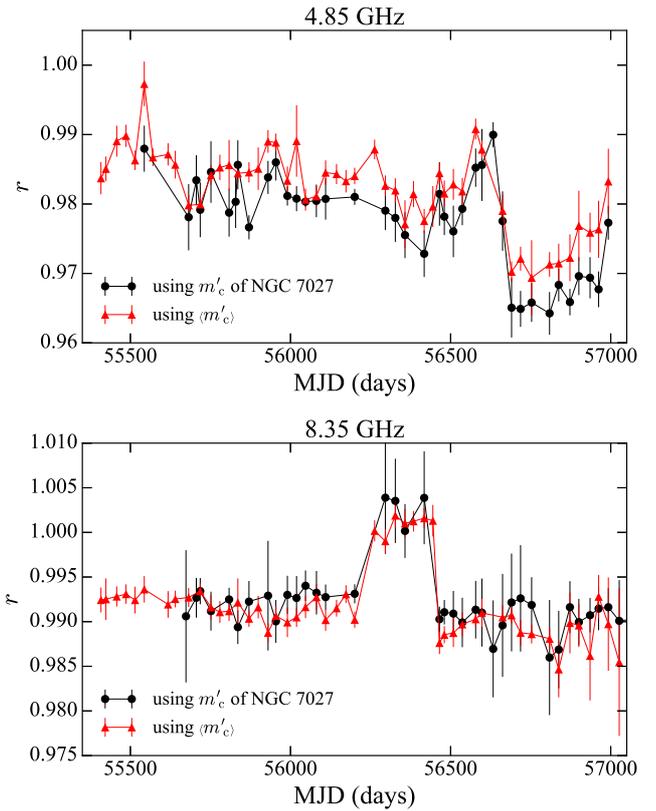} \\ 
    \caption{The ratio $r$ estimated using both measures of the circular polarization zero-level as described in 
    method A.}
  \label{fig:LRgcs}
\end{figure}

As shown in Fig.\ref{fig:LRgcs}, the $r_{j}$ values calculated by either of the two measures are in
excellent agreement. Depending on the data availability, the one or the other measure was used. For the
subset of sessions where both measures were available the average $r_{j}$ was used.

\paragraph{Method B: Singular value decomposition (SVD)}\mbox{}\\\\
The second method requires the presence of a number of stable circular polarization sources within our
sample, independently of whether they are polarized or not. If we divide Eqs.~\ref{eq:Lobs} and 
\ref{eq:Robs} we get
\begin{equation}
\frac{L'_{ij}}{R'_{ij}} = r_{j} \frac{L_{ij}}{R_{ij}} \Rightarrow Q'_{ij} = r_j Q_{ij}\\
\label{eq:Qobs}
\end{equation}
For a source $i$ with constant circular polarization, $Q_{ij}$ does not depend on the session $j$, so that
$Q_{ij}=Q_{i}$. Consequently, any variability seen in the measured $Q'_{ij}$ for these sources can only be
induced by the system though $r_{j}$. For the sources of constant circular polarization it can be written
\begin{equation}
Q'_{ij} = Q_{i} r_{j}
\label{eq:Qij_cals}
\end{equation}
or in matrix-vector form
\begin{equation}
\mathbf{Q}'= \vec{q} \vec{r}^\mathrm{T} ~.
\label{eq:Q_cals}
\end{equation}
Using the singular value decomposition method \citep[SVD, e.g.][]{Golub2013}, we can express matrix
$\mathbf{Q}'$ as a sum of matrices each of which has rank one.  For $i=1,\ldots, n$ sources with stable
circular polarization observed over $j=1,\ldots, m$ sessions and $n\le m$, it will be
\begin{equation}
\mathbf{Q'} =\sigma_1\vec{u}_1\vec{v}_1^{\mathrm{H}}+\dots+\sigma_n\vec{u}_n\vec{v}_n^{\mathrm{H}} ~,
\label{eq:Qsvd}
\end{equation}
where $\sigma_i$ are the singular values of the matrix $\mathbf{Q'}$ in decreasing magnitude and $\vec{u_i}$ and
$\vec{v_i}$ are its left- and right-singular vectors, respectively. $\vec{u_i}$ is of length $n$ and
$\vec{v_i}$ of length $m$.
If $\sigma_1 / \sigma_2 \gg 1$, $\mathbf{Q'}$ can be approximated by the first term in Eq.~\ref{eq:Qsvd} and one can write,
\begin{equation}
\mathbf{Q'} = \vec{q} \vec{r}^\mathrm{T} \simeq \sigma_1\vec{u}_1\vec{v}_1^{\mathrm{H}} ~.
\label{eq:SVDapprox}
\end{equation}
This implies that the unknown vectors $\vec{q}$ and $\vec{r}$ are parallel to $\vec{u}_1$ and $\vec{v}_1$, 
respectively:
\begin{align}
\vec{q} &= a \vec{u}_1 \label{eq:q_fin}\\
\vec{r}  &= b \vec{v}_1 \label{eq:r_fin}
\end{align}
with $a b = \sigma_{1}$. In order to solve for $a$ and $b$, we need at least one source of known 
circular polarization included in the list of $n$ stable sources. Assuming that this
is the first source of the set ($i=1$), $Q_{1}$ can be calculated from
\begin{equation}
Q_{1} = \frac{1 - m_{\mathrm{c,1}}}{m_{\mathrm{c,1}} + 1}
\label{eq:Q_known}
\end{equation}
where $m_{\mathrm{c,1}}$ is its known circular polarization degree, and using Eq.~\ref{eq:Qij_cals} we can write
\begin{equation}
\sum_{j} Q'_{1j}=\sum_{j} Q_{1} r_{j} = Q_{1} \sum_{j} r_{j}
\label{eq:sumQ_unpol}
\end{equation}
Additionally, from Eq.~\ref{eq:r_fin}, we have
\begin{equation}
\sum_{j} r_{j} = \sum_{j} b v_{j}\bigg|_1 = b \sum_{j} v_{j}\bigg|_1
\label{eq:sum_r}
\end{equation}
where $\sum_{j} v_{j}|_{1}$ is the sum of all the elements of vector $\vec{v}_{1}$. Therefore, using 
Eqs.~\ref{eq:sumQ_unpol} and \ref{eq:sum_r}, the factor $b$ can be computed as:
\begin{equation}
b = \frac{\sum_{j} r_{j}}{\sum_{j} v_{j}\big|_1} = \frac{\sum_{j} Q'_{1j}}{Q_{1} \sum_{j} v_{j}\big|_1}
\end{equation}
and the factor $a = \sigma_{1} / b$. Once $b$ has been computed, Eq.~\ref{eq:r_fin} will give us vector
$\vec{r}$, the elements of which are the circular polarization correction factors $r_j$ to be used in
Eq.~\ref{eq:mc_corr}.

The SVD methodology was implemented using three sources: NGC\,7027, 3C\,48 and 3C\,286. This subset of 
sources was selected as the best candidates with stable circular polarization based on the following 
criteria:
\begin{enumerate}
\item Stability of the observed circular polarization (even being unpolarized). Assuming that the observed variability 
is a superposition of the instrumental and the intrinsic polarization variability, the sources with the lowest 
$m'_{\mathrm{c}}$ variability are the best candidates to be intrinsically stable.
\item More frequently observed. This criterion ensures that we can apply Method B to as many sessions as possible 
and account for the instrumental polarization that can show pronounced variability even in short timescales (Fig.~\ref{fig:LRgcs}).
\end{enumerate}
For the analyzed datasets, the sources 3C\,286 and 3C\,48 best fulfill both of the above criteria. NGC\,7027 was 
selected as the source assumed to have known circular polarization. Its free-free emission is expected to be 
circularly unpolarized ($m_{\mathrm{c}}=0$) and hence its $Q=1$ according to Eq.~\ref{eq:Q_known}.
For a session with no NGC\,7027 data, we adopted a mock source of zero circular polarization to which we assigned 
as observed value the $m'_{\mathrm{c}}$ averaged over all sources in that session. A minimum of 20 sources was 
required in those cases. The circular polarization degree values for the other sources assumed stable 
(3C\,286 and 3C\,48 in our case) are not required for Method B, which is one of the main advantages of this calibration 
technique.

\begin{figure}[!ht]
  \centering 
    \includegraphics[width=\hsize]{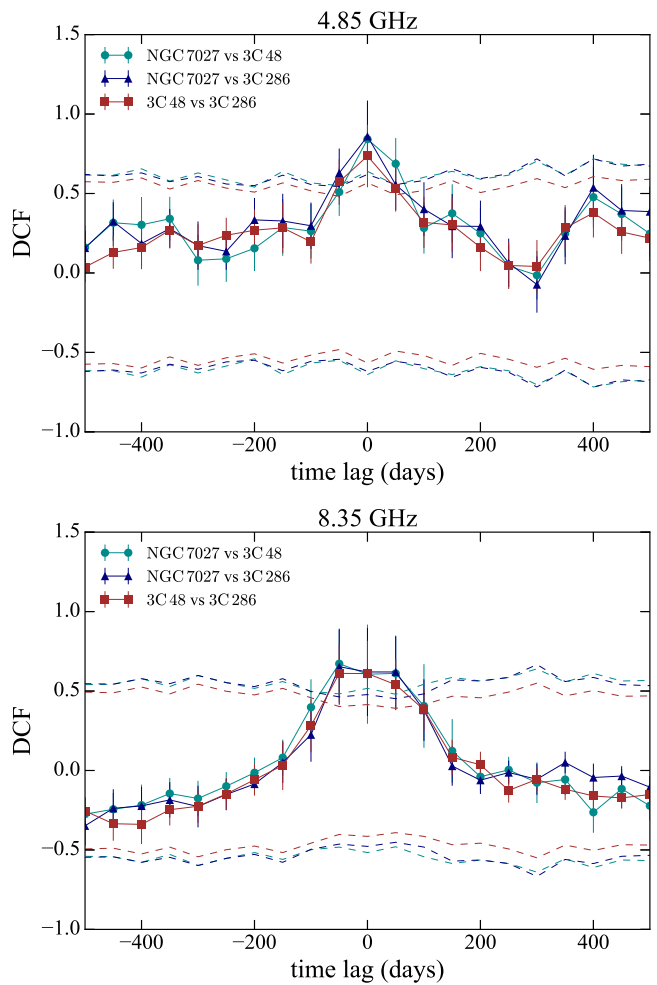} \\ 
    \caption{Discrete correlation function (DCF) between the $m'_{\mathrm{c}}$ light curves of all source
      pairs including NGC\,7027 and 3C\,48 and 3C\,286. The most prominent maxima above the the 5$\sigma$
      significance levels, indicated by the corresponding dashed lines, appear around zero time lag where the
      correlation factor is $\sim0.8\pm0.2$ and $\sim0.6\pm0.2$ for the 4.85~GHz and 8.35~GHz data,
      respectively. The dashed lines indicate the 5$\sigma$ significance level.}
  \label{fig:dcfs_cals}
\end{figure}

The circular polarization stability of the selected sources is also advocated by:  
\begin{enumerate}
\item The fact that they display the most significantly correlated, concurrent variability in circular polarization 
(Fig.~\ref{fig:dcfs_cals}). Their low circular polarization variability is supported by the coincidence of all lines there, 
as well as by the fact that for them the zero time lag correlation exceeds the 5$\sigma$ threshold.
\item The $\sigma_{1} / \sigma_{2}$ ratios for the 4.85~GHz and 8.35~GHz data using these sources were 
$\sim 521$ and $\sim 540$ ($\sim27$dB), respectively justifying the approximation of $Q'_{ij}$ with a 
single rank-one matrix.
\end{enumerate}

\subsubsection{Comparison between methods A, B and the UMRAO database}
\label{subsubsec:umrao_comparison}

In Fig.~\ref{fig:dcfs} we show the DCF of corrected circular polarization data for 4C\,+38.41 and
CTA\,102. The data corrected with methods A and B are shown separately. The improvement is directly evident in
the radical decrease of the correlation factors at zero time lag. In fact, the zero
time lag correlation does not exceed the 1$\sigma$ level.

\begin{figure}[!ht]
  \centering 
    \includegraphics[width=\hsize]{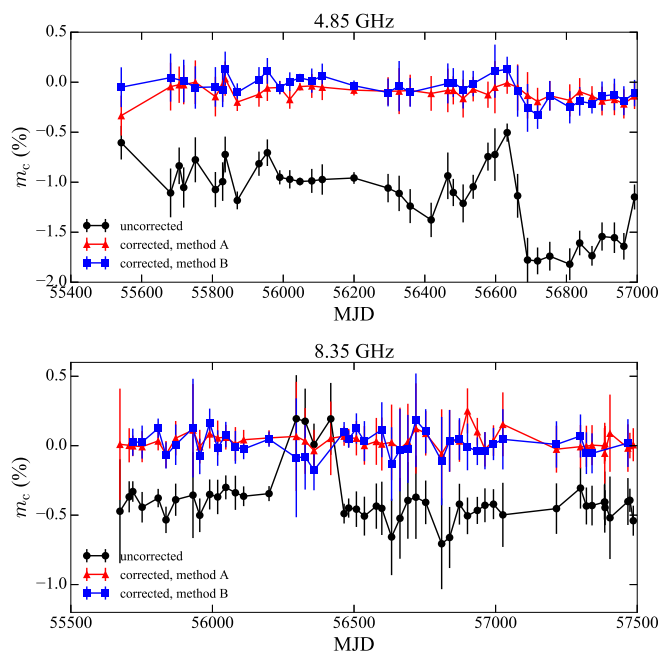} \\ 
    \caption{The $m_{\mathrm{c}}$ measurements of NGC\,7027 at 4.85~GHz and 8.35~GHz before and after the
      instrumental circular polarization correction using both methods A and B.}
  \label{fig:NGC7027_mc}
\end{figure}

Figure~\ref{fig:NGC7027_mc} now shows the $m_{\mathrm{c}}$ measurements of NGC\,7027 before and after the
correction for instrumental circular polarization with methods A and B. For the uncorrected data sets, we
systematically measure non-zero $m_{\mathrm{c}}$ with standard deviations of 0.4~\% and 0.2~\% at
4.85~GHz and 8.35~GHz, respectively. The corrected $m_{\mathrm{c}}$ on the other hand, from
methods A and B: are in excellent agreement; appear very close to zero; and their standard deviations are
reduced to 0.1~\%.

For the sources with the most stable behavior, Table~\ref{tab:mc_candidates} lists their circular polarization
measurements. Methods A and B agree well within the errors. Method B gives on average 0.016~\% smaller
standard deviations than method A does. This is most likely caused because method B assumes more sources with
constant polarization than method A.

Finally, we compared the circular polarization measurements from method A and B with measurements from the
UMRAO monitoring program \citep{Aller2013,Aller2016}. The comparison was performed for five sources with overlapping
data sets from both monitoring programs ($\sim$2010.5--2012.3), namely 3C\,84, OJ\,287, 3C\,279,
BL\,Lacertae and 3C\,454.3. Specifically, we compared all concurrent data points within a maximum separation 
of 2 weeks. There are 110 and 59 such data points of the UMRAO data set overlapping with the results of methods A and 
B, respectively. In both cases we found a median absolute difference in the circular polarization degree measurements of 
only 0.2~\%. The corresponding data sets for the time range embracing the overlapping period are shown 
in Fig.~\ref{fig:umrao_comparison}. The comparison of such contemporaneous measurements is particularly important because 
it can be used to detect or put strict limits on the very rapid variations usually observed in circular polarization.

\section{Comparison with the M\"uller matrix method}
\label{sec:M-matrix_comparison}

Traditionally the M\"uller matrix method has been the one adopted for treating the instrumental
polarization. Here we wish to carry out a comparison with our method.

The M\"uller matrix method is based on estimating the elements of the M\"uller matrix $\mathbf{M}$, which 
is a transfer function between the incident $S_{\mathrm{real}}$ and the measured $S_{\mathrm{obs}}$ Stokes 4-vectors:
\begin{equation}
S_{\mathrm{obs}} = \mathbf{M} \cdot S_{\mathrm{real}}
\label{eq:M_corr}
\end{equation}

A set of four independent measurements of sources with known $S_{\mathrm{real}}$ are enough to solve the
system of equations~\ref{eq:M_corr} and compute the $\mathbf{M}$ matrix elements. In case more measurements
are available a fit can determine the best-guess values. The inverse $\mathbf{M}$ matrix is then applied to
$S_{\mathrm{obs}}$ to correct for instrumental effects.

As we discussed in Sect.~\ref{subsec:instrument model}, our methodology models and corrects for the instrumental 
linear polarization across the whole beam before extracting the Stokes $Q$ and $U$ data, while the
M\"uller method has no handle on this. This can be essential for cases of low $Q$ or $U$ amplitudes which can
be corrupted to the point that the telescope response pattern cannot be seen in the data. For the majority of
such cases our methodology was able to recover the telescope response pattern (e.g. $Q_{\mathrm{elv}}$ in
Fig.~\ref{fig:artifact_corr}). In milder cases our methodology resulted in peak offsets closer to the source
position and FWHM values closer to the actual ones.

Another advantage of our approach is the milder conditions it requires. As we discussed in
Sect.~\ref{subsec:instrumental_org_pol_remove}, the only requirement for method A is the observation of a
circularly unpolarized source and for method B $n$ sources with constant circular polarization with the
need to know the exact value of $m_\mathrm{c}$ only for one of them. The M\"uller method on
the other hand, requires a good coverage of the Stokes parameter space and particularly for $V$.

To perform a quantitative comparison of the two techniques, we focused on the linear polarization
results. First, we calculated $I$, $Q$ and $U$ for all sources and observing sessions using both our
methodology and the M\"uller method, accounting for all post-measurement corrections described in
Sect.~\ref{subsec:corrections}. For each session, the 3x3 M\"uller matrix was determined using all
observations on the polarization calibrators shown in Table~\ref{tbl:pol_cals}.
\begin{table}[!ht]
\caption{The polarization calibrators used for the M\"uller matrix method. Their parameters were provided by Dr. A. Kraus (priv.comm.).}
  \label{tbl:pol_cals} 
  \centering
  \begin{tabular}{lrrrr}
\hline\hline
Source  & Freq.  & $I$   & $m_{\mathrm{l}}$ & $\chi$    \\
        & (GHz)  & (Jy)  &  (\%)            & ($\degr$) \\
\hline\\
3C\,286   & 4.85  & 7.48   & 11.19   &  33.0 \\
          & 8.35  & 5.22   & 11.19   &  33.0 \\
3C\,295   & 4.85  & 6.56   &  0.00   &   0.0 \\
          & 8.35  & 3.47   &  0.93   &  28.9 \\
3C\,48    & 4.85  & 5.48   &  4.19   & 106.6 \\
          & 8.35  & 3.25   &  5.39   & 114.5 \\
NGC\,7027 & 4.85  & 5.48   &  0.00   &   0.0 \\
          & 8.35  & 5.92   &  0.00   &   0.0 \\
\hline
  \end{tabular}
\end{table}

As a figure of merit for the comparison, we used the intra-session variability of $I$, $Q$ and $U$ in
terms of their standard deviation $\sigma_{I,Q,U}$. Since our sources are not expected to vary within a
session (not longer than 3 days), any variability in $I$, $Q$ and $U$ can naturally be attributed to
instrumental effects. Consequently, the technique leading to lower variability must be providing a better
handling of the instrumental effects. In our study we included measurements with linearly polarized flux of at
least 15~mJy.

We performed two-sample Kolmogorov-Smirnov (KS) tests to compare the corresponding $\sigma_{I,Q,U}$ distributions 
between the two techniques for three polarized flux ranges: all, high polarization ($\ge 100$~mJy), low 
polarization (15~mJy -- 100~mJy). The results for the 4.85~GHz receiver are presented in Table~\ref{tbl:MP_comparison} 
along with the median $\sigma_{I,Q,U}$ values for either of the two techniques. The cases where the KS test rejects 
the null hypothesis that the two distributions are the same at a level greater than $5\sigma$ are marked with an asterisk.

For Stokes $I$, both methods perform equally well since the corresponding KS-test results show no significant 
difference between the $\sigma_{I}$ distributions of the M\"uller method and our methodology. That is also the 
case for Stokes $Q$ and $U$ of the high polarization data.

Our approach performs significantly better 
for Stokes $Q$ and $U$ when we consider either the complete dataset (all) or the low polarization data. 
The corresponding KS-test results show that the $\sigma_{Q,U}$ distributions of the M\"uller method and our 
methodology are significantly different above the level of $5\sigma$. A direct comparison of the median 
$\sigma_{Q,U}$ values shows that our method delivers $\sim8\%$ and $\sim28\%$ more stable results for the 
complete dataset (all) and the low polarization data, respectively. The main reason for this 
improvement is the instrumental linear polarization correction scheme of our methodology that treats each 
sub-scan separately, accounting for the instrumental polarization contribution across the whole beam 
(Sect.~\ref{subsec:instrument model}).

\begin{table*}[!ht]
\caption{The KS-test results (KS statistic, $D$, and $p$-value) for the comparison between the $\sigma_{I,Q,U}$ 
distributions for the 4.85~GHz data calibrated by the M\"uller method and our methodology. The results are presented for three polarized 
flux ranges: all, high polarization ($\ge 100$~mJy) and low polarization (15~mJy -- 100~mJy). In column (4) we list 
the significance level at which the KS test null hypothesis can be rejected and in columns (5) and (6) we provide the 
median values of the corresponding $\sigma_{I,Q,U}$ distributions}
  \label{tbl:MP_comparison} 
  \centering
  \begin{tabular}{lrr@{$\times$}lcr@{$\pm$}lr@{$\pm$}l}
\hline\hline
Stokes & $D$  & \mc{2}{c}{$p$} & Significance & \mc{2}{c}{median $\sigma_{I,Q,U}$} & \mc{2}{c}{median $\sigma_{I,Q,U}$} \\
       &      & \mc{2}{c}{}    &  level       & \mc{2}{c}{[M\"uller method]}       & \mc{2}{c}{[this work]}             \\
       &      & \mc{2}{c}{}    &              & \mc{2}{c}{(mJy)}                   & \mc{2}{c}{(mJy)}                   \\
\hline\\
All \\
$I$       & 0.05 & 1.6 & $10^{-1}$  & 1.4$\sigma$ & 11.829 & 0.149 & 12.780 & 0.074 \\
$^{*}Q$   & 0.15 & 1.2 & $10^{-10}$ & 6.4$\sigma$ & 1.619  & 0.007 & 1.467  & 0.007 \\
$^{*}U$   & 0.13 & 4.7 & $10^{-8}$  & 5.5$\sigma$ & 1.431  & 0.008 & 1.324  & 0.006 \\
\\
High polarization \\
$I$       & 0.06 & 2.8 & $10^{-1}$  & 1.1$\sigma$ & 24.911 & 0.161 & 22.988 & 0.105 \\
$Q$       & 0.06 & 2.8 & $10^{-1}$  & 1.1$\sigma$ & 2.376  & 0.007 & 2.357  & 0.007 \\
$U$       & 0.08 & 8.8 & $10^{-2}$  & 1.7$\sigma$ & 2.322  & 0.008 & 2.247  & 0.008 \\
\\
Low polarization \\
$I$       & 0.09 & 2.4 & $10^{-2}$  & 2.3$\sigma$ & 6.417  & 0.132 & 7.379  & 0.019 \\
$^{*}Q$   & 0.24 & 1.3 & $10^{-14}$ & 7.7$\sigma$ & 1.099  & 0.006 & 0.730  & 0.004 \\
$^{*}U$   & 0.20 & 4.7 & $10^{-10}$ & 6.2$\sigma$ & 1.009  & 0.006 & 0.776  & 0.004 \\
\hline
  \end{tabular}
\end{table*}

\section{Sources with stable polarization}
\label{sec:standards_polarization}

The methodology described in Sect.~\ref{sec: FullS_polarimetry} was used to compute the linear and circular
polarization parameters of the observed sources at 4.85~GHz and 8.35~GHz. Once all four Stokes parameters 
have been computed, the degree of linear and circular polarization, $m_{\mathrm{l}}$ and $m_{\mathrm{c}}$ 
and the polarization angle $\chi$, were calculated as:
\begin{equation}
\label{eq:ml}
m_{\mathrm{l}} = \frac{\sqrt{Q^{2} + U^{2}}}{I} ~, \\
\end{equation}
\begin{equation}
\label{eq:mc}
m_{\mathrm{c}} = \frac{V}{I} ~,\\
\end{equation}
\begin{equation}
\label{eq:chi}
\chi = \frac{1}{2} \arctan \frac{U}{Q} ~.
\end{equation}
The corresponding errors were computed as the Gaussian error propagation of the uncertainties in the LCP, RCP,
COS, and SIN amplitudes through Eqs.~\ref{eq:I}--\ref{eq:V}, \ref{eq:lp_domains} and \ref{eq:ml}--\ref{eq:chi}. 
Finally, for each observing session, we computed the weighted average and standard deviation of the polarization 
parameters for all sub-scans on a given source, using their errors as weights.

Our data set includes a total of 155 sources and was searched for cases of stable linear and circular
polarization characteristics to be listed as reference sources for future polarization observations.
Because we are interested in identifying only cases with stable polarization parameters, we restricted our
search to a sub-sample of 64 sources that were observed:
\begin{itemize}
\item for at least three years, and 
\item with a cadence of one measurement every 1 to 3 months.
\end{itemize}

\begin{figure*}[!ht]
  \centering 
    \includegraphics[width=\hsize]{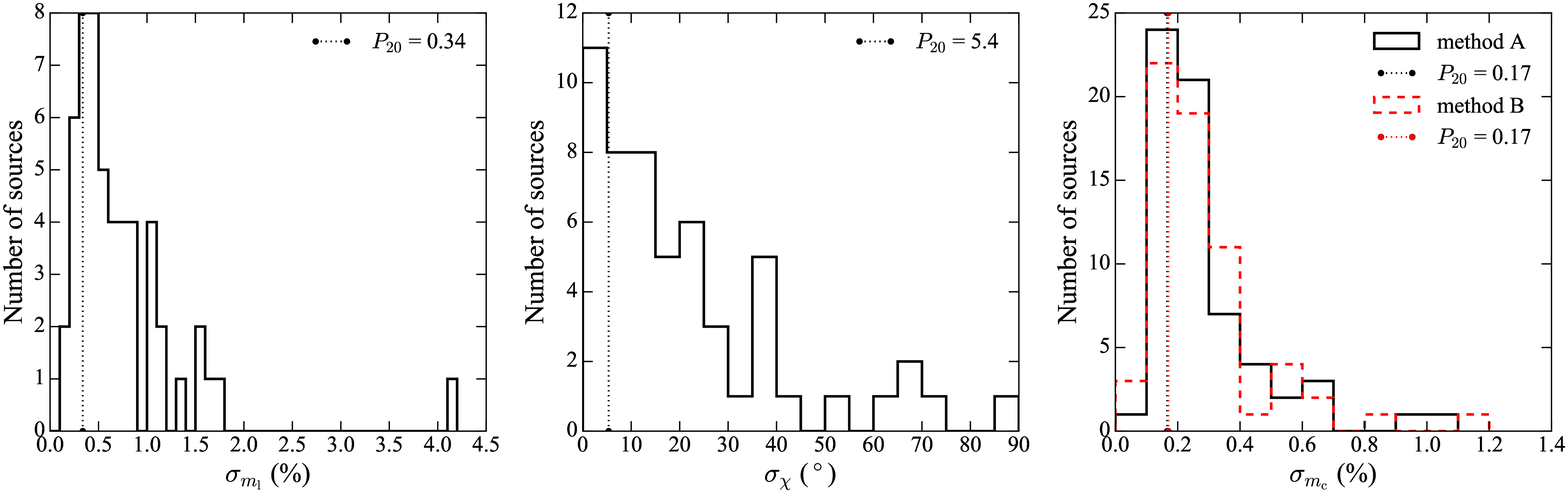} \\ 
    \includegraphics[width=\hsize]{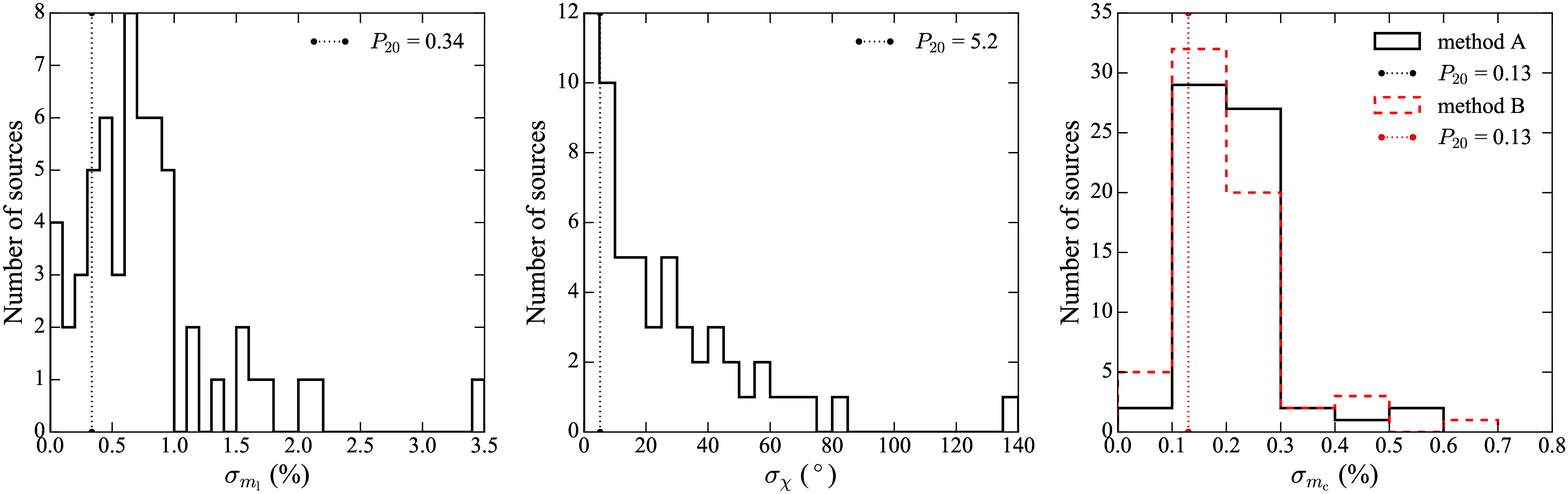} 
  \caption{Standard deviation distributions of the linear polarization degree, $\sigma_{m_{\mathrm{l}}}$,
    polarization angle, $\sigma_{\chi}$, and circular polarization degree, $\sigma_{m_{\mathrm{c}}}$,
    measurements at 4.85~GHz (top row) and 8.35~GHz (lower row). The 20th percentile, $P_{20}$, of each standard deviation distribution is marked by a dotted line. The circular polarization data contain the measurements corrected with both methods A (continuous, black line) and B (dashed, red line) as described in Sect.~\ref{subsec:instrumental_org_pol_remove}.}
  \label{fig:pol_std_distros_60}
\end{figure*}

In Figure~\ref{fig:pol_std_distros_60} we show the distributions of standard deviations $\sigma_{m_{\mathrm{l}}}$, 
$\sigma_{m_{\mathrm{c}}}$ and $\sigma_{\chi}$ at 4.85~GHz and 8.35~GHz. The sources exhibit a broad range of 
variability in all polarization parameters. The threshold for our search was set to the 20th percentile, 
$P_{20}$, i.e. one fifth of the corresponding standard deviation distribution, marked by the dotted lines in 
those plots.

In Tables~\ref{tab:ml_candidates} and \ref{tab:chi_candidates}, we list the sources with the most stable
$m_{\mathrm{l}}$ and $\chi$ values at 4.85~GHz and 8.35~GHz.  The names of sources for which both
$m_{\mathrm{l}}$ and $\chi$ were found to be stable at at least one observing frequency are marked in bold
face. The reported sources, exhibited significant linear polarization at least 95~\% of the times they were
observed. Significant linear polarization measurements are consider those with weighted mean of $m_{\mathrm{l}}$
at least three times larger than the weighted standard deviation in the corresponding session.
In Table~\ref{tab:no_ml_candidates} we list linearly unpolarized sources, that is sources with no significant 
linear polarization measurements. For the latter, we report the average values of their $m_{\mathrm{l}}$ 
3$\sigma$ upper limits.

\begin{table}[!ht]
  \caption{Sources with stable linear polarization degree, $m_{\mathrm{l}}$.  For each entry the upper row corresponds to 4.85~GHz and the lower to 8.35~GHz. The rows corresponding to frequencies at which the source was found to be either unpolarized or variable are filled with ``$\ldots$''. Sources with both stable $m_{\mathrm{l}}$ and $\chi$ are marked in bold face.}
  \label{tab:ml_candidates}
  \centering
  \begin{tabular}{l@{\hskip 0.3cm}cccccc}
    \hline\hline
Source name & $\Delta t_{\mathrm{obs}}$ & $N_{\mathrm{obs}}$ & $N_{m_{\mathrm{l}}}$ & $\left< I \right>$ &$\left< m_{\mathrm{l}} \right>$ & $\sigma_{m_{\mathrm{l}}}$ \\ 
& (yrs) & & & (Jy) & (\%) & (\%) \\ 
   \hline\\
   \textbf{3C\,286} & 5.7    &   78    &    78   & 7.42   &    11.26   &    0.22 \\
                    & 5.7    &   75    &    75   & 5.11   &    11.88   &    0.07 \\
   \textbf{3C\,295} & \ldots &  \ldots &  \ldots & \ldots &  \ldots    & \ldots  \\
                    & 5.5    &   47    &    47   & 3.37   &     0.93   &    0.07 \\
   \textbf{3C\,48}  & 5.6    &   68    &    68   & 5.47   &     4.24   &    0.13 \\
                    & 5.6    &   69    &    69   & 3.20   &     5.61   &    0.07 \\
     NRAO\,150      & 4.3    &   32    &    32   & 7.45   &     1.71   &    0.16 \\
                    & 4.3    &   33    &    33   & 9.14   &     1.27   &    0.22 \\
    MKN\,501        & \ldots &  \ldots &  \ldots & \ldots &  \ldots    & \ldots  \\
                    & 4.4    &   43    &   41    & 1.29   &     1.63   &    0.33 
  \\\hline                                  
  \end{tabular}
  \tablefoot{The entry in each column is as follows: (1) The source survey name, 
  (2) the period that the source was observed, (3) the number of sessions that the source was 
  observed, (4) the number of sessions that significant $m_{\mathrm{l}}$ was detected, (5) the 
  average value of Stokes $I$, (6) the average value of $m_{\mathrm{l}}$ and (7) the standard 
  deviation of $m_{\mathrm{l}}$ over the entire data set.}
\end{table}

\begin{table}[!ht]
  \caption{Sources with stable polarization angle, $\chi$. For each entry the upper row corresponds to 4.85~GHz and the lower to 8.35~GHz. The rows corresponding to frequencies at which the source was found to be either unpolarized or variable are filled with ``$\ldots$''.  Sources with both stable $m_{\mathrm{l}}$ and $\chi$ are marked in bold face.}
  \label{tab:chi_candidates}
  \centering
  \begin{tabular}{l@{\hskip 0.3cm}cccccc}
    \hline\hline
Source name & $\Delta t_{\mathrm{obs}}$ & $N_{\mathrm{obs}}$ & $N_{\chi}$ & $\left< I \right>$ & $\left< \chi \right>$ & $\sigma_{\chi}$ \\ 
  & (yrs) & & & (Jy) & ($\degr$) & ($\degr$) \\ 
    \hline\\
 \textbf{3C\,286}  & 5.7   &   78  &    78 & 7.42   &  32.16 &    0.52 \\
                   & 5.7   &   75  &    75 & 5.11   &  33.07 &    0.23 \\
 \textbf{3C\,295}  &\ldots &\ldots &\ldots & \ldots & \ldots & \ldots  \\
                   & 5.5   &   47  &    47 & 3.37   &  31.90 &    3.68 \\
 \textbf{3C\,48}   & 5.6   &   68  &    68 & 5.47   & -73.65 &    0.98 \\
                   & 5.6   &   69  &    69 & 3.20   & -64.25 &    0.32 \\
 PKS\,0528+134     & 4.0   &   36  &    36 & 1.75   & -26.72 &    2.69 \\
                   &\ldots &\ldots &\ldots & \ldots & \ldots & \ldots  \\
 S5\,0836+71       & 4.0   &   33  &    32 & 2.46   & -79.53 &    1.39 \\
                   & 4.0   &   31  &    31 & 2.57   & -83.15 &    2.27 \\
 PKS\,1127-14      & 4.0   &   38  &    37 & 3.10   & -27.29 &    2.34 \\
                   & 4.0   &   38  &    38 & 2.71   & -28.96 &    2.44 \\
 3C\,273           & 4.3   &   52  &    52 & 35.84  & -26.90 &    1.66 \\
                   & 5.2   &   51  &    51 & 27.09  & -38.26 &    0.93 \\
 3C\,454.3         & 5.5   &   52  &    52 & 11.70  &   4.01 &    2.03 \\
                   &\ldots &\ldots   &\ldots    & \ldots  & \ldots & \ldots 
 \\\hline                                  
  \end{tabular}
  \tablefoot{The entry in each column is as follows: (1) The source survey name, 
  (2) the period that the source was observed, (3) the number of sessions that the source was 
  observed, (4) the number of sessions that significant $m_{\mathrm{l}}$ was detected, 
  (5) the average value of Stokes $I$, (6) the average value of $\chi$ and (7) the standard 
  deviation of $\chi$ over the entire data set.}
\end{table}

\begin{table}[!ht]
  \caption{Linearly unpolarized sources. For each entry the upper row corresponds to 4.85~GHz and the lower to 8.35~GHz. The rows corresponding to frequencies at which the source was found to be polarized or variable are filled with ``$\ldots$''.}
  \label{tab:no_ml_candidates}
  \centering
  \begin{tabular}{l@{\hskip 0.3cm}cccc}
    \hline\hline
Source name & $\Delta t_{\mathrm{obs}}$ & $N_{\mathrm{obs}}$ & $\left< I \right>$ & $m_{\mathrm{l}}$ \\ 
  & (yrs) & & (Jy) & (\%) \\ 
    \hline\\
3C\,295          & 5.5    &   49    & 6.54   &  $<  0.12$ \\
                 & \ldots &  \ldots & \ldots & \ldots     \\
NGC\,1052        & 4.0    &   36    & 1.24   &  $<  1.10$ \\
                 & 4.0    &   36    & 1.41   &  $<  0.56$ \\
NGC\,7027        & 5.3    &   53    & 5.37   &  $<  0.16$ \\
                 & 5.3    &   49    & 5.75   &  $<  0.14$ 
  \\\hline                                  
  \end{tabular}
  \tablefoot{The entry in each column is as follows: (1) The source survey name, 
  (2) the period that the source was observed, (3) the number of sessions that the source was 
  observed, (4) the average value of Stokes $I$ and (5) the upper limit of $m_{\mathrm{l}}$ over the 
  entire data set.}
\end{table}

Finally, in Table~\ref{tab:mc_candidates}, we list the sources with the most stable $m_{\mathrm{c}}$. For
comparison, we list the average and standard deviation values of $m_{\mathrm{c}}$, corrected with both methods
A and B as described in Sect.~\ref{subsec:instrumental_org_pol_remove}. Most of the sources in Table~\ref{tab:mc_candidates} 
exhibit mean $m_{\mathrm{c}}$ values very close to zero and therefore are considered circularly unpolarized. 
However, 3C\,286, 3C\,295, 3C\,48 and CTA\,102, show significant circular polarization 
($\left< m_{\mathrm{c}} \right>/\sigma_{m_{\mathrm{c}}} \ge 3$) at 4.85~GHz with at least one of the two correction methods.

\citet{Komesaroff1984} observed two of the sources presented in Table~\ref{tab:mc_candidates} between December 
1976 and March 1982, namely CTA\,102 and PKS\,1127-14. At that time CTA\,102 showed variable circular polarization 
degree, which suggests that its $m_{\mathrm{c}}$ cannot be considered stable over such long time scales. On the other 
hand, PKS\,1127-14 was found stable in $m_{\mathrm{c}}$, with a synchronous decrease in both Stokes $I$ and $V$. The 
average circular polarization degree of PKS\,1127-14 over that period was $m_{\mathrm{c}} \approx -0.1\pm0.03$, which 
is very close to the value listed in Table~\ref{tab:mc_candidates}. This finding places additional bounds on its 
$m_{\mathrm{c}}$ stability, suggesting that it may have remained unchanged for $\sim$40 years.

\begin{table*}[]
  \caption{Sources with stable circular polarization degree, $m_{\mathrm{c}}$. For each entry the upper row corresponds to 4.85~GHz and the lower to 8.35~GHz. The rows corresponding to frequencies at which the source was found to be variable are filled with ``$\ldots$''.} 
  \label{tab:mc_candidates}
  \centering
  \begin{tabular}{l@{\hskip 0.3cm}ccccccccc}
    \hline\hline
Source name & $\Delta t_{\mathrm{obs}}$ & $N_{\mathrm{obs}}$ & $\left< I \right>$ & $N_{m_{\mathrm{c,A}}}$ & $\left< m_{\mathrm{c,A}} \right>$ & $\sigma_{m_{\mathrm{c,A}}}$ & $N_{m_{\mathrm{c,B}}}$ & $\left< m_{\mathrm{c,B}} \right>$ & $\sigma_{m_{\mathrm{c,B}}}$ \\ 
&  (yrs) & & (Jy) & & (\%) & (\%) & & (\%) & (\%) \\ 
    \hline\\
3C\,286         & 5.7    & 78     & 7.42   &  53    &  -0.26 &   0.17 &  35    &  -0.36 &   0.09 \\
                & 5.7    & 75     & 5.11   &  61    &  -0.11 &   0.12 &  39    &  -0.13 &   0.09 \\ 
3C\,295         & 5.5    & 49     & 6.54   &  24    &  -0.57 &   0.13 &  16    &  -0.68 &   0.11 \\
                & 5.5    & 47     & 3.37   &  32    &  -0.16 &   0.08 &  20    &  -0.19 &   0.08 \\ 
3C\,48          & 5.6    & 68     & 5.47   &  50    &  -0.49 &   0.17 &  36    &  -0.60 &   0.09 \\
                & 5.6    & 69     & 3.20   &  56    &  -0.12 &   0.12 &  39    &  -0.15 &   0.08 \\ 
B2\,0218+35     & \ldots & \ldots & \ldots & \ldots & \ldots & \ldots & \ldots & \ldots & \ldots \\ 
                & 3.2    & 20     & 1.26   &  18    &   0.04 &   0.12 &  13    &   0.02 &   0.07 \\ 
4C\,+28.07      & \ldots & \ldots & \ldots & \ldots & \ldots & \ldots & \ldots & \ldots & \ldots \\ 
                & 4.2    & 40     & 3.20   &  38    &   0.14 &   0.12 &  23    &   0.15 &   0.10 \\
NRAO\,150       & \ldots & \ldots & \ldots & \ldots & \ldots & \ldots & \ldots & \ldots & \ldots \\ 
                & 4.3    & 33     & 9.14   &  33    &  -0.19 &   0.09 &  21    &  -0.21 &   0.11 \\
PKS\,1127-14    & 4.0    & 38     & 3.10   &  38    &  -0.20 &   0.17 &  29    &  -0.27 &   0.17 \\
                & \ldots & \ldots & \ldots & \ldots & \ldots & \ldots & \ldots & \ldots & \ldots \\ 
3C\,345         & \ldots & \ldots & \ldots & \ldots & \ldots & \ldots & \ldots & \ldots & \ldots \\ 
                & 4.0    & 36     & 6.54   &  35    &   0.05 &   0.10 &  21    &   0.04 &   0.13 \\
MKN\,501        & 4.4    & 43     & 1.44   &  42    &   0.02 &   0.14 &  27    &   0.10 &   0.15 \\ 
                & 4.4    & 43     & 1.29   &  42    &   0.04 &   0.12 &  28    &   0.02 &   0.12 \\
CTA\,102        & 4.3    & 44     & 4.57   &  44    &  -0.23 &   0.14 &  31    &  -0.32 &   0.08 \\
                & \ldots & \ldots & \ldots & \ldots & \ldots & \ldots & \ldots & \ldots & \ldots \\ 
NGC\,7027       & 5.3    & 53     & 5.37   &  39    &   0.10 &   0.08 &  36    &  0.06  &   0.11 \\
                & 5.3    & 49     & 5.75   &  49    &   0.03 &   0.05 &  39    &  0.01  &   0.08 
\\\hline                                  
  \end{tabular}
  \tablefoot{The entry in each column is as follows:  (1) The source survey name, 
  (2) the period that the source was observed, (3) the number of sessions that the source was 
  observed, (4) the average value of Stokes $I$ over the entire data set, (5) the number of 
  sessions that the $m_{\mathrm{c}}$ measurements were corrected with method A, (6) the 
  average value of $m_{\mathrm{c}}$, corrected with method A, (7) the standard deviation of 
  $m_{\mathrm{c}}$, corrected with method A, (8) the number of sessions 
  that the $m_{\mathrm{c}}$ measurements were corrected with method B, (9) the average value of 
  $m_{\mathrm{c}}$, corrected with method B, (10) the standard deviation 
  of $m_{\mathrm{c}}$, corrected with method B, over the entire data set.}
\end{table*}

The sources reported in Tables~\ref{tab:ml_candidates}--\ref{tab:mc_candidates} show stable 
behavior in different polarization properties. Nevertheless there is a small subgroup, 
namely 3C\,286, 3C\,295, 3C\,48 and NGC\,7027, which remain stable in both linear and circular 
polarization throughout the period we examined (2010.5--2016.3). \citet{Perley2013} and 
\citet{Zijlstra2008} show that these sources exhibit stable or well-predicted behavior 
also in Stokes $I$. Therefore they are best suited for the simultaneous calibration of all 
Stokes parameters.

\subsection{Long-term stability of the circular polarization handedness}
\label{subsec:hand_stability}

Previous studies have revealed a general tendency of the circular polarization handedness to remain stable over many 
years \citep[e.g.][]{Komesaroff1984,Homan1999,Homan2001}. The consistency of the circular polarization sign may indicate 
either a consistent underlying ordered jet magnetic field component (e.g. toroidal or helical) or a general 
property of AGN with the sign of circular polarization set by the SMBH/accretion disk system \citep[e.g.][]{Ensslin2003}.

We compared our dataset with the one presented several decades ago in \citet{Komesaroff1984} to investigate the long-term stability of the 
circular polarization handedness, independently of the stability in the amplitude of $m_{\mathrm{c}}$. There are ten common
sources between the two samples. Three of them -- namely PKS\,1127-14, 3C\,273 and 3C\,279 -- show stable circular 
polarization handedness in both datasets with the same sign of $m_{\mathrm{c}}$. 3C\,161 shows stable circular 
polarization handedness in both datasets but with the sign of $m_{\mathrm{c}}$ reversed. The rest of the common sources 
-- namely PKS\,0235+164, OJ\,287, PKS\,1510-089, PKS\,1730-130, CTA\,102 and 3C\,454.3 -- show short term variability 
of the circular polarization handedness in at least one of the two datasets. A thorough presentation of our circular 
polarization dataset and its comparison with previous works will be presented in a future publication.

\section{Discussion and Conclusions} 
\label{sec:discussion}

We presented the analysis of the radio linear and circular polarization of more than 150 sources observed
with the Effelsberg 100-m telescope at 4.85~GHz and 8.35~GHz. The observations cover the period from July 2010
to April 2016 with a median cadence of around 1.2 months. We developed a new methodology for recovering all
four Stokes parameters from the Effelsberg telescope observables. Although our method has been implemented for
an observing system with circularly polarized feeds, it is easily generalizable to systems with linearly polarized 
feeds.

The novelty of our approach relies chiefly on the thorough treatment of the instrumental effects. In fact, our
method aims at correcting the observables already prior to the computation of the Stokes parameters. In
contrast, conventional methodologies -- like the M\"uller matrix -- operate on the Stokes
vector. Consequently, cases of instrumentally corrupted observables that would be conventionally unusable, can
be recovered by the careful treatment of their raw data. Additionally, for the correction of the circular
polarization the M\"uller matrix method requires a good coverage of the parameter space. Our method on the
other hand requires a small number of stable reference sources with the explicit knowledge of $m_{\mathrm{c}}$ 
only for one of them. Finally, the M\"uller matrix method lacks the capacity to 
treat the shortest observation cycle (sub-scan) operating on mean values.

In our method, the instrumental linear polarization -- which is most likely caused by the slight ellipticity of
the circular feed response -- is modeled across the whole beam on the basis of 
the telescope response to unpolarized sources. Each sub-scan is then cleaned of the instrumental contribution, 
separately. Our results indicate that the instrumental linear polarization of the systems we used 
remained fairly stable throughout the period of 5.5 years we examined.

For the treatment of the instrumental circular polarization we introduced two independent methods: the
zero-leveling of the $m_\mathrm{c}$ and the powerful singular value decomposition (SVD) method. Both rely on
very few and easy to satisfy requirements and give very similar results. The results of both methods are also
in agreement with the UMRAO data set.

The clean data are then subjected to a series of operations including the opacity and elevation-gain corrections
which we found to be immune to the incident radiation's polarization state. Moreover, the Airy disk beam pattern
delivers amplitude estimates precise enough to accommodate reliable low circular polarization measurements.

All in all, our methodology allows us to minimize the uncertainties in linear and circular polarization degree at the
level of 0.1--0.2~\%. The polarization angle can be measured with an accuracy of the order of $1\degr$.

We have estimated the instrumental rotation potentially caused by our apparatus by observing the Moon which
has a simple radial configuration of the polarization angle. It provides then an excellent reference for (a)
estimating the instrumental rotation and (b) conducting absolute angle calibration. We found that the
instrument introduces a minute rotation of $1.26\degr$ and $-0.5\degr$ for 4.85~GHz and 8.35~GHz, respectively.
What is however worth noting is that we found evidence that there must be an, at least, mild dependence of the
rotation on the source elevation. For completeness further investigation is worthy despite the marginal
magnitude of the effect.

Despite the conceptual differences between our method and the M\"uller matrix one, we conducted a
quantitative comparison of their effectiveness. We examined the intra-session variability in the linear
polarization parameters. Those should remain unchanged over such short time scales even for intrinsically
variable sources. Our methodology performs significantly better, particularly for low linear polarization 
observations (15--100 mJy polarized flux), where it delivers 28~\% more stable Stokes $Q$ and $U$ results than 
the M\"uller method.

After having reconstructed as accurately as possible the polarization state of our sample, we
searched for sources with stable polarization characteristics. We found five sources with significant and stable
linear polarization. A list of three sources remain constantly unpolarized over the entire period of almost 5.5
years we examined. A total of 11 sources were found to have stable circular polarization degree four of which
with non-zero $m_\mathrm{c}$. One of the sources with stable circular polarization degree, namely PKS\,1127-14, 
was found to be stable and at the same level several decades ago by \citet{Komesaroff1984}, suggesting that its 
$m_\mathrm{c}$ may have remained unchanged for $\sim$40 years. Additionally, we found eight sources that maintain 
a stable polarization angle over the examined period. All this is provided to the community for future polarization 
observations reference.

Finally, we investigated the long-term stability of the circular polarization handedness for the ten common 
sources between our sample and the dataset presented in \citet{Komesaroff1984}. Three sources show stable circular 
polarization handedness in both datasets with the same sign of $m_{\mathrm{c}}$, one with the opposite sign and the 
other six sources show short term variability of the circular polarization handedness in at least one of the two datasets.

\begin{acknowledgements}
  This research is based on observations with the 100-m telescope of the MPIfR (Max-Planck-Institut f\"ur Radioastronomie) at Effelsberg. I.M. and V.K.were funded by the International Max Planck Research School (IMPRS) for Astronomy and Astrophysics at the Universities of Bonn and Cologne. This research was supported in part by funds from NSF grant AST-0607523. The authors thank  A. Roy, the internal MPIfR referee, for his useful comments.
\end{acknowledgements}

\bibliographystyle{aa}
\bibliography{im}

\begin{thebibliography}{39}
\expandafter\ifx\csname natexlab\endcsname\relax\def\natexlab#1{#1}\fi

\bibitem[{Aller \& Aller(2013)}]{Aller2013}
Aller, H.~D. \& Aller, M.~F. 2013, in The Innermost Regions of Relativistic
  Jets and Their Magnetic Fields, Granada, Spain

\bibitem[{Aller {et~al.}(2016)Aller, Aller, Hughes, \& Latimer}]{Aller2016}
Aller, M.~F., Aller, H.~D., Hughes, P.~A., \& Latimer, G.~E. 2016, in HAP
  Workshop: Monitoring the non-thermal Universe, Cochem, Germany

\bibitem[{{Angelakis}(2007)}]{Angelakis2007}
{Angelakis}, E. 2007, PhD thesis, Max-Planck-Institut f{\"u}r Radioastronomie,
  \url{http://hss.ulb.uni-bonn.de/2007/0968/0968.htm}

\bibitem[{{Angelakis} {et~al.}(2015){Angelakis}, {Fuhrmann}, {Marchili},
  {Foschini}, {Myserlis}, {Karamanavis}, {Komossa}, {Blinov}, {Krichbaum},
  {Sievers}, {Ungerechts}, \& {Zensus}}]{2015A&A...575A..55A}
{Angelakis}, E., {Fuhrmann}, L., {Marchili}, N., {et~al.} 2015, \aap, 575, A55

\bibitem[{{Angelakis} {et~al.}(2009){Angelakis}, {Kraus}, {Readhead}, {Zensus},
  {Bustos}, {Krichbaum}, {Witzel}, \& {Pearson}}]{2009A&A...501..801A}
{Angelakis}, E., {Kraus}, A., {Readhead}, A.~C.~S., {et~al.} 2009, \aap, 501,
  801

\bibitem[{Baars {et~al.}(1977)Baars, Genzel, Pauliny-Toth, \&
  Witzel}]{Baars1977}
Baars, J. W.~M., Genzel, R., Pauliny-Toth, I. I.~K., \& Witzel, A. 1977,
  Astronomy and Astrophysics, 61, 99

\bibitem[{Cenacchi {et~al.}(2009)Cenacchi, Kraus, Orfei, \&
  Mack}]{Cenacchi2009}
Cenacchi, E., Kraus, A., Orfei, A., \& Mack, K.-H. 2009, Astronomy and
  Astrophysics, 498, 591

\bibitem[{Chandrasekhar(1950)}]{Chandrasekhar1950}
Chandrasekhar, S. 1950, Oxford, Clarendon Press

\bibitem[{Cohen(1958)}]{Cohen1958}
Cohen, M.~H. 1958, Proceedings of the IRE, 46, 172

\bibitem[{Edelson \& Krolik(1988)}]{Edelson1988}
Edelson, R.~A. \& Krolik, J.~H. 1988, The Astrophysical Journal, 333, 646

\bibitem[{{En{\ss}lin}(2003)}]{Ensslin2003}
{En{\ss}lin}, T.~A. 2003, \aap, 401, 499

\bibitem[{{Fuhrmann} {et~al.}(2016){Fuhrmann}, {Angelakis}, {Zensus},
  {Nestoras}, {Marchili}, {Pavlidou}, {Karamanavis}, {Ungerechts}, {Krichbaum},
  {Larsson}, {Lee}, {Max-Moerbeck}, {Myserlis}, {Pearson}, {Readhead},
  {Richards}, {Sievers}, \& {Sohn}}]{Fuhrmann2016}
{Fuhrmann}, L., {Angelakis}, E., {Zensus}, J.~A., {et~al.} 2016, \aap, 596, A45

\bibitem[{{Golub} \& {Van Loan}(2013)}]{Golub2013}
{Golub}, G.~H. \& {Van Loan}, C.~F. 2013, Matrix computations, 4th edn. (JHU
  Press)

\bibitem[{Heiles(2002)}]{Heiles2002}
Heiles, C. 2002, Single-Dish Radio Astronomy: Techniques and Applications, 278,
  131

\bibitem[{{Heiles} \& {Drake}(1963)}]{Heiles1963}
{Heiles}, C.~E. \& {Drake}, F.~D. 1963, \icarus, 2, 281

\bibitem[{{Homan} {et~al.}(2001){Homan}, {Attridge}, \& {Wardle}}]{Homan2001}
{Homan}, D.~C., {Attridge}, J.~M., \& {Wardle}, J.~F.~C. 2001, \apj, 556, 113

\bibitem[{{Homan} \& {Lister}(2006)}]{Homan2006}
{Homan}, D.~C. \& {Lister}, M.~L. 2006, \aj, 131, 1262

\bibitem[{Homan {et~al.}(2009)Homan, Lister, Aller, Aller, \&
  Wardle}]{Homan2009}
Homan, D.~C., Lister, M.~L., Aller, H.~D., Aller, M.~F., \& Wardle, J. F.~C.
  2009, The Astrophysical Journal, 696, 21

\bibitem[{{Homan} \& {Wardle}(1999)}]{Homan1999}
{Homan}, D.~C. \& {Wardle}, J.~F.~C. 1999, \aj, 118, 1942

\bibitem[{Huang \& Shcherbakov(2011)}]{Huang2011}
Huang, L. \& Shcherbakov, R.~V. 2011, Monthly Notices of the Royal Astronomical
  Society, 416, 2574

\bibitem[{{IEEE Standards Board}(1979)}]{IEEEpol}
{IEEE Standards Board}. 1979, ANSI/IEEE Std 149-1979, 61

\bibitem[{{Jackson}(1998)}]{Jackson1998}
{Jackson}, J.~D. 1998, {Classical Electrodynamics, 3rd Edition} (Wiley-VCH),
  832

\bibitem[{Jones \& O'Dell(1977)}]{Jones1977a}
Jones, T. \& O'Dell, S. 1977, The Astrophysical Journal, 214, 522

\bibitem[{{Klein} {et~al.}(2003){Klein}, {Mack}, {Gregorini}, \&
  {Vigotti}}]{Klein2003}
{Klein}, U., {Mack}, K.-H., {Gregorini}, L., \& {Vigotti}, M. 2003, \aap, 406,
  579

\bibitem[{{Komesaroff} {et~al.}(1984){Komesaroff}, {Roberts}, {Milne},
  {Rayner}, \& {Cooke}}]{Komesaroff1984}
{Komesaroff}, M.~M., {Roberts}, J.~A., {Milne}, D.~K., {Rayner}, P.~T., \&
  {Cooke}, D.~J. 1984, \mnras, 208, 409

\bibitem[{{Kraus}(1966)}]{Kraus1966}
{Kraus}, J.~D. 1966, {Radio astronomy} (Cygnus-Quasar Books)

\bibitem[{Laing(1980)}]{Laing1980a}
Laing, R.~A. 1980, Monthly Notices of the Royal Astronomical Society, 193, 439

\bibitem[{{Lehar} {et~al.}(1992){Lehar}, {Hewitt}, {Burke}, \&
  {Roberts}}]{Lehar1992}
{Lehar}, J., {Hewitt}, J.~N., {Burke}, B.~F., \& {Roberts}, D.~H. 1992, \apj,
  384, 453

\bibitem[{Marscher {et~al.}(2008)Marscher, Jorstad, D'Arcangelo, Smith,
  Williams, Larionov, Oh, Olmstead, Aller, Aller, McHardy,
  L{\"{a}}hteenm{\"{a}}ki, Tornikoski, Valtaoja, Hagen-Thorn, Kopatskaya, Gear,
  Tosti, Kurtanidze, Nikolashvili, Sigua, Miller, \& Ryle}]{Marscher2008}
Marscher, A.~P., Jorstad, S.~G., D'Arcangelo, F.~D., {et~al.} 2008, Nature,
  452, 966

\bibitem[{McKinnon(1992)}]{McKinnon1992}
McKinnon, M.~M. 1992, Astronomy and Astrophysics (ISSN 0004-6361), 260, 533

\bibitem[{{Myserlis}(2015)}]{Myserlis2015}
{Myserlis}, I. 2015, PhD thesis, Max-Planck-Institut f{\"u}r Radioastronomie,
  \url{http://kups.ub.uni-koeln.de/6967/}

\bibitem[{Myserlis {et~al.}(2014)Myserlis, Angelakis, Fuhrmann, Pavlidou,
  Nestoras, Karamanavis, Kraus, \& Zensus}]{Myserlis2014}
Myserlis, I., Angelakis, E., Fuhrmann, L., {et~al.} 2014, eprint
  arXiv:1401.2072

\bibitem[{Ott {et~al.}(1994)Ott, Witzel, Quirrenbach, Krichbaum, Standke,
  Schalinski, \& Hummel}]{Ott1994}
Ott, M., Witzel, A., Quirrenbach, A., {et~al.} 1994, Astronomy and Astrophysics
  (ISSN 0004-6361), 284, 331

\bibitem[{Pacholczyk(1970)}]{Pacholczyk1970a}
Pacholczyk, A.~G. 1970, Series of books in astronomy and astrophysics, -1, xxi,
  269

\bibitem[{Perley \& Butler(2013{\natexlab{a}})}]{Perley2013}
Perley, R.~A. \& Butler, B.~J. 2013{\natexlab{a}}, The Astrophysical Journal
  Supplement Series, 204, 19

\bibitem[{Perley \& Butler(2013{\natexlab{b}})}]{Perley2013a}
Perley, R.~A. \& Butler, B.~J. 2013{\natexlab{b}}, The Astrophysical Journal
  Supplement Series, 206, 16

\bibitem[{{Poppi} {et~al.}(2002){Poppi}, {Carretti}, {Cortiglioni}, {Krotikov},
  \& {Vinyajkin}}]{Poppi2002}
{Poppi}, S., {Carretti}, E., {Cortiglioni}, S., {Krotikov}, V.~D., \&
  {Vinyajkin}, E.~N. 2002, in American Institute of Physics Conference Series,
  Vol. 609, Astrophysical Polarized Backgrounds, ed. S.~{Cecchini},
  S.~{Cortiglioni}, R.~{Sault}, \& C.~{Sbarra}, 187--192

\bibitem[{Wardle {et~al.}(1998)Wardle, Homan, Ojha, \& Roberts}]{Wardle1998}
Wardle, J., Homan, D., Ojha, R., \& Roberts, D. 1998, Nature, 395, 457

\bibitem[{Zijlstra {et~al.}(2008)Zijlstra, van Hoof, \& Perley}]{Zijlstra2008}
Zijlstra, A.~A., van Hoof, P. A.~M., \& Perley, R.~A. 2008, The Astrophysical
  Journal, 681, 1296

\end{thebibliography}

\begin{appendix}

\section{Instrument model for the 4.85~GHz and 8.35~GHz Effelsberg receivers}
\label{app:instr_models}

The functional forms of the instrument models -- one for each Stokes parameter 
and scanning direction -- for the 4.85~GHz receiver are:
\begin{align}
 M_{Q,\mathrm{azi}} &= \alpha_1 I e^{\frac{-[x-(\mu-\beta_1)]^{2}}{2(\gamma_1 \sigma)^{2}}} + \alpha_2 I e^{\frac{-[x-(\mu-\beta_2))]^{2}}{2(\gamma_2 \sigma)^{2}}} \label{eq:Q60_azimuth} \\
 M_{Q,\mathrm{elv}} &= \alpha_1 I e^{\frac{-[x-(\mu-\beta_1)]^{2}}{2(\gamma_1 \sigma)^{2}}} + \alpha_2 I e^{\frac{-[x-(\mu-\beta_2)]^{2}}{2(\gamma_2 \sigma)^{2}}} + \alpha_3 I e^{\frac{-[x-(\mu-\beta_3)]^{2}}{2(\gamma_3 \sigma)^{2}}} \label{eq:Q60_elevation} \\
 M_{U,\mathrm{azi}} &= \frac{\alpha_1 I \left[x-(\mu-\beta_1)\right] e^\frac{-[x-(\mu-\beta_1)]^{2}}{2(\gamma_1 \sigma)^{2}}}{(\gamma_1 \sigma)^{2}} + \alpha_2 I e^{\frac{-[x-(\mu-\beta_2)]^{2}}{2(\gamma_2 \sigma)^{2}}} \label{eq:U60_azimuth} \\
 M_{U,\mathrm{elv}} &= \frac{\alpha_1 I \left[x-(\mu-\beta_1)\right] e^\frac{-[x-(\mu-\beta_1)]^{2}}{2(\gamma_1 \sigma)^{2}}}{(\gamma_1 \sigma)^{2}} + \alpha_2 I e^{\frac{-[x-(\mu-\beta_2)]^{2}}{2(\gamma_2 \sigma)^{2}}} \label{eq:U60_elevation}
\end{align}
and for the 8.35~GHz receiver:
\begin{align}
 M_{Q,\mathrm{azi}} &= \alpha_1 I e^{\frac{-[x-(\mu-\beta_1)]^{2}}{2(\gamma_1 \sigma)^{2}}} + \alpha_2 I e^{\frac{-[x-(\mu-\beta_2)]^{2}}{2(\gamma_2 \sigma)^{2}}} + \alpha_3 I e^{\frac{-[x-(\mu-\beta_3)]^{2}}{2(\gamma_3 \sigma)^{2}}} \label{eq:Q36_azimuth} \\
 M_{Q,\mathrm{elv}} &= \alpha_1 I e^{\frac{-[x-(\mu-\beta_1)]^{2}}{2(\gamma_1 \sigma)^{2}}} + \alpha_2 I e^{\frac{-[x-(\mu-\beta_2)]^{2}}{2(\gamma_2 \sigma)^{2}}} \label{eq:Q36_elevation} \\
 M_{U,\mathrm{azi}} &= \alpha_1 I e^{\frac{-[x-(\mu-\beta_1)]^{2}}{2(\gamma_1 \sigma)^{2}}} \label{eq:U36_azimuth} \\
 M_{U,\mathrm{elv}} &= \frac{\alpha_1 I \left[x-(\mu-\beta_1)\right] e^\frac{-[x-(\mu-\beta_1)]^{2}}{2(\gamma_1 \sigma)^{2}}}{(\gamma_1 \sigma)^{2}} + \alpha_2 I e^{\frac{-[x-(\mu-\beta_2)]^{2}}{2(\gamma_2 \sigma)^{2}}} \label{eq:U36_elevation}
\end{align}
where,
\[
\begin{array}{lp{0.8\linewidth}}
  \mathrm{azi},\mathrm{elv} & the scanning direction \\
  I                & is the measured mean amplitude of the LCP and RCP signals \\
  \mu            & is the measured mean offset of the LCP and RCP signals \\
  \sigma	   & is the measured mean FWHM of the LCP and RCP signals \\
  \alpha_j,\beta_j, \gamma_j & are the fitted parameters for each model with $j=1,2,3$\\
\end{array}
\]
In Table~\ref{tab:QU_artifact_par_init_values} we show a set of initial parameter values that we use as starting point for the fitting algorithm.

\begin{table*}[!ht]
  \caption{The initial values of the Stokes $Q$ and $U$ instrument model parameters for the 4.85~GHz and 8.35~GHz receivers. The functional forms of the model are given in Eqs.\ref{eq:Q60_azimuth}--\ref{eq:U36_elevation}.}
  \label{tab:QU_artifact_par_init_values}
  \centering
  \begin{tabular}{lcccccccccc}
    \hline\hline
Model & $\nu_{\mathrm{obs}}$ & $\alpha_1$ & $\beta_1$ & $\gamma_1$ & $\alpha_2$ & $\beta_2$ & $\gamma_2$ & $\alpha_3$ & $\beta_3$ & $\gamma_3$  \\
      & (GHz) &  &  &  &  & & &  & &  \\
    \hline\\
$M_{Q,\mathrm{azi}}$ & 4.85 & $-$0.005 & $-$70 & 0.5 & $-$0.002 & $-$5   & 0.5    & \ldots & \ldots & \ldots \\
$M_{Q,\mathrm{elv}}$ & 4.85 & 0.003    & 110   & 0.2 & $-$0.010 & 0      & 0.2    & 0.002  & $-$110 & 0.2    \\
$M_{U,\mathrm{azi}}$ & 4.85 & 0.3      & 0     & 0.9 & $-$1     & 40     & 0.5    & \ldots & \ldots & \ldots \\
$M_{U,\mathrm{elv}}$ & 4.85 & $-$0.3   & 5     & 0.3 & $-$0.5   & 0      & 0.2    & \ldots & \ldots & \ldots \\
$M_{Q,\mathrm{azi}}$ & 8.35 & 0.001    & 67    & 0.2 & $-$0.003 & $-$6   & 0.3    & 0.001  & $-$77  & 0.2    \\
$M_{Q,\mathrm{elv}}$ & 8.35 & $-$0.002 & 22    & 0.4 & $-$0.002 & $-$35  & 0.3    & \ldots & \ldots & \ldots \\
$M_{U,\mathrm{azi}}$ & 8.35 & 0.005    & $-$5  & 0.4 & \ldots   & \ldots & \ldots & \ldots & \ldots & \ldots \\
$M_{U,\mathrm{elv}}$ & 8.35 & 0.1      & $-$8  & 0.9 & 0.002    & 44     & 0.3    & \ldots & \ldots & \ldots \\
    \hline
  \end{tabular}
\end{table*}

\section{Feed ellipticity and the measurement of Stokes parameters}
\label{app:feed_ellipticity}

In this appendix we estimate the effect of feed ellipticity on the measurement of Stokes parameters. In the following, 
we provide an elementary approach where several aspects have been oversimplified, e.g. parameters $a$ and $b$ in Eqs.~\ref{eq:E_l_rec} 
and \ref{eq:E_r_rec} are considered real instead of complex numbers. This approach was selected in order to derive a rough estimate of
the effect on the measured parameters. A thorough study of the effect can be found in e.g. \citet{McKinnon1992} or \citet{Cenacchi2009}.

The feed ellipticity can be parameterized as a cross-talk between the left- and right-hand circularly polarized electric field components 
recorded by the system:
\begin{align}
E'_{\mathrm{l}}(t) &= E_{\mathrm{l}}(t) + aE_{\mathrm{r}}(t)= E_{\mathrm{L}} e^{i \omega t} + aE_{\mathrm{R}} e^{i (\omega t + \delta)}, \label{eq:E_l_rec}\\
E'_{\mathrm{r}}(t) &= E_{\mathrm{r}}(t) + bE_{\mathrm{l}}(t)= E_{\mathrm{R}} e^{i (\omega t + \delta)} + bE_{\mathrm{L}} e^{i \omega t}, \label{eq:E_r_rec}
\end{align}
where,
\[
\begin{array}{lp{0.8\linewidth}}
  E_{\mathrm{L,R}}  & the amplitudes of the two incident (orthogonal) circularly polarized electric field components \\ 
  \omega            & the angular frequency of the electromagnetic wave \\
  \delta            & the phase difference between $E_{\mathrm{l}}(t)$ and $E_{\mathrm{r}}(t)$\\
  a, b              & the percentage of the incident left-hand polarized electric field component recorded by the right-hand polarized channel and vice versa.
\end{array}
\]
The primed and unprimed quantities in Eqs.~\ref{eq:E_l_rec}~and~\ref{eq:E_r_rec} refer to the recorded and incident signals, respectively. 
The terms which contain parameters $a$ and $b$ appear due to the ellipticity of the circular feed response (see Eqs.~\ref{eq:E_l} and 
\ref{eq:E_r} for comparison).

Using Eqs.~\ref{eq:E_l_rec}~and~\ref{eq:E_r_rec}, the LCP, RCP, COS and SIN signals can be written as:
\begin{small}
\begin{align}
\mathrm{LCP} &= \left< E'^{*}_{\mathrm{l}} E'_{\mathrm{l}}\right> = \left< E^{2}_{\mathrm{L}} + aE_{\mathrm{L}}E_{\mathrm{R}}e^{i\delta} + aE_{\mathrm{L}}E_{\mathrm{R}}e^{-i\delta} + a^{2}E^{2}_{\mathrm{R}} \right>, \label{eq:LCP_rec}\\
\mathrm{RCP} &= \left< E'^{*}_{\mathrm{r}} E'_{\mathrm{r}}\right> = \left< E^{2}_{\mathrm{R}} + bE_{\mathrm{L}}E_{\mathrm{R}}e^{-i\delta} + bE_{\mathrm{L}}E_{\mathrm{R}}e^{i\delta} + b^{2}E^{2}_{\mathrm{L}} \right>, \label{eq:RCP_rec}\\
\mathrm{COS} &= \left< E'^{*}_{\mathrm{l}} E'_{\mathrm{r}}\right> = \left< E_{\mathrm{L}}E_{\mathrm{R}}e^{i\delta} + bE^{2}_{\mathrm{L}}+ aE^{2}_{\mathrm{R}} + abE_{\mathrm{L}}E_{\mathrm{R}}e^{-i\delta} \right>, \label{eq:COS_rec}\\
\mathrm{SIN} &= \left< E'^{*}_{\mathrm{l}} E'_{\mathrm{r}}\right>_{90\degr} = \left< E_{\mathrm{L}}E_{\mathrm{R}}e^{i(\delta-90\degr)} + bE^{2}_{\mathrm{L}}+ aE^{2}_{\mathrm{R}} + abE_{\mathrm{L}}E_{\mathrm{R}}e^{-i(\delta-90\degr)} \right>, \label{eq:SIN_rec}
\end{align}
\end{small}
where the ``*'' denotes the complex conjugate and the subscript ``$90\degr$'' of Eq.~\ref{eq:SIN_rec} 
denotes that an additional phase difference of $90\degr$ is introduced between $E'_{\mathrm{l}}(t)$ and 
$E'_{\mathrm{r}}(t)$. The real part of Eqs.~\ref{eq:LCP_rec}--\ref{eq:SIN_rec}, which is recorded by the 
system, is:
\begin{align}
\mathrm{LCP} &= \left< E^{2}_{\mathrm{L}} + a^{2}E^{2}_{\mathrm{R}} + 2aE_{\mathrm{L}}E_{\mathrm{R}}\cos{\delta} \right>, \label{eq:LCP_rec_R}\\
\mathrm{RCP} &= \left< E^{2}_{\mathrm{R}} + b^{2}E^{2}_{\mathrm{L}} + 2bE_{\mathrm{L}}E_{\mathrm{R}}\cos{\delta} \right>, \label{eq:RCP_rec_R}\\
\mathrm{COS} &= \left< (ab+1)E_{\mathrm{L}}E_{\mathrm{R}}\cos{\delta} + aE^{2}_{\mathrm{R}} + bE^{2}_{\mathrm{L}} \right>, \label{eq:COS_rec_R}\\
\mathrm{SIN} &= \left< (ab+1)E_{\mathrm{L}}E_{\mathrm{R}}\sin{\delta} + aE^{2}_{\mathrm{R}} + bE^{2}_{\mathrm{L}} \right>, \label{eq:SIN_rec_R}
\end{align}

To derive a rough estimate of the effect on the measurement of the four Stokes parameters, we can further 
simplify the above expressions by assuming that $a \approx b = k$:
\begin{align}
\mathrm{LCP} &= \left< E^{2}_{\mathrm{L}} + k^{2}E^{2}_{\mathrm{R}} + 2kE_{\mathrm{L}}E_{\mathrm{R}}\cos{\delta} \right>, \label{eq:LCP_rec_R_k}\\
\mathrm{RCP} &= \left< E^{2}_{\mathrm{R}} + k^{2}E^{2}_{\mathrm{L}} + 2kE_{\mathrm{L}}E_{\mathrm{R}}\cos{\delta} \right>, \label{eq:RCP_rec_R_k}\\
\mathrm{COS} &= \left< (k^{2}+1)E_{\mathrm{L}}E_{\mathrm{R}}\cos{\delta} + k\left(E^{2}_{\mathrm{R}} + E^{2}_{\mathrm{L}}\right) \right>, \label{eq:COS_rec_R_k}\\
\mathrm{SIN} &= \left< (k^{2}+1)E_{\mathrm{L}}E_{\mathrm{R}}\sin{\delta} + k\left(E^{2}_{\mathrm{R}} + E^{2}_{\mathrm{L}}\right) \right>, \label{eq:SIN_rec_R_k}
\end{align}
The last terms in Eqs.~\ref{eq:COS_rec_R_k}~and~\ref{eq:SIN_rec_R_k} describe a contribution of Stokes 
$I$ $\left(= \left< E^{2}_{\mathrm{R}} \right> + \left< E^{2}_{\mathrm{L}} \right>\right)$ to the measured 
linearly polarized flux density as recorded by the COS and SIN channels. We can estimate the parameter $k$ 
from the observations of linearly unpolarized sources. Such sources have a random phase difference $\delta$, 
which means that $\left<\cos{\delta}\right>=0$ and hence the first terms of Eqs.~\ref{eq:COS_rec_R_k}~and~\ref{eq:SIN_rec_R_k} 
are vanished. The last terms, on the other hand, describe the spurious instrumental linear polarization signals 
that we correct for using the instrument model (see Sect.~\ref{subsec:instrument model}). Thus the parameter $k$ 
is described (across the whole beam) by the instrument model (e.g. Fig.~\ref{fig:all_models}) which is at maximum 
0.005 for the systems we used. Therefore -- in this simplified approach --  we can exclude the second order terms 
of $k$ from Eqs.~\ref{eq:LCP_rec_R_k}--\ref{eq:SIN_rec_R_k} since $k^{2}\rightarrow0$ (it is at maximum 
$2.5 \cdot 10^{-5}$) which results in:
\begin{align}
\mathrm{LCP} &= \left< E^{2}_{\mathrm{L}} + 2kE_{\mathrm{L}}E_{\mathrm{R}}\cos{\delta} \right>, \label{eq:LCP_rec_R_k_final}\\
\mathrm{RCP} &= \left< E^{2}_{\mathrm{R}} + 2kE_{\mathrm{L}}E_{\mathrm{R}}\cos{\delta} \right>, \label{eq:RCP_rec_R_k_final}\\
\mathrm{COS} &= \left< E_{\mathrm{L}}E_{\mathrm{R}}\cos{\delta} + k\left(E^{2}_{\mathrm{R}} + E^{2}_{\mathrm{L}}\right) \right>, \label{eq:COS_rec_R_k_final}\\
\mathrm{SIN} &= \left< E_{\mathrm{L}}E_{\mathrm{R}}\sin{\delta} + k\left(E^{2}_{\mathrm{R}} + E^{2}_{\mathrm{L}}\right) \right>, \label{eq:SIN_rec_R_k_final}
\end{align}

Equations~\ref{eq:LCP_rec_R_k_final}--\ref{eq:SIN_rec_R_k_final} can be used to estimate the effect of the feed 
ellipticity to the measurement of the Stokes parameters. The last terms in Eqs.~\ref{eq:COS_rec_R_k_final}~and~\ref{eq:SIN_rec_R_k_final} 
are removed by the instrumental linear polarization correction scheme as described in Sect.~\ref{subsec:instrument model}. 
Therefore, Stokes $Q$ and $U$ should not affected by the feed ellipticity after this correction step. Stokes $V$ remains also 
unaffected because the subtraction of LCP from RCP removes the last terms of Eqs.~\ref{eq:LCP_rec_R_k_final}~and~\ref{eq:RCP_rec_R_k_final} 
which appear due to the feed ellipticity effect.

Finally, Stokes $I$ seems to be affected by the feed ellipticity effect even in this elementary approach. The recorded Stokes $I$ 
can deviate from the incident one by $4kE_{\mathrm{L}}E_{\mathrm{R}}\cos{\delta}$, which is proportional to the linear polarization 
of the source ($E_{\mathrm{L}}E_{\mathrm{R}}\cos{\delta}=\frac{Q}{2}$). Using the average linearly polarized flux density over all 
our measurements (0.125 Jy) as an estimate of Stokes $Q$ in the above expression we get:
\begin{equation}
4kE_{\mathrm{L}}E_{\mathrm{R}}\cos{\delta} = 4k \frac{Q}{2} \approx 4\cdot0.005\cdot\frac{0.125}{2}=0.00125~~\mathrm{Jy}
\end{equation}
Therefore, the measured Stokes $I$ deviates from the incident one by $\sim$1 mJy on average. This value is much smaller than 
the average uncertainty of our measurements (15--20 mJy, Table~\ref{tab:observations}). In fact, if we use the maximum linearly 
polarized flux density that we have ever measured (2.116 Jy) we calculate that the feed ellipticity effect on Stokes $I$ is at 
maximum 21 mJy, which is comparable to the average uncertainty.

The above discussion shows that, in most cases, the feed ellipticity doesn’t seem to have any measurable effect on the 
Stokes parameter measurements once we account for the instrumental linear polarization (e.g. using the correction scheme 
presented in Sect.~\ref{subsec:instrument model}). The parameter which is mostly affected is Stokes $I$ but the effect 
becomes significant only when the linearly polarized flux density of the observed source is particularly large, e.g. 
$\geq$2 Jy.

\section{Circular polarization curves}

\begin{figure*}[!ht] 
\centering
\begin{tabular}{cc}
\includegraphics[trim={20 20 20 20}, clip, width=0.4\textwidth,angle=0]{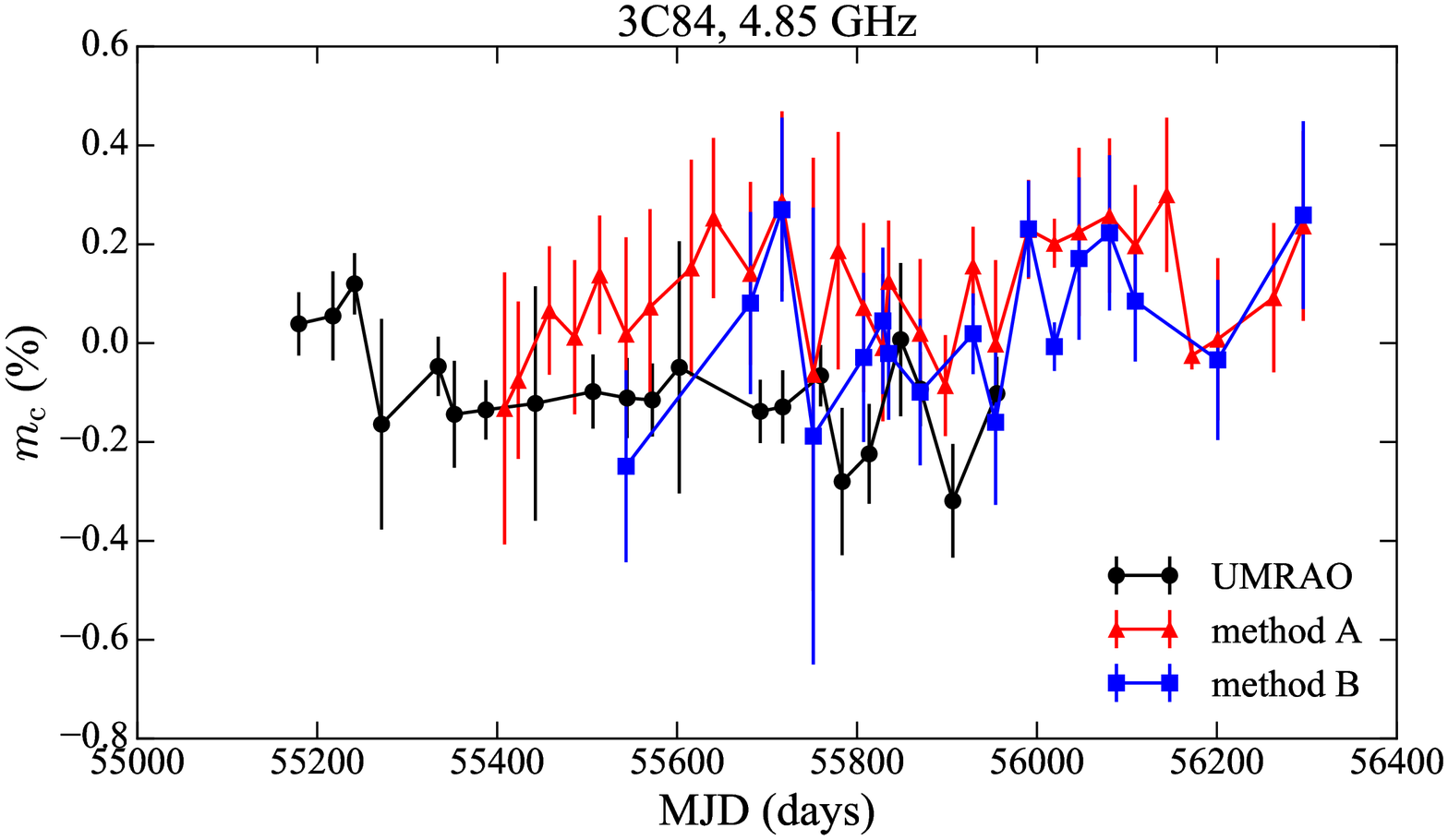} & \includegraphics[trim={20 20 20 20}, clip, width=0.4\textwidth,angle=0]{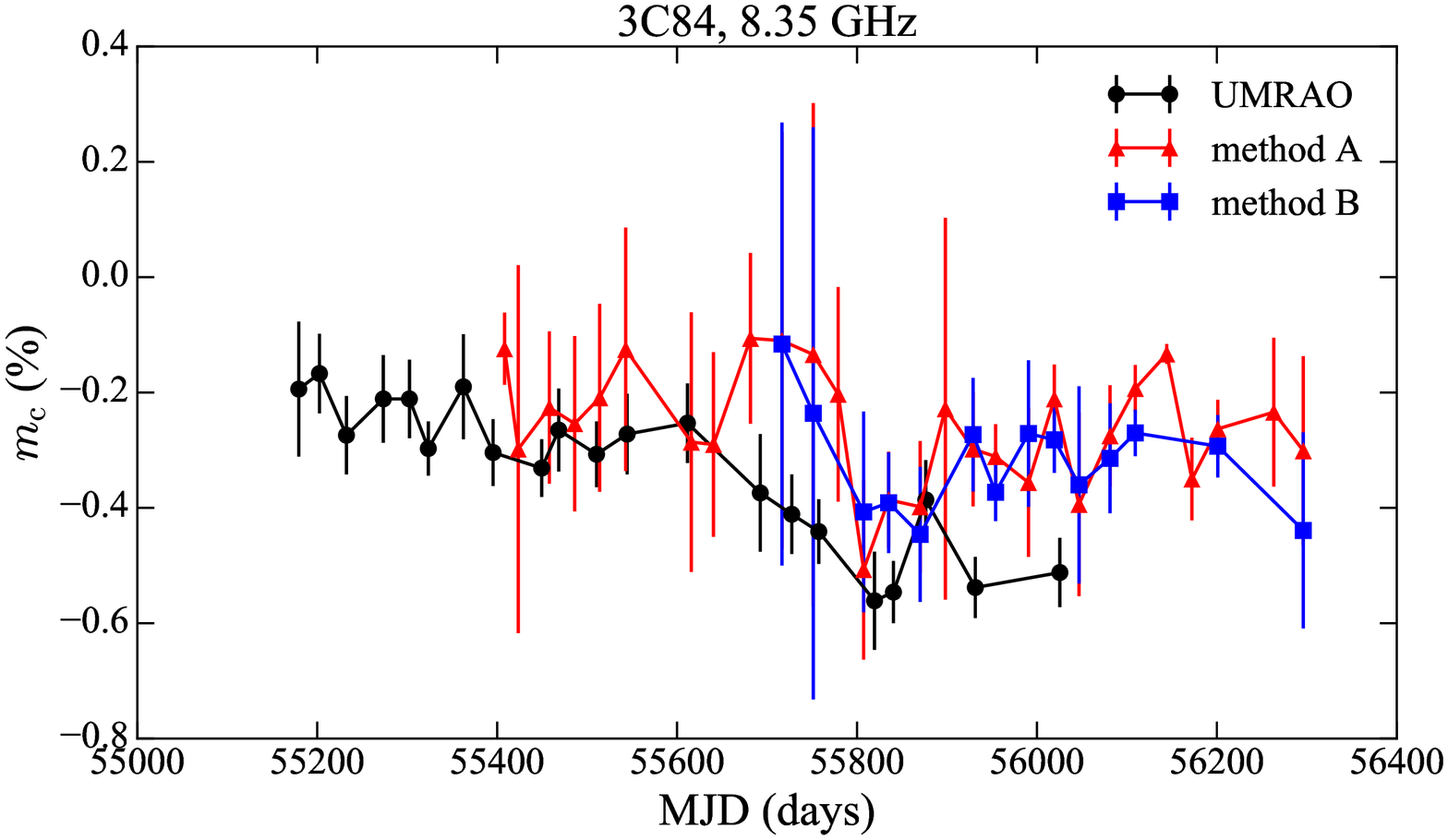} \\
\includegraphics[trim={20 20 20 20}, clip, width=0.4\textwidth,angle=0]{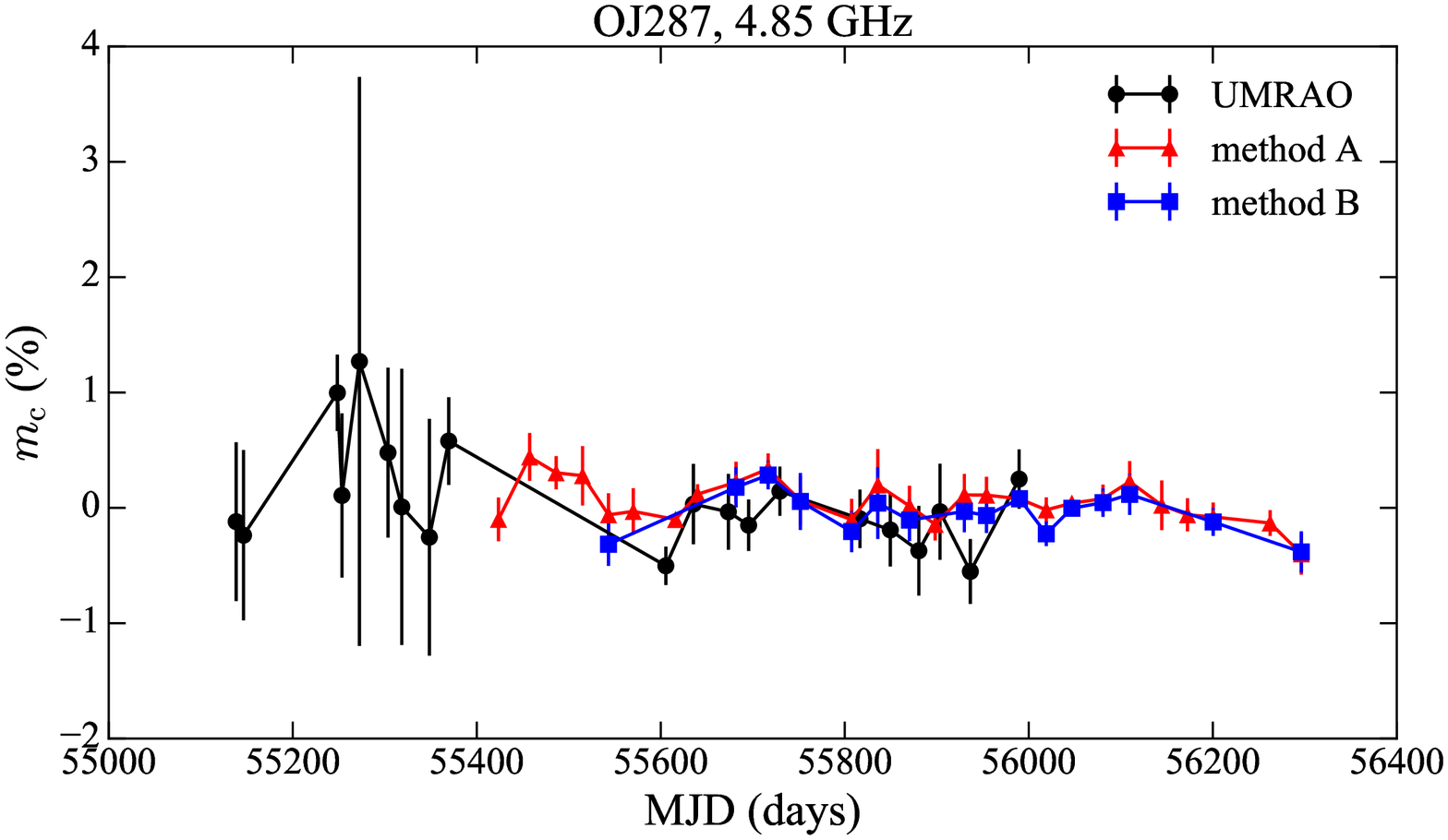} & \includegraphics[trim={20 20 20 20}, clip, width=0.4\textwidth,angle=0]{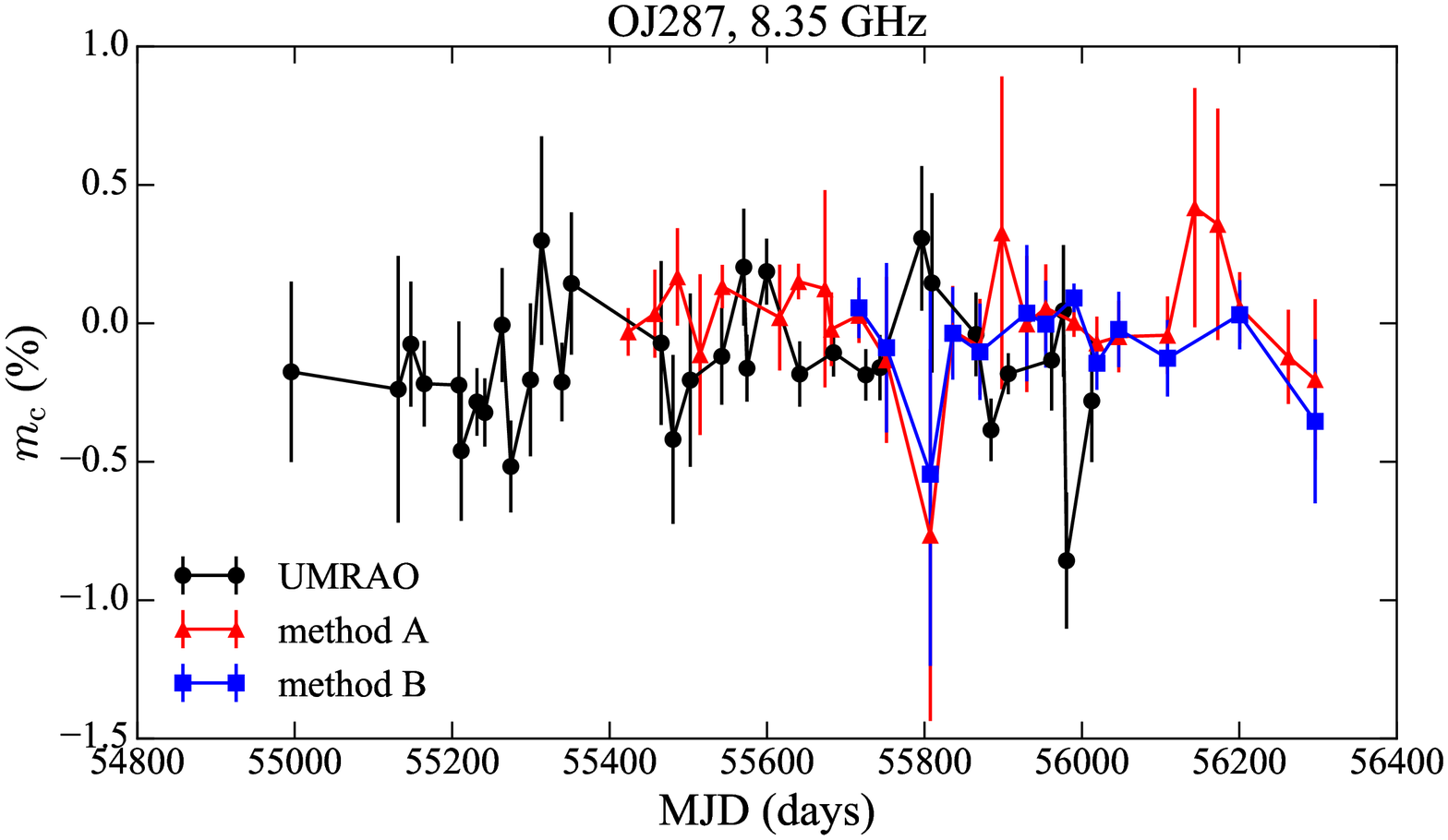} \\
\includegraphics[trim={20 20 20 20}, clip, width=0.4\textwidth,angle=0]{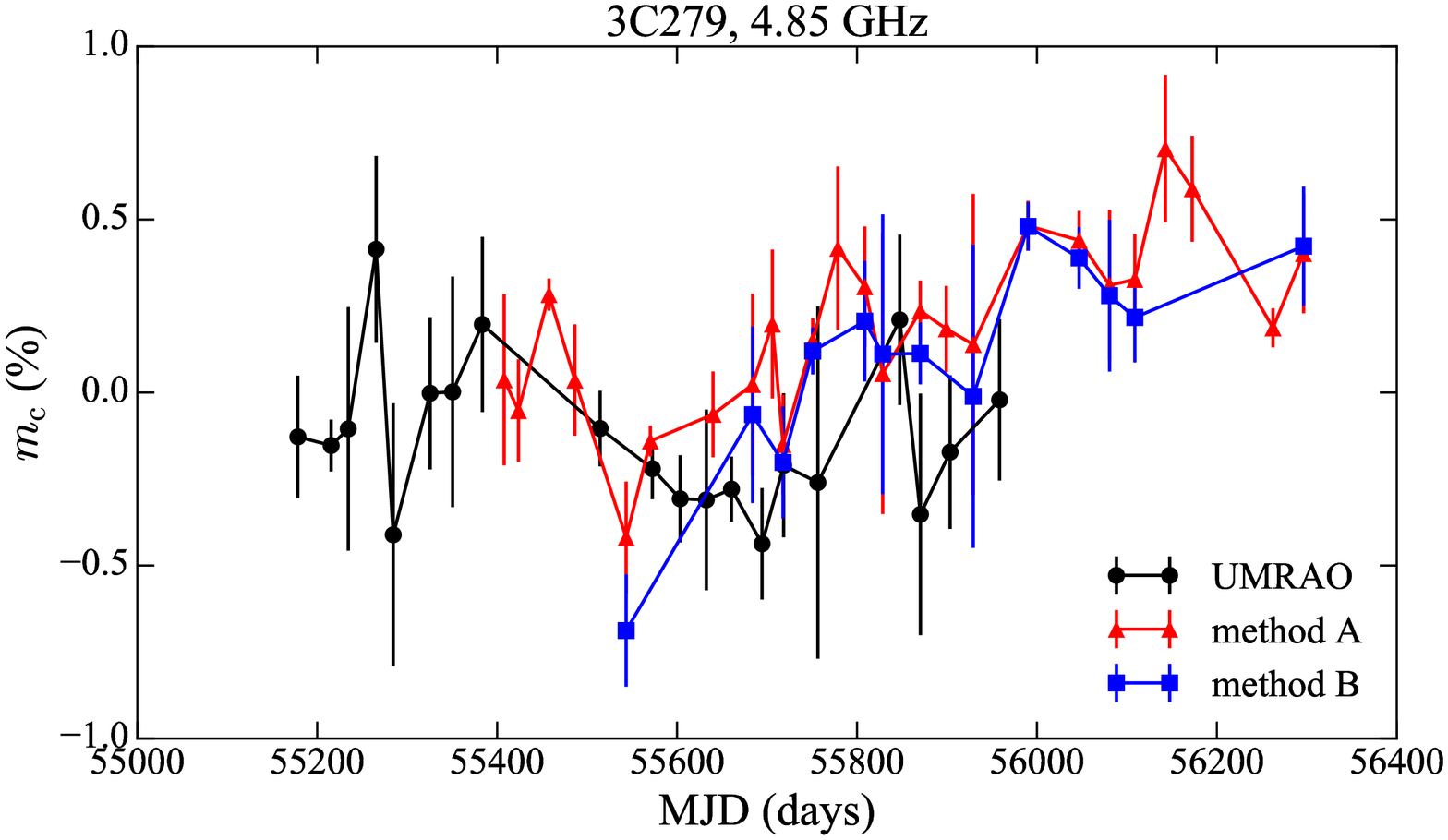} & \includegraphics[trim={20 20 20 20}, clip, width=0.4\textwidth,angle=0]{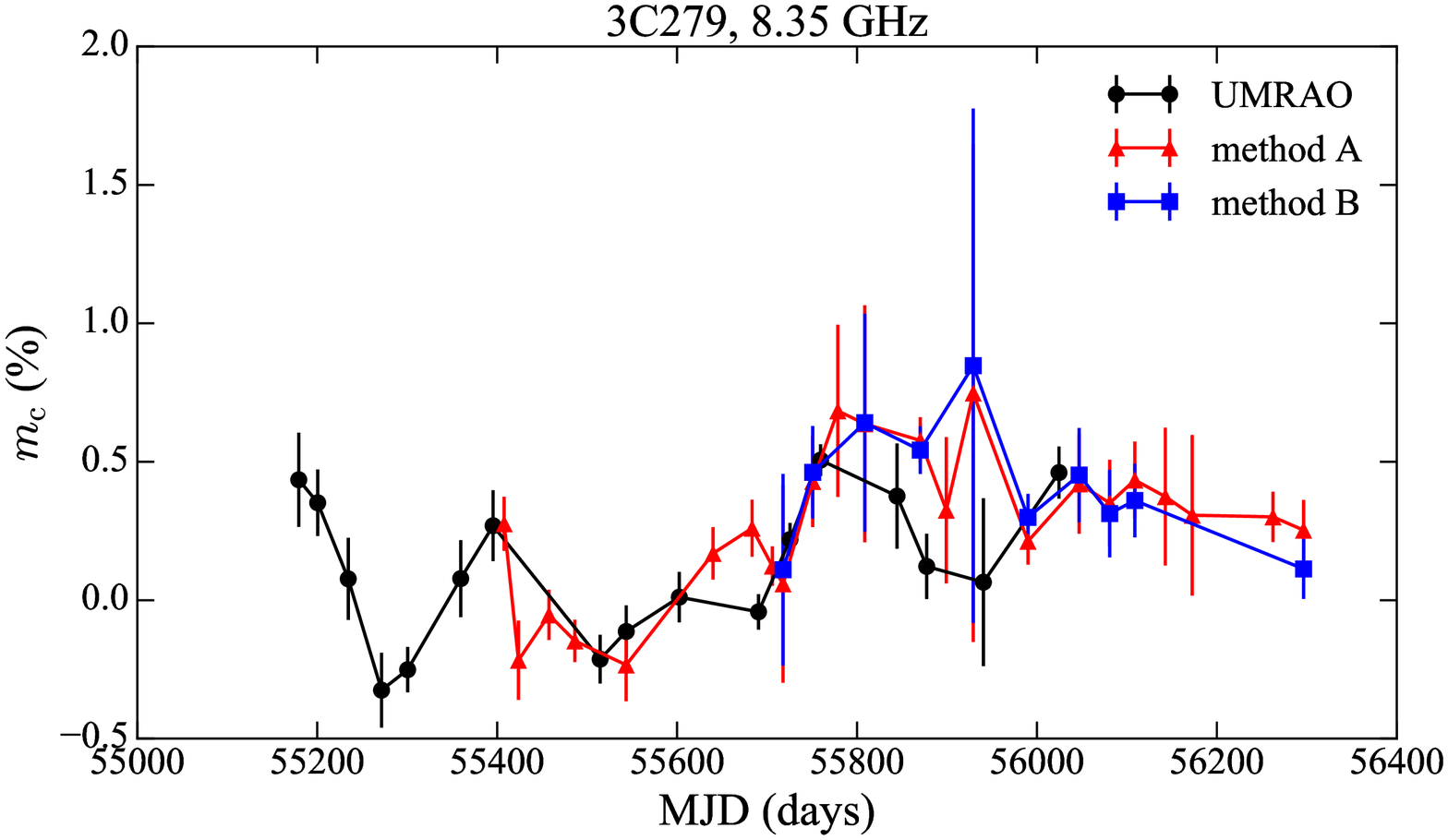} \\
\includegraphics[trim={20 20 20 20}, clip, width=0.4\textwidth,angle=0]{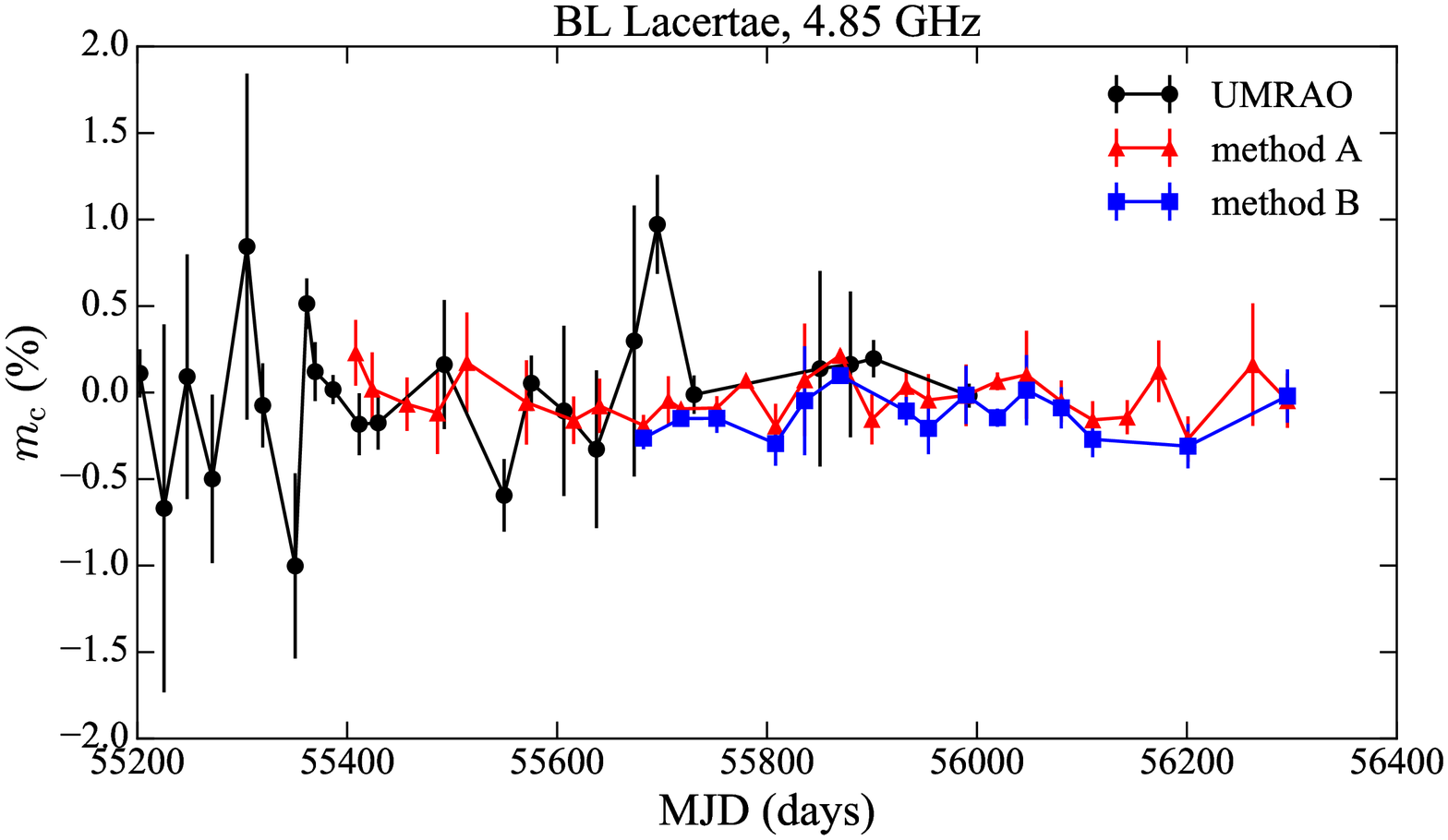} & \includegraphics[trim={20 20 20 20}, clip, width=0.4\textwidth,angle=0]{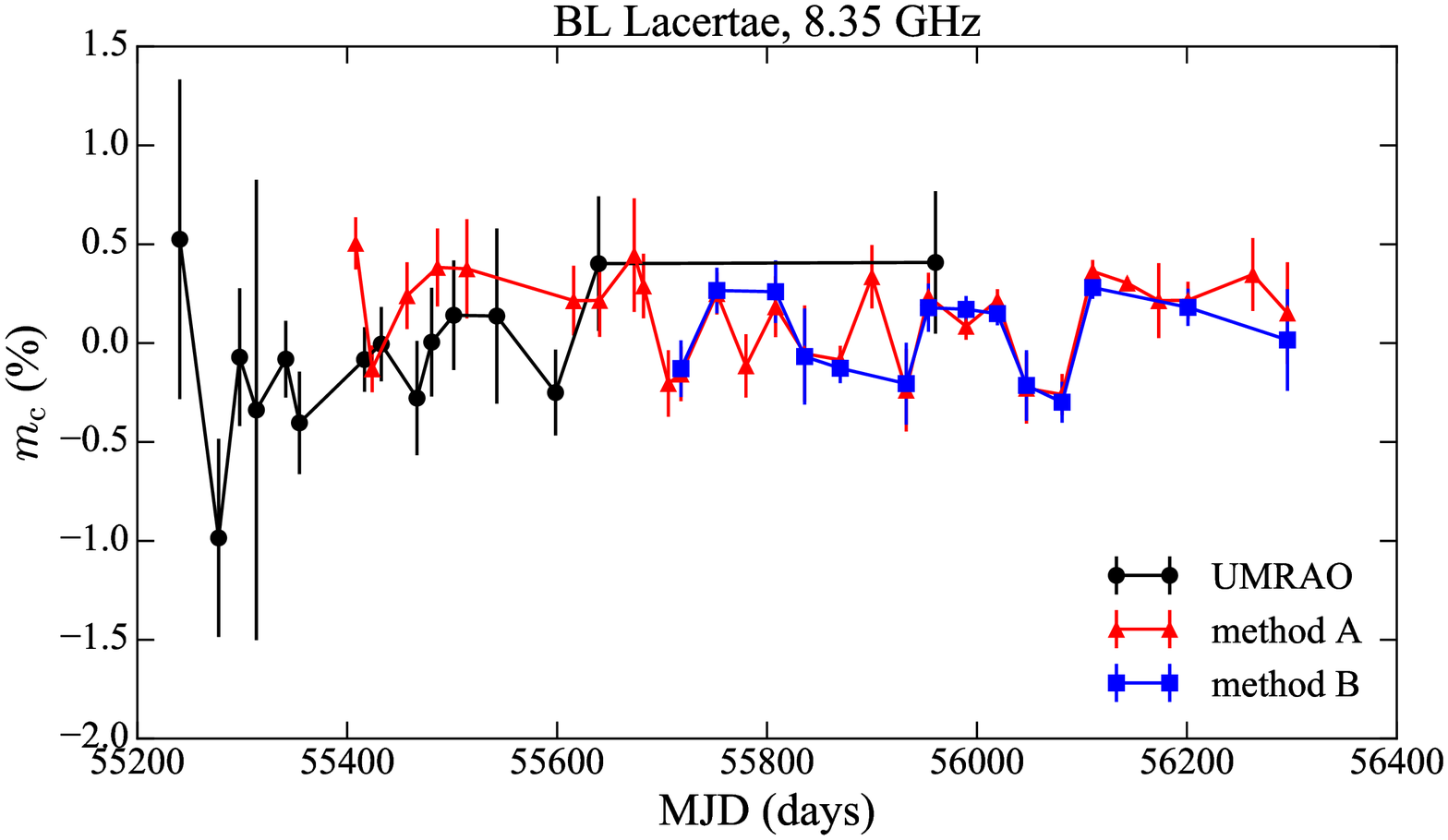} \\
\includegraphics[trim={20 20 20 20}, clip, width=0.4\textwidth,angle=0]{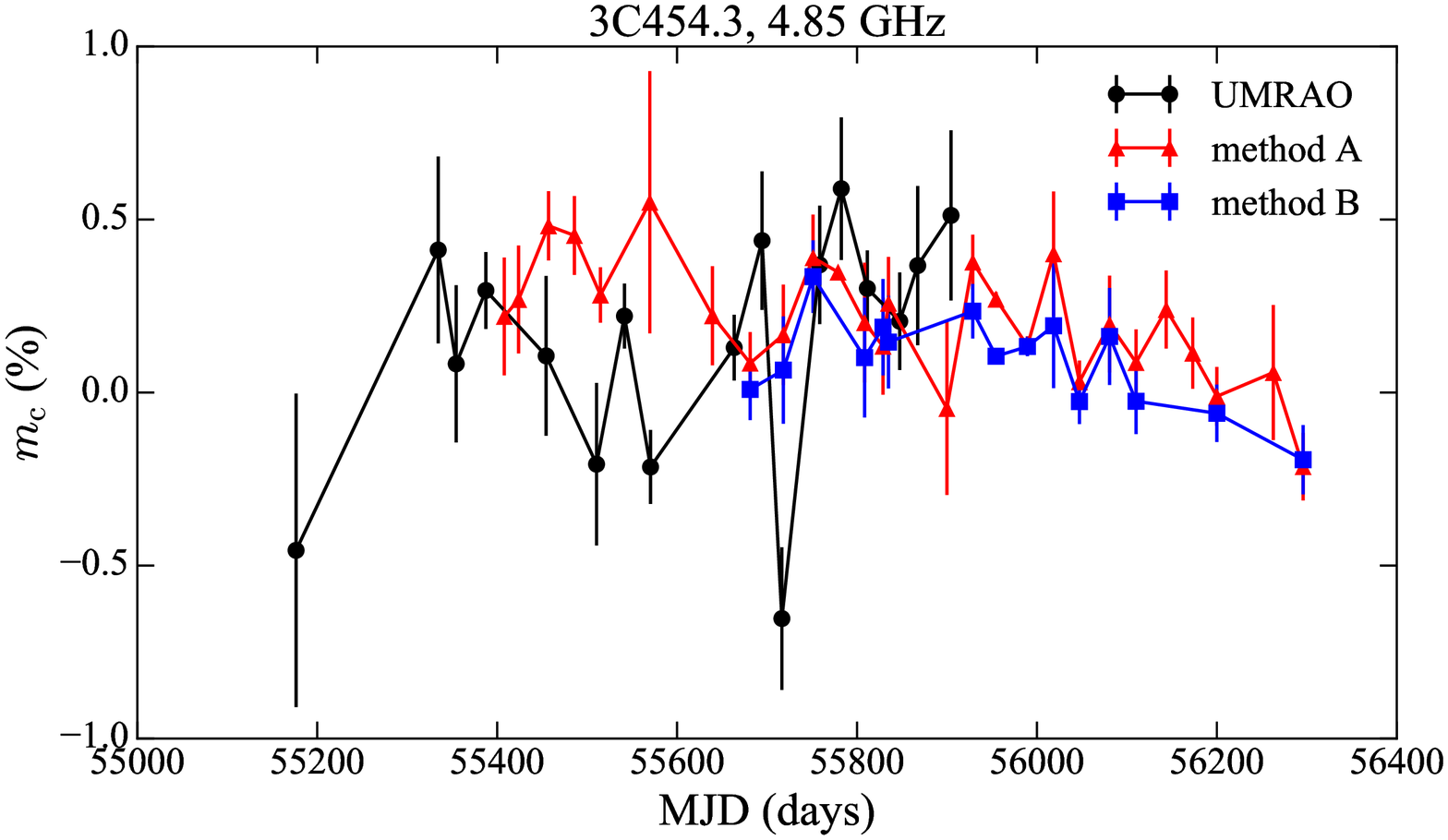} & \includegraphics[trim={20 20 20 20}, clip, width=0.4\textwidth,angle=0]{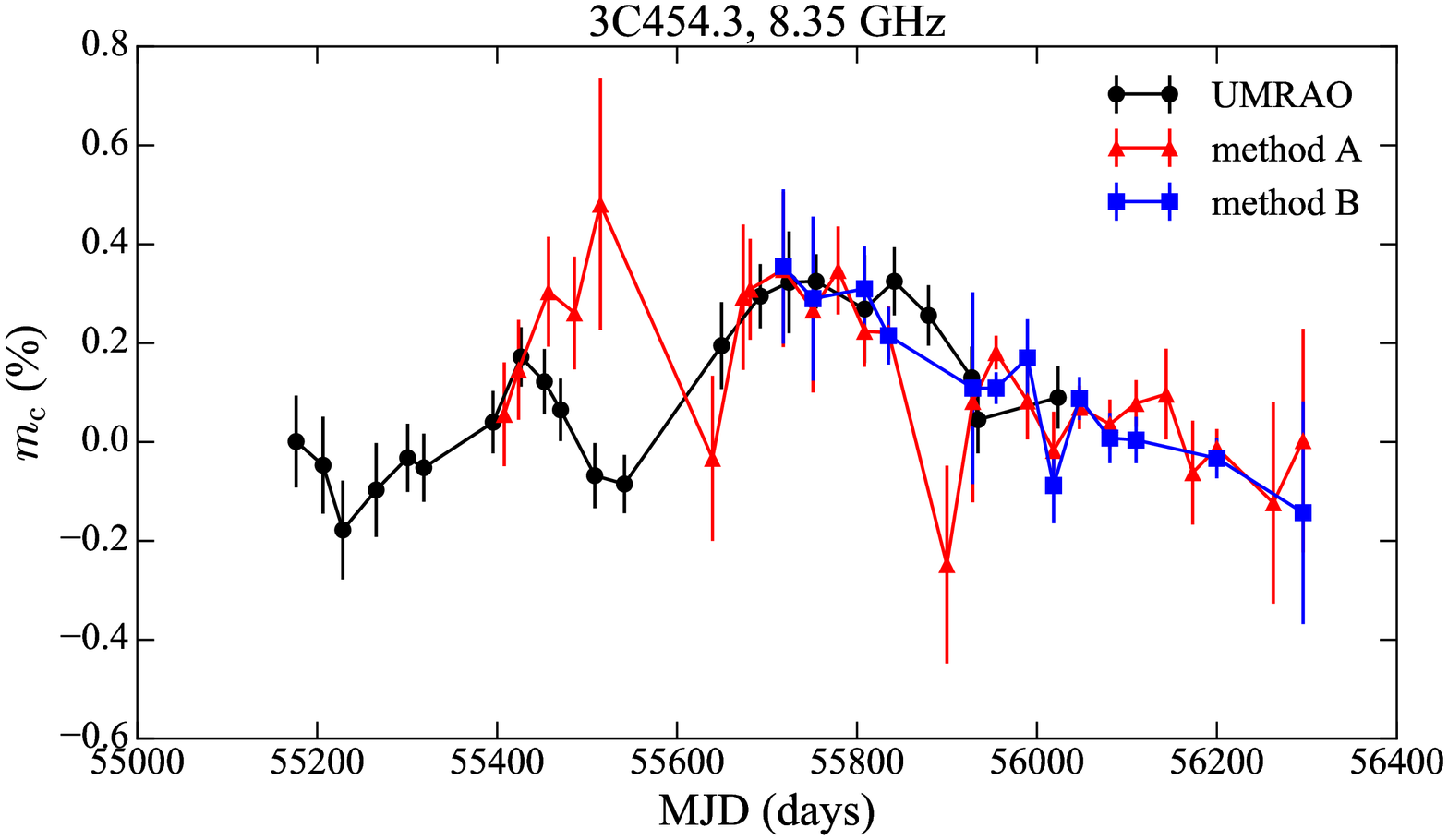} \\ 
\end{tabular}
\caption{Circular polarization curves of the UMRAO dataset over-plotted against our results using both 
calibration methods A and B as described in Sect.~\ref{subsec:LR_gain_corr}. The comparison is performed over five 
sources with overlapping data sets, which were observed at 4.85~GHz (left column) and 8.35~GHz (right column).}
\label{fig:umrao_comparison}
\end{figure*}

In Fig.~\ref{fig:umrao_comparison} we show the circular polarization degree curves for five common sources with the 
UMRAO monitoring program. The two datasets are in very good agreement as described in Sect.~\ref{subsubsec:umrao_comparison}.

\end{appendix}

\end{document}